\documentclass[aps,pre,reprint,longbibliography,showpacs,superscriptaddress,groupaddress,showkeys,floatfix,byrevtex3]{revtex4-1}
\usepackage{graphicx}
\usepackage{dcolumn}
\usepackage{bm}
\usepackage[utf8x]{inputenc}
\usepackage{srctex}
\usepackage{stackrel}
\usepackage{amsmath}
\usepackage{amsfonts}
\usepackage{amssymb}
\usepackage{graphicx}
\usepackage{srctex}
\usepackage{xcolor}
\usepackage{subfigure}
\usepackage{cancel}

\begin{document}
\title{Diffusion of active chiral particles}
\author{Francisco J. Sevilla}
\email[]{fjsevilla@fisica.unam.mx}
\affiliation{Instituto de F\'isica, Universidad Nacional Aut\'onoma de M\'exico,
Apdo.\ Postal 20-364, 01000, M\'exico D.F., Mexico}
\date{Today}

\begin{abstract}
The diffusion of chiral active Brownian particles in three-dimensional space is studied analytically, by consideration of the corresponding Fokker-Planck equation for the probability density of finding a particle at position $\boldsymbol{x}$ and moving along the direction $\hat{\boldsymbol{v}}$ at time $t$, and numerically, by the use of Langevin dynamics simulations. The analysis is focused on the marginal probability density of finding a particle at a given location and at a given time (independently of its direction of motion), which is found from an infinite hierarchy of differential-recurrence relations for the coefficients that appear in the multipole expansion of the probability distribution which contains the whole kinematic information. This approach allows the explicit calculation of the time dependence of the mean squared displacement and the time dependence of the kurtosis of the marginal probability distribution, quantities from which the effective diffusion coefficient and the ``shape'' of the positions distribution are examined. Oscillations between two characteristic values were found in the time evolution of the kurtosis, namely, between the value that corresponds to a Gaussian and the one that corresponds to a distribution of spherical shell shape. In the case of an ensemble of particles, each one rotating around an uniformly-distributed random axis, it is found evidence of the so called effect ``anomalous, yet Brownian, diffusion'', for which particles follow a non-Gaussian distribution for the positions yet the mean squared displacement is a linear function of time.
\end{abstract}


\maketitle

\section{Introduction} 

The transport properties of active or self-propelled particles have received particular attention over the past several years. On the one hand, physicists, both theoreticians and experimentalists, have found a fertile ground to probe and explore ideas regarding the out-of-equilibrium conditions at which active motion occurs. On the other, there are potential applications for the designing and/or controlling the self-propulsion mechanisms which would make possible to manipulate the diffusive properties of such particles at will \cite{GolestanianNJP2007,abbot,MirkovicACSNano2010,SanchezNature2012,kosa,SotoPRL2014,GaoNano2014}.

The out-of-equilibrium element of active systems relies undoubtedly on the single-particle mechanism that give rise to self-propulsion. Such a mechanisms, breaks the fluctuation-dissipation relation \cite{CatesRepProgPhys2012}, which otherwise characterizes the motion of passive Brownian particles by linking in a direct way, the diffusion properties of the particle to the temperature of the surrounding fluid. In practice, the detailed microscopic dynamics of the self-propelling mechanism occurs at a smaller time scales than the corresponding one of the observed pattern of motion. This time-scales disparity allows us to employ a reductionist approach for which the complexity of the self-propelling mechanism can be simplified. 

Such simplification considers the over-damped dynamics for time evolution of the particle's speed, so one can assume that the particle moves at constant speed over a coarse-scale of time at which the pattern of motion is described (see Ref. \cite{DossettiPRL2015} for instance). This approximation is well supported by experimental studies in many real biological systems \cite{BazaziCurrBio2008,BazaziProcRSocLondon2011,BodekerELett2010,EdwardsNature2007,GautraisJMathBio2009,LiPhysBiol2011} where fluctuations around the average value are small.

In regards to the study of pattern of motion observed in active systems, two wide lines of research can be identified, on the one hand, there has been a great interest on the emergent patterns of collective motion of collections of a large number of interacting self-propelled particles. Indeed collections of self-propelled particles are ubiquitous in nature, from micro- to macro-organisms in biology \cite{ViswanathanBook} and more recently in man-made systems where micron-sized particles self-propel by conversion of chemical energy into mechanical one as has been demonstrated in a variety of example \cite{HowsePRL2007, JiangPRL2010}.

On the other hand, the diversity of patterns of motion of single active particles, either biological or synthetic, is wide, particularly in the biological realm, where there are as many of such patterns as species of organisms in nature. Thus, no wonder why the other main line of research focuses on developing the theoretical frameworks to describe such, most of the times complex, patterns of motion exhibited by single active particles \cite{SchnitzerPRE1993,Schienbein1993,BartumeusJTB2008,CodlingJRoySocInterface2008,SevillaPRE2014,TaktikosPlos2014,SevillaPRE2015}. One aspect of interest corresponds to those swimmers, either alive or passive, that show chiral motion, i.e., a well defined state (clockwise or anticlockwise) of the circular motion component of the particle trajectories. As a matter of fact, a plethora of biological organisms \cite{BergBioPhysJ1990,CrenshawAmZoo1996,Crenshaw1BullMathBio1993,Crenshaw2BullMathBio1993,Crenshaw3BullMathBio1993,WoolleyReproduction2003,DiLuzioNature2005,RiedelScience2005,LaugaBioPhysJ2006,HillPRL2007,ShenoyPNAS2007,FriedrichNJP2008,FriedrichPRL2009,SuPNAS2012} and synthetic particles as well \cite{NakataLangmuir1997,DreyfusNatureLetters2005,DharNanoLet2006,SchmidtEBioPhysJ2008,WaltherSoftMat2008,MarinePRE2013} exhibit chiral motion exhibited as helical motion in three dimensions and circular in two.

The processes that lead to chiral motion of active articles may be diverse \cite{ShumPRSocLonA2010,LeoniEPL2010,Ledesma-AguilarEPJE2012,DunstanPhysFluids2012}, the simplest situation in two dimensions corresponds to a geometric effect, that is to say, to the misaligning of the direction of the propelling force and the orientation of the particle axis \cite{VanTeeffelenPRE2008,KummelPRL2013}. A simple effective-force model, that leads to circular patterns of motion, is the inclusion of an effective constant ``torque'' in the \emph{Langevin} equations that drive the orientation of the self-propulsive force \cite{KummelReplyPRL2014}. Such constant torque exerts the particle to rotate with constant angular velocity \cite{VanTeeffelenPRE2008,FriedrichNJP2008,FriedrichPRL2009,NourhaniPRE2013}, leading to circular trajectories in two dimensions and to helical ones in three dimensions. Such torque, for instance, may be externally caused by a magnetic field that act over the magnetic moment of magnetic bacteria or used over nanorods to steer them \cite{KlineAnChem2005}.
This theoretical framework is now standard and has been used in a variety of studies as in the study of the diffusion properties of active particles moving in two dimensions \cite{WeberPRE2011, WeberPRE2012}, of motors having a component of circular motion \cite{MarinePRE2013}, and of the effects of confinement in the diffusion properties of chiral-active particles where directed motion has been observed \cite{RadtkePRE2012,AoEPL2015}. Another approach that has been used to study two-dimensional chiral motion is the \emph{rotationally persistent} random walks, where the introduction of a clockwise or counter-clockwise angular bias at each new step the walker takes \cite{LarraldePRE1997}. A connection, if any, among all these analytical approaches is still missing in the literature and deserves a future analysis.

Analytical studies of diffusion of active particles in three dimensions has been received more less attention than its two-dimensional counterpart. In Ref \cite{SandovalPRE2013}, for instance, the diffusion of torqued, active particles in three-dimensional space is analyzed through overdamped-Langevin equations, which are solved for the time dependence of the first two moments of the particle positions, namely, the average position and the mean square displacement from which the effective diffusion coefficient is computed. In Ref. \cite{LarraldeJPhysA2015} the diffusion properties of swimmers that move in three dimensions with fixed, mean curvature and torsion, are studied by the use of stochastic Frenet-Serret equations which generalizes the deterministic description of helical motion given in \cite{Crenshaw1BullMathBio1993}. A more general instance is studied in Ref. \cite{WittkowskiPRE2012} where a self-propelled Brownian spinning top is considered through the analysis of overdamped-Langevin equations. 

A complete description in three dimensions in terms of the Smoluchowski-like equations is challenging and deserves a thorough analysis even in the absence of chirality. This approach leads us directly to the time evolution of the probability distribution of the particle positions, and from it, to relevant information regarding the characteristic features of the pattern of motion as its non-Gaussian nature \cite{ZhengPRE2013,SevillaPRE2014,SevillaPRE2015}. 

In this work we study the diffusion of active Brownian particles that move freely with constant speed in infinite three-dimensional space subject to an effective torque. We derive  Smoluchowski-like equations that take into account the persistence effects of active Brownian motion and of chirality as well. The equations are derived from the Fokker-Planck equation for the total probability density of finding a particle at position $\boldsymbol{x}$ moving in the direction $\hat{\boldsymbol{v}}$ at time $t$, $P(\boldsymbol{x},\hat{\boldsymbol{v}},t)$ by coarse-graining over the direction of motion. From $P\left(\boldsymbol{x},t\right)$ analytical expression for the mean square displacement and the kurtosis are given. A comparison of our prescription formulas with numerical simulations was carried out by solving the corresponding Langevin-like equations of active particles subject to torques. Our analysis reveals oscillations on time-dependence of the kurtosis in the ballistic regime and for large values of the torque strength. These oscillations points the helical pattern of motion. We also compute the stationary value of the kurtosis for an ensemble of active articles, each particle moving under the effects of an instance of an effective torque uniformly distributed on the sphere. Interestingly this situation exhibit the ``anomalous, yet Brownian, diffusion'' effect, also known as \emph{weakly anomalous diffusion}, where the probability distribution is not Gaussian but the diffusion is normal with a mean squared displacement that grows linearly with time.

This paper continues as follows: In section \ref{SectII} we present the Langevin equations for the trajectories of particles that move with constant velocity and their corresponding Fokker-Planck equation for the probability density $P(\boldsymbol{x},\varphi,t)$ of a particle being at point $\boldsymbol{x}$, moving in the direction $\varphi$ at time $t$ is stated in \ref{SectIII}. In sect the method of analysis is presented \ref{SectIV}. Results are discussed in \ref{SectV}. We finally give our conclusion and final remarks in section \ref{SectVI}.    

\section{\label{SectII}Helical trajectories of chiral active Brownian particles}
We consider a self-propelled microscopic particle, for which the influence of thermal fluctuations due to the surrounding fluid can not be neglected. The interaction with the fluid accounts for both, the Brownian component of the particle motion and the disspative mechanism due to the fluid viscosity. The active component of the particle's motion is accounted as the result of an \emph{active} or \emph{swimming force} \cite{TakatoriPRL2014}, which is defined to be proportional to the particle's swimming velocity, i.e. $\boldsymbol{F}^{\text{swim}}(t)=\zeta \boldsymbol{v}^{\text{swim}}(t)$ where $\zeta$ is the hydrodynamic resistance that couples translational velocity to force given by
$6\pi\eta a$ for a sphere, with $\eta$ the fluid viscosity and $a$ the particle radius. Thus the time evolution of the particle kinematic velocity, $\boldsymbol{v}(t)$ is given by the Langevin equation
\begin{equation*}
\frac{d}{dt}\boldsymbol{v}(t)=-\zeta \boldsymbol{v}(t)+\zeta\boldsymbol{v}^{\text{swim}}(t)+\overline{\boldsymbol{\xi}}(t).
\end{equation*}
For low Reynolds numbers the approximated, overdamped dynamics is valid and last equation is replaced with 
\begin{equation*}
 \frac{d}{dt}\boldsymbol{x}(t)=v_{s}(t)\hat{\boldsymbol{v}}^{\text{swim}}(t)+\boldsymbol{\xi}_{\mathcal{T}}(t),
\end{equation*} 
where $\boldsymbol{\xi}_{\mathcal{T}}(t)=\overline{\boldsymbol{\xi}}(t)$. Thus, the change in time of the particle position is due to the particles internal drive (self-propulsion) and to the influence of stochastic passive fluctuations, $\boldsymbol{\xi}_{\mathcal{T}}(t)$, that randomize the translational motion of the particle.

Last equation is supplemented by additional stochastic differential equations for the swimming velocity $\boldsymbol{v}^{\text{swim}}(t)=v_{s}(t)\hat{\boldsymbol{v}}^{\text{swim}}(t)$, from which the explicit time dependence of the swimming speed $v_{s}(t)$ and the swimming direction $\hat{\boldsymbol{v}}^{\text{swim}}(t)$, are determined \cite{DossettiPRL2015,RomanczukPRL2011}. In the overdamped-speed limit, i.e. when the dynamics that drives the time evolution of $v_{s}(t)$ (around a characteristic, fixed value $v_{0}$), is faster than others in the system, the particle speed can directly be set to $v_{0}$. This leaves to consideration of only one stochastic differential equation that provides the evolution in time of the direction of the swimming velocity, from now on simply denoted with $\hat{\boldsymbol{v}}(t)$. We assume that such evolution in time is only due to \emph{active} fluctuations $\boldsymbol{\xi}_{\mathcal{R}}(t)$, which in many cases surpass thermal ones. Chirality is taken into account by assuming that rotational, active fluctuations does not average zero but a constant, finite value $\boldsymbol{\tau}=\tau_{0}\hat{\boldsymbol{\tau}}$, which gives a fixed direction $\hat{\boldsymbol{\tau}}$ in three dimensional space around the particles rotate with constant angular acceleration $\tau_{0}$.

Under these considerations, the time evolution of the particles kinematic state is therefore given by the following stochastic differential equations 
\begin{subequations}\label{modelo}
 \begin{align}
\frac{d{}}{dt}\boldsymbol{x}(t)&=v_{0}\, \hat{\boldsymbol{v}}(t)+\boldsymbol{\xi}_{\mathcal{T}}(t),\label{position}\\
\frac{d}{dt}\hat{\boldsymbol{v}}(t)&=\boldsymbol{\xi}_{\mathcal{R}}(t)\times\hat{\boldsymbol{v}}(t).\label{direction}
 \end{align}
\end{subequations}
$\boldsymbol{\xi}_{\mathcal{T}}(t)$ and $\boldsymbol{\xi}_{\mathcal{R}}(t)$ are modeled as three-dimensional vectors with Gaussian white noise components, i.e. their entries satisfy $\langle\xi_{\mathcal{T}\mu}(t)\rangle=0$ and $\langle\xi_{\mathcal{T}\mu}(t)\xi_{\mathcal{T}\nu}(s)\rangle=2D_{B}\delta(t-s)\delta_{\mu,\nu}$, for the translational ones, and $\langle\xi_{\mathcal{R}\mu}(t)\rangle=\tau_{\mu}$ and $\langle\xi_{\mathcal{R}\mu}(t)\xi_{\mathcal{R}\nu}(s)\rangle=2D_{\Omega}\delta(t-s)\delta_{\mu\nu}$.
Greek sub-indices denote vector components;  $D_{B}=k_{B}T/6\pi\eta a$, where $T$ and $\eta$ correspond to the temperature and viscosity of the fluid respectively, $k_{B}$ is the Boltzmann constant and $a$ the radius of the particle that has been assumed spherical. $D_{\Omega}$ denotes the active-rotational diffusion constant (temperature independent) that characterizes active noise, and as usual, $\delta(x)$ and $\delta_{\mu\nu}$ denote the Dirac delta and Kronecker delta respectively.

\begin{figure}[h]
 \includegraphics[width=0.3\textwidth,clip=true]{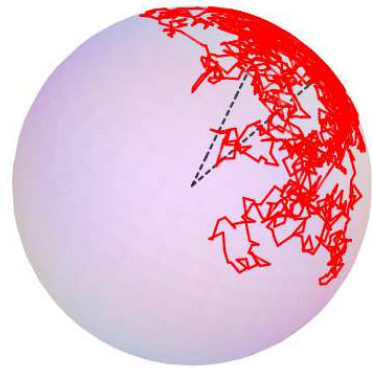}
 \caption{(Color online) A trajectory on the sphere traced by the tip of the unitary vector $\hat{\boldsymbol{v}}(t)$, computed from Eq. \eqref{direction}. The trajectory corresponds to an instance of Brownian motion on the surface of the sphere of radius one.}
 \label{unitsphere}
\end{figure}

The proper integration of equation \eqref{direction} requires the consideration of the explicit multiplicative process involved and that $\vert\hat{\boldsymbol{v}}(t)\vert\equiv\sqrt{\hat{v}_{x}^{2}(t)+\hat{v}_{y}^{2}(t)+\hat{v}_{z}^{2}(t)}=1$ at each time, where $\hat{v}_{\mu}(t)$, $\mu=x,y,z$ are the components of the unitary vector $\hat{\boldsymbol{v}}(t)$. Both aspects are taken into account if the process described by Eq. \eqref{direction} is acknowledged to be equivalent to the Brownian motion of the tip of the unit vector $\hat{\boldsymbol{v}}(t)$ on the unit sphere (see Fig. \ref{unitsphere}). In the interpretation of It\'o \cite{GardinerBook}, Eq. \eqref{direction} can be transformed, with the use of spherical coordinates, into the following pair of stochastic differential equations for the azimuthal, $\varphi(t)$, and polar, $\theta(t)$, angles
\begin{subequations}\label{LangevinIto}
 \begin{multline}\label{polar}
  d\theta(t)=\tau_{0}\sin\theta_{\tau}\sin[\varphi_{\tau}-\varphi(t)]dt+\\
  \frac{D_{\theta}}{\tan\theta(t)}dt+\xi_{\theta}(t)dt
 \end{multline}
 \begin{multline}\label{azimuthal}
  d\varphi(t)=\tau_{0}\{\cos\theta_{\tau}-\sin\theta_{\tau}\cot\theta(t)\cos[\varphi_{\tau}-\varphi(t)]\}dt\\
  +\frac{\xi_{\varphi}(t)}{\sin\theta(t)}dt,
 \end{multline}
\end{subequations}
where $\hat{\boldsymbol{v}}(t)=[\sin\theta(t)\cos\varphi(t),\sin\theta(t)\sin\varphi(t),\cos\theta(t)]$ and the components of the constant vector $\boldsymbol{\tau}$ has been written using the spherical angles, $(\varphi_{\tau},\theta_{\tau})$, as $\boldsymbol{\tau}=$ $\tau_{0}(\sin\theta_{\tau}\cos\varphi_{\tau},\sin\theta_{\tau}\sin\varphi_{\tau},\cos\theta_{\tau})$ with $\tau_{0}$ its magnitude. The stochastic processes $\xi_{\theta}(t)$ and $\xi_{\varphi}(t)$ are Gaussian white noises with zero mean and autocorrelation function $\langle\xi_{\theta}(t)\xi_{\theta}(s)\rangle=2D_{\Omega}\delta(t-s)$ and $\langle\xi_{\varphi}(t)\xi_{\varphi}(s)\rangle=2D_{\Omega}\delta(t-s)$, respectively. One advantage of this formalism is that simple integration schemes, as the Euler one, are numerically stable when applied to equations \eqref{LangevinIto} than when applied directly to \eqref{direction}. 

We reserve the use of variable with an explicit time dependence to denote those stochastic processes that appear in the Langevin Eqs. \eqref{modelo} and \eqref{LangevinIto}, reserving the use of the same symbols, but without the explicit temporal dependence, to the corresponding variables that appear in the Fokker-Planck equation.

\begin{figure}[htb]
\includegraphics[width=0.75\columnwidth,trim=0 0 0 0,clip=true]{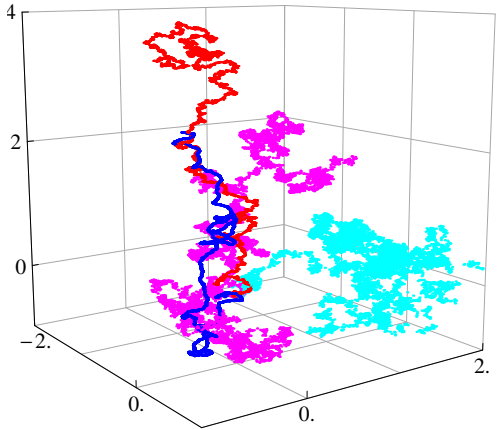}
\caption{(Color online) Helical trajectories of active particles moving with  direction pointing along the $\hat{\boldsymbol{z}}$  and chiral intensity $\bar{\tau}=10$, for different values of the P\'eclet number: $Pe=10^{2}$  (blue), $10^{2}$ (red), 10 (magenta), and 1 (cyan). Axes correspond to the Cartesian coordinates $x$, $y$, $z$ in units of $v_{0}/D_{\Omega}$.}
\label{trajectories}
\end{figure}

Thus our analysis considers the isotropic diffusion process on the sphere, with rotational diffusion coefficient $D_{\Omega}$, which allows us to choose it as a time scale $t_{0}=D_{\Omega}^{-1}$ and the length scale $l_{0}=v_{0}t_{0}$. This choice leads to two free dimensionless parameters, namely: the P\'eclet number $Pe=v_{0}^{2}/D_{B}D_{\Omega}$, which measures the effects of active motion in relation to diffusion, i.e., the larger the P\'eclet number the larger are the persistence effect due to activity (see Fig. \ref{trajectories}); and the strength of the chirality $\tilde{\tau}=\tau_{0}/t_{0}$. Regarding the chirality we consider two cases: i) when this is constant and the same for each particle and ii) when the chirality depends on each particle, i.e. different trajectories realizations correspond to different realizations of noise and chiral direction, in this last case the chirality direction for each particle is chosen from a uniform probability distribution.

Numerical calculations have been carried out by integration of Eqs. \eqref{LangevinIto} using a simple Euler scheme with a time step $10^{-3}t_{0}$, in figure \ref{trajectories} some trajectories are shown for different values of $Pe=10^{3}$ (blue), $Pe=10^{2}$ (red), $Pe=10$ (magenta)  and $Pe=1$ (cyan) with a torque pointing along the $\hat{\boldsymbol{z}}$ direction with magnitude $\tilde{\tau}=10$. Numerical results presented in the following sections were obtained by averaging over $10^{5}$ trajectories.

\section{\label{SectIII}The Fokker-Planck Equation}
In this section we present a Fokker-Planck equation that accounts for the evolution in time of the one-particle probability density, $P({\boldsymbol{x}},\hat{\boldsymbol{v}},t)$, of finding an active, chiral particle diffusing freely in three-dimensional space, at position $\boldsymbol{x}$ and moving in the direction $\hat{\boldsymbol{v}}$ at time $t$. Such an equation can be derived in a simple manner by use of Novikov's theorem \cite{gang,HorsthemkeBook} (see Appendix \ref{appA}). We follow this procedure and not the alternate one of deriving the Fokker-Planck equation from It\'o's interpretation of Eq. \eqref{direction}, since the later would give rise to extra terms not present in the former derivation, terms that usually makes the analysis more difficult. Later on in this paper, the results obtained from the analysis of the Fokker-Planck obtained are compared with the numerical solutions of the Langevin equation \eqref{LangevinIto} in the It\'o interpretation.

In addition, Eq. \eqref{direction} describes the standard diffusion of a point-particle on the surface of the unitary sphere as mentioned before, its corresponding Smoluchowski equation is an instance of the general theory of \emph{Brownian motion on a manifold} developed by van Kampen in Ref. \cite{vanKampenJStatPhys1986}. There, the author analyses the consequences of \emph{geometrical} constraints as long as of \emph{symmetry induced} constraints, on the diffusion of a point particle.

Thus, we start with the Fokker-Planck equation
\begin{widetext}
\begin{multline}\label{FPE}
\frac{\partial}{\partial t}P({\boldsymbol{x}},\hat{\boldsymbol{v}},t)+v_{0}\hat{\boldsymbol{v}}\cdot \nabla P({\boldsymbol{x}},\hat{\boldsymbol{v}},t)=D_{B}\nabla^{2}P({\boldsymbol{x}},\hat{\boldsymbol{v}},t)
+\frac{1}{\sin\theta}\frac{\partial}{\partial\varphi}\left[\left(\hat{\boldsymbol{v}}\times\boldsymbol{\tau}\right)\cdot\hat{\boldsymbol{\varphi}}\, P({\boldsymbol{x}},\hat{\boldsymbol{v}},t)\right]\\
+\frac{1}{\sin\theta}\frac{\partial}{\partial\theta}\left[\sin\theta\left(\hat{\boldsymbol{v}}\times\boldsymbol{\tau}\right)\cdot\hat{\boldsymbol{\theta}}\, P({\boldsymbol{x}},\hat{\boldsymbol{v}},t)\right]
+\mathcal{L}(\hat{\boldsymbol{v}})P({\boldsymbol{x}},\hat{\boldsymbol{v}},t),
\end{multline}
\end{widetext}
where $\nabla=(\partial/\partial x,\partial/\partial y,\partial/\partial z)$, 
\begin{equation}\label{unitv}
\hat{\boldsymbol{v}}=(\sin\theta\cos\varphi,\sin\theta\sin\varphi,\cos\theta),
\end{equation}
and 
\begin{align}
 \hat{\boldsymbol{\theta}}&=(\cos\theta\cos\varphi,\cos\theta\sin\varphi,-\sin\theta)\label{unittheta}\\
 \hat{\boldsymbol{\varphi}}&=(-\sin\theta\sin\varphi,\sin\theta\cos\varphi,0)\label{unitphi}
\end{align}
form the standard set of local covariant vectors that span the tangent space at the surface of the unitary sphere $S^{2}$. $\mathcal{L}(\hat{\boldsymbol{v}})$ is the Laplace-Beltrami operator which is given explicitly by
\begin{equation}
 \mathcal{L}(\hat{\boldsymbol{v}})=D_{\Omega}\left[\frac{1}{\sin\theta}\frac{\partial}{\partial\theta}\left(\sin\theta\frac{\partial}{\partial\theta}\right)+\frac{1}{\sin^{2}\theta}\frac{\partial^{2}}{\partial\varphi^{2}}\right].
\end{equation}

Exact, closed, analytical solutions to equation \eqref{FPE} are not known for the whole time evolution, not even in the long-time or diffusive regime. In unbounded space, natural boundary conditions allow a simplified analysis of Eq. \eqref{FPE} by transforming the spatial coordinates to Fourier ones, $\boldsymbol{x}\rightarrow\boldsymbol{k}$, namely
\begin{multline}\label{FPEFourier}
\frac{\partial}{\partial t}\hat{P}({\boldsymbol{k}},\hat{\boldsymbol{v}},t)+iv_{0}\hat{\boldsymbol{v}}\cdot\boldsymbol{k}\, \hat{P}({\boldsymbol{k}},\hat{\boldsymbol{v}},t)=-D_{B}\boldsymbol{k}^{2}\hat{P}({\boldsymbol{k}},\hat{\boldsymbol{v}},t)\\
+\frac{1}{\sin\theta}\frac{\partial}{\partial\varphi}\left[\left(\hat{\boldsymbol{v}}\times\boldsymbol{\tau}\right)\cdot\hat{\boldsymbol{\varphi}}\, \hat{P}({\boldsymbol{k}},\hat{\boldsymbol{v}},t)\right]\\
+\frac{1}{\sin\theta}\frac{\partial}{\partial\theta}\left[\sin\theta	\left(\hat{\boldsymbol{v}}\times\boldsymbol{\tau}\right)\cdot\hat{\boldsymbol{\theta}}\, \hat{P}({\boldsymbol{k}},\hat{\boldsymbol{v}},t)\right]\\
+\mathcal{L}(\hat{\boldsymbol{v}})\hat{P}({\boldsymbol{k}},\hat{\boldsymbol{v}},t) 
\end{multline}
being
\begin{equation}
 \hat{P}({\boldsymbol{k}},\hat{\boldsymbol{v}},t)=\int \frac{d^{3}\boldsymbol{x}}{(2\pi)^{3/2}}\, e^{-i\boldsymbol{k}\cdot\boldsymbol{x}}P({\boldsymbol{x}},\hat{\boldsymbol{v}},t),
\end{equation}
the unitary Fourier transform of $P({\boldsymbol{x}},\hat{\boldsymbol{v}},t)$ with respect the spatial variable $\boldsymbol{x}$.

Without loss of generality we choose a system of Cartesian coordinates such that $\boldsymbol{\tau}=\tau_{0}\hat{\boldsymbol{z}}=\tau_{0}(\cos\theta\, \hat{\boldsymbol{v}}-\sin\theta\, \hat{\boldsymbol{\theta}})$, thus equation \eqref{FPEFourier} reduces to 
\begin{multline}\label{FPEFourier2}
\frac{\partial}{\partial t}\hat{P}({\boldsymbol{k}},\hat{\boldsymbol{v}},t)+iv_{0}\hat{\boldsymbol{v}}\cdot\boldsymbol{k}\, \hat{P}({\boldsymbol{k}},\hat{\boldsymbol{v}},t)=-D_{B}\boldsymbol{k}^{2}\hat{P}({\boldsymbol{k}},\hat{\boldsymbol{v}},t)\\
-\tau_{0}\frac{\partial}{\partial\varphi}\hat{P}({\boldsymbol{k}},\hat{\boldsymbol{v}},t)\\
+\mathcal{L}(\hat{\boldsymbol{v}})\hat{P}({\boldsymbol{k}},\hat{\boldsymbol{v}},t).  
\end{multline}

We now expand $\hat{P}(\boldsymbol{k},\hat{\boldsymbol{v}},t)$ on the set of eigenfunctions of equation \eqref{FPEFourier2} when $v_{0}$ is set to zero, specifically, on the set of functions $e^{-D_{B}\boldsymbol{k}^{2}t}e^{-\lambda_{n,m}t}\, Y_{n}^{m}(\hat{\boldsymbol{v}}),$ where $\lambda_{n,m}=D_{\Omega}n(n+1)+i\tau_{0}m$, $n=0,\,1,2,\,\ldots$, $m=-n,\,\ldots,\,n$, and $Y_{m}^{n}(\hat{\boldsymbol{v}})$ denotes the spherical harmonic functions that are standardly defined as $(-1)^{m}\sqrt{\frac{(2n+1)}{4\pi}\frac{(n-m)!}{(n+m)!}}  \, P_n^m (\cos{\theta}) \, e^{im\varphi}$, $P_n^m (\cos{\theta})$ being the associated Laguerre polynomial, notice that the explicit dependence on $\theta$ and $\varphi$ is made clear through expression \eqref{unitv}, thus we have
\begin{multline}\label{Expansion}
\hat{P}(\boldsymbol{k},\hat{\boldsymbol{v}},t)=e^{-D_{B}\boldsymbol{k}^{2}t}\sum_{n=0}^{\infty}\sum_{m=-n}^{n}\hat{P}_{n}^{m}(\boldsymbol{k},t)\\
e^{-\lambda_{n,m}t}\, Y_{n}^{m}(\hat{\boldsymbol{v}}),
\end{multline}
where we recognize in the first factor the Fourier transform of the Gaussian probability density
\begin{equation}\label{TranslationalPropagator}
 G_{B}(\boldsymbol{x},t)=\frac{1}{(2 D_{B}t)^{3/2}}\exp\left\{-\frac{\boldsymbol{x}^{2}}{4D_{B}t}\right\}
\end{equation}
due to translational Brownian motion solution of the three-dimensional Diffusion equation, while the second factor encompasses the dynamics that corresponds to the effects due to self-propulsion. Further, expansion \eqref{Expansion} explicitly shows up the time-scale associated with the damping factor of each multipole distribution (spherical harmonic), that contributes to \eqref{Expansion}, the higher the multipole order the faster it decays with time. In fact, in the asymptotic limit, when high multipole distributions have damped, only the rotationally symmetric distribution is expected to remain.

The coefficients $\hat{P}_{n}^{m}({\boldsymbol{k}},t)$, in expression \eqref{Expansion} satisfy $\hat{P}_{n}^{m}({\boldsymbol{k}},t)=(-1)^{m}{\hat{P}}_{n}^{-m*}(-\boldsymbol{k},t)$ and are given explicitly by 
\begin{multline}\label{ExpansionCoeff}
e^{[D_{B}\boldsymbol{k}^{2}+\lambda_{n,m}]t}\times\\
 \int \frac{ d^{3}\boldsymbol{x}}{(2\pi)^{3/2}}\int d\Omega\, e^{-i\boldsymbol{k}\cdot\boldsymbol{x}}\, 
 {Y_{n}^{m}}^{*}(\hat{\boldsymbol{v}})\, P(\boldsymbol{x},\hat{\boldsymbol{v}},t),
\end{multline}
where $d\Omega$ is the infinitesimal element of solid angle on the sphere $\sin\theta\, d\theta d\varphi$. In the spatial coordinates, i.e., in consideration of the inverse Fourier transform of \eqref{Expansion}, the convolution of $G_{B}(\boldsymbol{x},t)$ with the coefficient $P_{n}^{m}(\boldsymbol{x},t)$,
\begin{equation}\label{Pconvolution}
 \mathcal{P}_{n}^{m}(\boldsymbol{x},t)=e^{-\lambda_{n,m}t}
  \int \frac{d^{3}\boldsymbol{x}^{\prime}}{(2\pi)^{3/2}}G_{B}(\boldsymbol{x}-\boldsymbol{x}^{\prime},t)P_{n}^{m}(\boldsymbol{x}^{\prime},t),
\end{equation}
corresponds to the space-dependent multipole of the decomposition, into spherical harmonics, of the distribution of the direction of self-propulsion on the unit sphere. In this way, one should expect the monopole $\mathcal{P}_{0}^{0}({\boldsymbol{x}},t)$ to be the dominant term in the long time limit, for which the distribution $\hat{\boldsymbol{v}}$ is uniform on the unit sphere and leads to the well-known diffusive behavior, at a shorter time regime the dipole distribution denoted as an order one rank tensor, $\left[\mathcal{P}_{1}^{-1}({\boldsymbol{x}},t),\, \mathcal{P}_{1}^{0}({\boldsymbol{x}},t),\, \mathcal{P}_{1}^{1}({\boldsymbol{x}},t)\right]$, that characterizes the polar order of the distribution of $\hat{\boldsymbol{v}}$, must be taken into account. In this regime the effects of persistence are apparent and, at an even shorter time regime, the quadrupole distribution that corresponds to a traceless, symmetric second order rank tensor, which can be written in terms only of $\mathcal{P}_{2}^{\pm2}({\boldsymbol{x}},t),\, \mathcal{P}_{2}^{\pm1}({\boldsymbol{x}},t),\, \mathcal{P}_{2}^{0}({\boldsymbol{x}},t)$ (see appendix), is related to the nematic order of the distribution of $\hat{\boldsymbol{v}}$. Further, one can notice that the expansion \eqref{Expansion} is akin to the expansion in powers of the unit vector $\hat{\boldsymbol{v}}$ \cite{DuderstadtTransportTheory,CatesEPL2013}, namely
\begin{equation}\label{Expansion2}
P(\boldsymbol{x},\hat{\boldsymbol{v}},t)=\varrho(\boldsymbol{x},t)+\boldsymbol{J}(\boldsymbol{x},t)\cdot\hat{\boldsymbol{v}}+\hat{\boldsymbol{v}}\cdot\boldsymbol{Q}(\boldsymbol{x},t)\cdot\hat{\boldsymbol{v}}+\ldots
\end{equation}
Commonly, such expansion is approximately closed at the first two terms (also known as $P_{1}$ approximation \cite{DuderstadtTransportTheory}) that involved the probability density of  $\varrho(\boldsymbol{x},t)$ that equals $\mathcal{P}_{0}^{0}(\boldsymbol{x},t)/\sqrt{4\pi}$ and the current field $\boldsymbol{J}(\boldsymbol{x},t)$ whose components in terms of the dipole distribution are given explicitly in the appendix. This approximation takes into account the persistence effects of motion in various contexts and generally leads to telegrapher-like equations whose validity is restricted to the long-time regime. Description of phenomena at shorter time regimes requires the consideration of higher order terms than the dipole one, which results in a difficult task. Analogously, $\boldsymbol{Q}(\boldsymbol{x},t)$ can be written explicitly in terms of the five independent quadrupole coefficients as given in the appendix. 
To close this paragraph, we want to comment in passing that the transport equation \eqref{FPE} corresponds to the \emph{one-speed diffusion equation} (see Ref. \cite{DuderstadtTransportTheory}), used to describe, in the absence of chirality, the mono-energetic transport process of neutrons and photons in the simplified case for which the scattering of the direction of motion is considered independent of the particles kinetic energy. 
\begin{widetext}
\subsection{The hierarchy equations for $\hat{P}^{m}_{n}(\boldsymbol{k},t)$}
The relation between $\mathcal{P}_{n}^{m}(\boldsymbol{x},t)$ with the coefficients $P_{n}^{m}(\boldsymbol{x},t)$ in \eqref{Pconvolution} allows us to focus on these last ones. Substitution of expansion \eqref{Expansion} into Eq. \eqref{FPEFourier2}, results in a equation that after being multiplied by ${Y}_{n^{\prime}}^{m^{\prime}*}(\hat{\boldsymbol{v}})$ and  integrated over the solid angle $d\Omega$, the following hierarchy of equations for the coefficients $\hat{P}_{n}^{m}(\boldsymbol{k},t)$ are obtained
\begin{equation}\label{SphHarmCoefficients}
 \frac{d}{dt}\hat{P}_{n}^{m}(\boldsymbol{k},t)=-\sum_{n^{\prime}=0}^{\infty}\sum_{m^{\prime}=-n^{\prime}}^{n^{\prime}}\hat{P}_{n^{\prime}}^{m^{\prime}}(\boldsymbol{k},t)\, e^{-(\lambda_{n^{\prime},m^{\prime}}-\lambda_{n,m})t}\int d\Omega\, Y_{n^{\prime}}^{m^{\prime}}(\hat{\boldsymbol{v}})\left[iv_{0}\hat{\boldsymbol{v}}\cdot\boldsymbol{k}\right]{Y}_{n}^{m*}(\hat{\boldsymbol{v}}),
\end{equation}
where the orthogonality property of the spherical harmonics has been used. The integral in \eqref{SphHarmCoefficients} gives the explicit coupling factors among the coefficients $\hat{P}_{n}^{m}(\boldsymbol{k},t)$ owing to the advection term related to self-propulsion, $iv_{0}\hat{\boldsymbol{v}}\cdot\boldsymbol{k}$ in \eqref{FPEFourier}, and is reminiscent of the integrals that commonly appear in quantum mechanics regarding the calculation, to first order in perturbation theory, of the transition dipole moments for an electron of a hydrogenoid atom in an external electromagnetic field. In our case we define $\boldsymbol{k}\cdot\boldsymbol{I}^{m,m^{\prime}}_{n,n^{\prime}}=k_{x}{I_{x}}^{m,m^{\prime}}_{n,n^{\prime}}+k_{y}{I_{y}}^{m,m^{\prime}}_{n,n^{\prime}}+k_{z}{I_{z}}^{m,m^{\prime}}_{n,n^{\prime}}$ where the matrix elements 
\begin{subequations}\label{MatrixElementsI}
 \begin{align}
{I_{x}}^{m,m^{\prime}}_{n,n^{\prime}}&=\int d\Omega\, Y_{n^{\prime}}^{m^{\prime}}(\hat{\boldsymbol{v}}){Y}_{n}^{m*}(\hat{\boldsymbol{v}})\sin\theta\cos\varphi,\\
{I_{y}}^{m,m^{\prime}}_{n,n^{\prime}}&=\int d\Omega\, Y_{n^{\prime}}^{m^{\prime}}(\hat{\boldsymbol{v}}){Y}_{n}^{m*}(\hat{\boldsymbol{v}})\sin\theta\sin\varphi,\\
{I_{z}}^{m,m^{\prime}}_{n,n^{\prime}}&=\int d\Omega\, Y_{n^{\prime}}^{m^{\prime}}(\hat{\boldsymbol{v}}){Y}_{n}^{m*}(\hat{\boldsymbol{v}})\cos\theta,
 \end{align}
\end{subequations}
are explicitly given in the appendix. There, one observes that the coupling factors vanish except when both, $\Delta n \equiv n-n^{\prime}=\pm1$ and $\Delta m\equiv m-m^{\prime}=0,\, \pm1$ are met. Thus we have that for $n\le1$, the coefficient with given pair of indices $(n,m)$ is coupled with only the six \textquotedblleft nearest\textquotedblright coefficient neighbors with indices $(n+1,m\pm1)$, $(n-1,m\pm1)$, $(n\pm1,m)$, and we have explicitly that
\begin{multline}\label{Pnm}
\frac{d}{dt}\hat{P}_{n}^{m}=\frac{v_{0}}{2}e^{-2D_{\Omega}(n+1)t}\left\{\hat{P}_{n+1}^{m+1}\left[\frac{(n+m+2)(n+m+1)}{(2n+1)(2n+3)}\right]^{1/2}e^{-i\tau_{0}t}(k_{y}+ik_{x})\right.\\
+\hat{P}_{n+1}^{m-1}\left[\frac{(n-m+2)(n-m+1)}{(2n+1)(2n+3)}\right]^{1/2}e^{i\tau_{0}t}(k_{y}-ik_{x})\\
-\left.\hat{P}_{n+1}^{m}\left[\frac{(n+m+1)(n-m+1)}{(2n+1)(2n+3)}\right]^{1/2}2ik_{z}\right\}\\
-\frac{v_{0}}{2}e^{2D_{\Omega}nt}\left\{\hat{P}_{n-1}^{m+1}\left[\frac{(n-m)(n-m-1)}{(2n-1)(2n+1)}\right]^{1/2}e^{-i\tau_{0}t}(k_{y}+ik_{x})\right.\\
+\hat{P}_{n-1}^{m-1}\left[\frac{(n+m)(n+m-1)}{(2n-1)(2n+1)}\right]^{1/2}e^{i\tau_{0}t}(k_{y}-ik_{x})\\
+\left.\hat{P}_{n-1}^{m}\left[\frac{(n-m)(n+m)}{(2n-1)(2n+1)}\right]^{1/2}2ik_{z}\right\}.
\end{multline}
For $n=0$ the coefficient with indices $n=0$, $m=0$ is coupled only with the three coefficients with pair of indices $(1,\pm1)$ and $(1,0)$, since $\hat{P}_{n}^{m}\equiv 0$ whenever $n<0$ and/or $\vert m\vert>n$.
\section{\label{SectIV}Equations for the Coarse-grained probability distribution $P(\boldsymbol{x},t)$}
We are interested in deriving the equation, and its corresponding solutions, that dictates the time evolution of the probability density of finding a particle at position $\boldsymbol{x}$ at time $t$ independent of the particle direction of motion, namely 
\begin{equation}\label{Pmarginal}
 P(\boldsymbol{x},t)=\int d\Omega\, P(\boldsymbol{x},\hat{\boldsymbol{v}},t)=\sqrt{4\pi}\, \mathcal{P}_{0}^{0}(\boldsymbol{x},t)=\sqrt{4\pi}\int [d^{3}\boldsymbol{x}^{\prime}/(2\pi)^{3/2}]\, G_{B}(\boldsymbol{x}-\boldsymbol{x}^{\prime},t)P_{0}^{0}(\boldsymbol{x}^{\prime},t),
\end{equation}
where definition \eqref{Pconvolution} has been used. Thus $P(\boldsymbol{x},t)$ is determined by the knowledge of the coefficient $P_{0}^{0}(\boldsymbol{x},t)$ that gives the contribution to the probability density distribution of particles being at $\boldsymbol{x}$ and at time $t$, due to self-propulsion and corresponds to the inverse Fourier transform of $\hat{P}_{0}^{0}(\boldsymbol{k},t)$ which satisfies the equation
\begin{equation}\label{P00}
 \frac{d}{dt}\hat{P}_{0}^{0}=\frac{v_{0}}{2}e^{-2D_{\Omega}t}\left[\left(\frac{2}{3}\right)^{1/2}e^{-i\tau_{0}t}(k_{y}+ik_{x})\hat{P}_{1}^{1}\right.
 +\left(\frac{2}{3}\right)^{1/2}e^{i\tau_{0}t}(k_{y}-ik_{x})\hat{P}_{1}^{-1}
 \left.-\left(\frac{1}{3}\right)^{1/2}2ik_{z}\hat{P}_{1}^{0}
 \right],
\end{equation}
where from now on, we omit arguments of the functions $\hat{P}_{n}^{n}$ whenever possible for the sake of writing-clarity. Equation \eqref{P00} is complemented by the condition $\hat{P}_{0}^{0}(\boldsymbol{k},t)\vert_{\boldsymbol{k}=0}=[\sqrt{2}(2\pi)^{2}]^{-1}$, that follows from the normalization condition for $P(\boldsymbol{x},\hat{\boldsymbol{v}},t)$.
Notice the explicit coupling to the coefficients $\hat{P}_{1}^{\pm1,0},$ these ones satisfy, respectively
\begin{subequations}\label{P1m}
\begin{multline}
 \frac{d}{dt}\hat{P}_{1}^{1}=\frac{v_{0}}{2}e^{-4D_{\Omega}t}\left[\left(\frac{4}{5}\right)^{1/2}e^{-i\tau_{0}t}(k_{y}+ik_{x})\hat{P}_{2}^{2}+\left(\frac{2}{15}\right)^{1/2}e^{i\tau_{0}t}(k_{y}-ik_{x})\hat{P}_{2}^{0}\right. \left.-\left(\frac{1}{5}\right)^{1/2}2ik_{z}\hat{P}_{2}^{1} \right]\\
 -\frac{v_{0}}{2}e^{2D_{\Omega}t}\left(\frac{2}{3}\right)^{1/2}e^{i\tau_{0}t}(k_{y}-ik_{x})\hat{P}_{0}^{0}
\end{multline}
\begin{multline}
 \frac{d}{dt}\hat{P}_{1}^{-1}=\frac{v_{0}}{2}e^{-4D_{\Omega}t}\left[\left(\frac{2}{15}\right)^{1/2}e^{-i\tau_{0}t}(k_{y}+ik_{x})\hat{P}_{2}^{0}+\left(\frac{4}{5}\right)^{1/2}e^{i\tau_{0}t}(k_{y}-ik_{x})\hat{P}_{2}^{-2}\right. \left.-\left(\frac{1}{5}\right)^{1/2}2ik_{z}\hat{P}_{2}^{-1} \right]\\
 -\frac{v_{0}}{2}e^{2D_{\Omega}t}\left(\frac{2}{3}\right)^{1/2}e^{-i\tau_{0}t}(k_{y}+ik_{x})\hat{P}_{0}^{0}
\end{multline}
\begin{multline}
 \frac{d}{dt}\hat{P}_{1}^{0}=\frac{v_{0}}{2}e^{-4D_{\Omega}t}\left[\left(\frac{2}{5}\right)^{1/2}e^{-i\tau_{0}t}(k_{y}+ik_{x})\hat{P}_{2}^{1}+\left(\frac{2}{5}\right)^{1/2}e^{i\tau_{0}t}(k_{y}-ik_{x})\hat{P}_{2}^{-1}\right. \left.-\left(\frac{4}{15}\right)^{1/2}2ik_{z}\hat{P}_{2}^{0} \right]\\
 -\frac{v_{0}}{2}e^{2D_{\Omega}t}\left(\frac{1}{3}\right)^{1/2}2ik_{z}\hat{P}_{0}^{0}
\end{multline}
\end{subequations}
and so on for higher order coefficients. 
Equations \eqref{P1m} can be combined with \eqref{P00} to get 
\begin{multline}\label{IHTE3D}
\frac{d^{2}}{dt^{2}}\hat{P}_{0}^{0}+2D_{\Omega}\frac{d}{dt}\hat{P}_{0}^{0}+\frac{v_{0}^{2}}{3}\boldsymbol{k}^{2}\hat{P}_{0}^{0}=\left(\frac{2}{3}\right)^{1/2}i\tau_{0}\frac{v_{0}}{2}e^{-2D_{\Omega}t}\left[e^{i\tau_{0}t}(k_{y}-ik_{x})\hat{P}_{1}^{-1}-e^{-i\tau_{0}t}(k_{y}+ik_{x})\hat{P}_{1}^{1}\right]+\\
\left(\frac{v_{0}}{2}\right)^{2}e^{-6D_{\Omega}t}\left(\frac{8}{15}\right)^{1/2}\left[e^{-2i\tau_{0}t}(k_{y}+ik_{x})^{2}\hat{P}_{2}^{2}+e^{2i\tau_{0}t}(k_{y}-ik_{x})^{2}\hat{P}_{2}^{-2}\right]-\\
\left(\frac{v_{0}}{2}\right)^{2}e^{-6D_{\Omega}t}\left(\frac{2}{15}\right)^{1/2}4ik_{z}\left[e^{-i\tau_{0}t}(k_{y}+ik_{x})\hat{P}_{2}^{1}+e^{i\tau_{0}t}(k_{y}-ik_{x})\hat{P}_{2}^{-1}\right]+\\
\left(\frac{v_{0}}{2}\right)^{2}e^{-6D_{\Omega}t}\left(\frac{4}{45}\right)^{1/2}2\left(k_{x}^{2}+k_{y}^{2}-2k_{z}^{2}\right)\hat{P}_{2}^{0}.
\end{multline}
\end{widetext}
For the sake of simplicity, initial distributions that corresponds to rotationally symmetric, single pulses, with zero net current are chosen, thus, each instance of the particle trajectory starts at the origin moving along a random direction drawn from a uniform distribution on the sphere, i.e., $P({\boldsymbol{x}},\hat{\boldsymbol{v}},0)=\delta^{(3)}({\boldsymbol{x}})/4\pi$,
where $\delta^{(3)}({\boldsymbol{x}})$ denote the 3-dimensional Dirac delta. The election of this initial condition is intended to explore the nature of the Green function of the related equation for $P(\boldsymbol{x},t)$, and since it results in Fourier space $\hat{P}_{n}^{m}(\boldsymbol{k},0)=\delta_{n,0}\delta_{m,0}[\sqrt{2}(2\pi)^{2}]^{-1}$, where $\delta_{n,m}$ denotes the Kronecker delta, it simplifies our analysis.

Notice that the right hand side of \eqref{IHTE3D} vanishes asymptotically with time, implying that in such a limit only the coefficient $\hat{P}_{0}^{0}(\boldsymbol{k},t)$ of the monopole term of the expansion \eqref{Expansion} (that weighs the uniform distribution of the swimming directions on the unitary sphere) couples to translational motion as was anticipated lines above when Eq. \eqref{Pconvolution} was discussed. Furthermore, in the same limit one can recognize that $\hat{P}_{0}^{0}$ satisfies the three-dimensional \emph{telegrapher's equation} (in Fourier domain $\boldsymbol{k}$) of particles propagating at speed $v_{0}/\sqrt{3}$ and subject to changes in the direction of motion at rate $2D_{\Omega}$, in spatial coordinates it reads  
\begin{equation}\label{TE3D}
\frac{\partial^{2}}{\partial t^{2}}P_{0}^{0}+2D_{\Omega}\frac{\partial}{\partial t}P_{0}^{0}=\frac{v_{0}^{2}}{3}\nabla^{2}P_{0}^{0}.
\end{equation}
Originally introduced by Goldstein \cite{GoldsteinQJMAM1951} in one dimension and later analyzed by Bourret \cite{BourretCJP1960, BourretCJP1961}, Eq. \eqref{TE3D} generalizes the diffusion equation in that it properly accounts for the finite speed signal propagation that results into a non-Gaussian probability density functions of the particle positions. In the situation studied in this paper, the physics that underlies the origen of Eq. \eqref{TE3D} in  the long-time regime, corresponds to the persistence effects induced by the isotropic distribution of swimming directions. Indeed, it is clear that during a time interval $\Delta t\ll D_{\Omega}^{-1}$, the particle displaces itself with a swimming direction that deviates uniformly, only in a small amount solid angle $\Delta S$ that depends on $\Delta t$, this process generalizes the one-dimensional model in the continuum, of particles  moving with constant speed and changing directions (left, right) at a constant rate \cite{KenkreSevilla2007}. The transport properties described by the telegrapher's equation have been discussed in different contexts and in various dimensions, however, except for the one dimensional case for which it gives a proper description of particles that move at constant speed and change direction of motion at a rate $D_{\Omega}$, in higher dimensions gives a correct description only in the long-time regime when the persistence effects are small. 

The solution to the homogeneous part of Eq. \eqref{IHTE3D} is given by
\begin{equation}
 \hat{P}_{0}^{0}(\boldsymbol{k},t)=\hat{P}_{0}^{0}(\boldsymbol{k},0)e^{-D_{\Omega}t}\left[D_{\Omega}\frac{\sin(\omega_{k}t)}{\omega_{k}}+\cos(\omega_{k}t)\right],
\end{equation}
where the dispersion relation for kinematic motion is 
\begin{equation}\label{omegak}
 \omega_{k}^{2}=c^{2}\boldsymbol{k}^{2}-D_{\Omega}^{2}
\end{equation}
$c=v_{0}/\sqrt{3}$ being the propagation speed.

In spatial coordinates the solution is given by 
\begin{multline}
 P_{0}^{0}(\boldsymbol{x},t)=e^{-D_{\Omega}t}\int d^{3}\boldsymbol{x}^{\prime}\left[D_{\Omega}+\frac{\partial}{\partial t}\right]\times\\
 G_{Tel}(\boldsymbol{x}-\boldsymbol{x}^{\prime},t)P_{0}^{0}(\boldsymbol{x}^{\prime},0)
\end{multline}
where $G_{Tel}(\boldsymbol{x},t)$ is the propagator defined by the inverse Fourier transform of $\sin(\omega_{k}t)/\omega_{k}$ given explicitly by
\begin{multline}
 G_{Tel}(\boldsymbol{x},t)=\frac{\pi^{1/2}}{2^{1/2}c\vert\boldsymbol{x}\vert}\left[\frac{D_{\Omega}}{c}\frac{\vert\boldsymbol{x}\vert}{\sqrt{\vert\boldsymbol{x}\vert^{2}-c^{2}t^{2}}}\right.\\
 J_{1}\left(\frac{D_{\Omega}}{c}\sqrt{\vert\boldsymbol{x}\vert^{2}-c^{2}t^{2}}\right)\,u\left(ct-\vert\boldsymbol{x}\vert\right)+\\
 \left.\delta\left(ct-\vert\boldsymbol{x}\vert\right)\right]
\end{multline}
$u\left(\tau\right)$ being the step or Heaviside function taking the value 1 for $\tau>0$ and zero otherwise. In the short-time regime, the telegrapher's equation \eqref{TE3D} describe wave-like solutions, which according to J. D. Barrow \cite{BarrowPhilTransRoySocLond1983}, favors three dimensions for signal fidelity transmission as a part of the anthropic principle. Barrow's argument is based on the fact that in three dimensions, the wave-equation has as solution the one given by Kirchoff, which in contrast to ones in one and two dimensions, has a domain of dependence consisting only be the surface of the sphere of radius $ct$, and therefore, concluding that all three-dimensional wave phenomena travel only at the wave speed $c$. We controvert this conclusion by comparison of the results obtained in the short-time regarding the propagation of self-propelled particles.

\section{\label{SectV} Results}
We first analyze the simplified case that corresponds to the absence of chirality, though being the most simple situation in the present study, the analytical expression obtained are of  
enough interest to be discussed in detail.

\paragraph{Isotropic case (no chirality effects) }In the isotropic case [put $\tau_{0}=0$ in Eq. \eqref{IHTE3D}], the time evolution of the probability density function $\hat{P}_{0}^{0}$ is directly coupled only to the $P_{2}^{\pm2,\pm1,0}$ coefficients [notice that the first term in the right hand side of Eq. \eqref{IHTE3D} is proportional to $\tau$ and therefore vanishes when $\tau=0$], any attempt to solve exactly equation \eqref{IHTE3D} seems meaningless since it requires the solution of the infinite hierarchy \eqref{Pnm}. However, an approximated solution for \eqref{IHTE3D} that is accurate up to the fourth moment can be obtained by cutting off the hierarchy, holding up to the $\hat{P}_{2}^{\pm2,\pm1,0}$ coefficients and neglecting higher ones. This approach goes beyond the standard $P_{1}$ or \emph{dipole} approximation in that it considers the quadrupole effects related to the nematic order of the distribution of the self-propelling direction of motion through the $P_{2}$ coefficients.	

Before attempting to obtain the isotropic solutions, we want to remark one aspect of Eq. \eqref{IHTE3D} when $\tau_{0}=0$, namely, one can show that the inhomogeneous term that involve the $\hat{P}_{2}^{\pm2,\pm1,0}$ coefficients do not contribute to the calculation of the second moment of $\hat{P}_{0}^{0}$ (this has bee the case in 2 dimensions, see \cite{SevillaPRE2015}), from which it can be concluded that the solution to the telegrapher's equation \eqref{TE3D} approximates the exact probability density function in that it gives the exact time dependence of the mean square displacement induced by rotational diffusion (second moment of $P_{0}^{0}$). Such approximated solution retains a finite signal speed propagation and therefore the shape is not Gaussian. The larger the time the better is the approximation as can be checked from the fact that in the long time regime the terms proportional to $e^{-6D_{\Omega}t}$ in \eqref{IHTE3D} can be  neglected. It is clear, thus, that the next higher moments are well approximated by the telegrapher equation only in the asymptotic limit, breaking down the short time regime. Situations like this are frequently encountered in transport theory \cite{KenkreSevilla2007} and deserve a more deep analysis.

From the definition of $P(\boldsymbol{x},t)$ [\eqref{Pmarginal}] we have that its second moment is given by 
\begin{equation}\label{TotalMSD}
 \langle\boldsymbol{x}^{2}(t)\rangle=6D_{B}t+\langle\boldsymbol{x}^{2}(t)\rangle_{0}
\end{equation}
where $\langle\boldsymbol\cdot\rangle_{0}$ denotes the average of $(\cdot)$ with respect to the distribution of the positions  $\sqrt{4\pi}\, P_{0}^{0}(\boldsymbol{x},t)$. Expression \eqref{TotalMSD} is valid in general, where the first term in the rhs gives simply the contribution from translational diffusion $\langle\boldsymbol{x}^{2}(t)\rangle_{B}=6D_{B}\, t,$ while the second term gives the contribution due to the persistence effects of active motion, in the present case, due to rotational active diffusion. The observation made in the previous paragraph, makes the second moment of $\sqrt{4\pi}\, P_{0}^{0}(\boldsymbol{x},t)$ be obtained	 directly from Eq. \eqref{TE3D}, which leads to the following equation for $\langle\boldsymbol{x}^{2}(t)\rangle_{0}$
\begin{equation}
 \frac{d^{2}}{dt^{2}}\langle\boldsymbol{x}^{2}(t)\rangle_{0}+2D_{\Omega}\frac{d}{dt}\langle\boldsymbol{x}^{2}(t)\rangle_{0}=2v_{0}^{2},
\end{equation}
whose solution for the initial condition $\langle\boldsymbol{x}^{2}(t)\rangle_{0}=0$ is
\begin{equation}\label{ActiveMSD}
\langle\boldsymbol{x}^{2}(t)\rangle_{0}=6D_{A}\left[t-\frac{1}{2D_{\Omega}}\left(1-e^{-2D_{\Omega}t}\right)\right].
\end{equation}
The diffusion coefficient corresponding to active motion is obtained by taking the long time limit of expression \eqref{ActiveMSD} and is given by $D_{A}=v_{0}^{2}/6D_{\Omega}$, which gives the rate at which the variance of the position distribution grows due to the rotational diffusion at rate $2D_{\Omega}$ of particles that move at speed $v_{0}/\sqrt{3}$. In the short time limit, the ballistic regime expression \eqref{ActiveMSD} reduces to $v_{0}^{2}t^{2}.$

The total mean-square displacement is then given by  
\begin{equation}\label{MSD}
 \langle\boldsymbol{x}^{2}(t)\rangle=6\left(D_{B}+D_{A}\right)t-\frac{v_{0}^{2}}{2D_{\Omega}^{2}}\left(1-e^{-2D_{\Omega}t}\right)
\end{equation}
from which the effective diffusion constant is obtained in the asymptotic limit, namely $D_{\text{eff}}^{0}=D_{B}+D_{A}$, expression that coincides with the one calculated from the Kubo formula \cite{HowsePRL2007}. This enhancement of diffusion due to self-propulsion over the passive value $D_{B}$ has been pointed out theoretically \cite{TailleurEPL2009, encu} and corroborated experimentally \cite{PalacciPRL2010,MaggiSoftMatt2013} in the case of non-interacting active particles and in situations where the effects of confinement are unimportant. Under this simplifications, an effective temperature $T_{e}$ can be correspondingly introduced through the relation $k_{B}T_{e}=k_{B}(T+T_{A})$, where the active temperature $T_{A}$ is defined as $6\pi\eta a\, v_{0}^{2}/D_{\Omega}$, and which expresses the fact that in the asymptotic regime, active Brownian motion can be thought as passive Brownian motion in a homogeneous, hotter bath. 

In the short time limit, $D_{\Omega}t\ll1$, on the other hand, the mean-square displacement has the expression $\langle\boldsymbol{x}^{2}\rangle\approx v_{0}^{2}t^{2}(1+6D_{B}/v_{0}^{2}t)$ that characterizes the ballistic regime in the time regime $1\gg D_{\Omega}t\gg6 Pe^{-1}$ and diffusive with diffusion constant $D_{B}$ in the time regime $D_{\Omega}t\ll6 Pe^{-1}$ as is shown in Fig. for $Pe=10^3$.
\begin{figure}
\includegraphics[width=0.45\textwidth,clip=true]{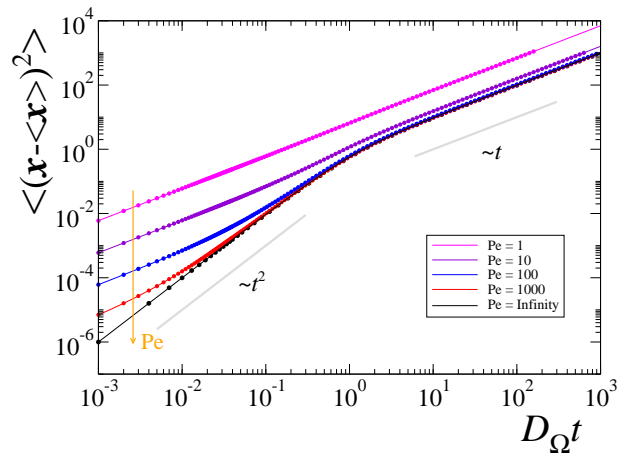}
\caption{\label{fig:2} (Color online) Time dependence of the total mean square displacement, in units of $D_{\Omega}^{-1}$ and $v_{0}^{2}/D_{\Omega}$ respectively, for different values of the Peclet number, namely, 1, 10, 100, 1000 and infinity. In these units the effective diffusion constant is given by $1+(Pe)^{-1}$.}
\end{figure}
The apparent resemblance of expression \eqref{MSD} with the corresponding one obtained from the Ornstein-Uhlenbeck (OU) process  has been noticed before \cite{BechingerArXiv2016}. Observe however that if the fluctuation-dissipation relation is assumed to be valid for the OU process, both expression can not correspond to each other. Indeed, the fluctuation-dissipation relation on the OU process implies that speed scale due to diffusive behavior $\sqrt{6D(\gamma/m)}$ equals the mean thermal propulsion-speed that emerge in the ballistic regime of the mean-square displacement $v_{T}$, where $D$ is the diffusion coefficient, $v_{T}\equiv\sqrt{6k_{B}T/m}$, $m$ the particle mass and $\gamma=6\pi\eta a$ the coefficient of the dragging force that appears in the corresponding Langevin equation for the Ornstein-Uhlenbeck process. In contrast, such equivalence can not be established from expression \eqref{MSD} since the speed scale  $\sqrt{6D(\gamma/m)}$ associated to diffusion behavior with diffusion constant $D_{\text{eff}}^{0}$, does not agree with $v_{0}$. This discrepancy  explicitly shows the departure from equilibrium measured by $(v_{0}^{2}/v_{T}^{2})(\gamma/m)/D_{\Omega}$, evidently the fluctuation-dissipation relation is restored whenever $v_{0}\ll v_{T}$ and/or $(\gamma/m)\ll D_{\Omega}$.

Another aspect of interest corresponds to the short-time behavior of the front propagation of self-propelled particles. As mentioned before, the transmission fidelity of signals (defined as the propagation without the effects of reverberation or wake), as discussed by  John D. Barrow in \cite{BarrowPhilTransRoySocLond1983}, favors three dimensions supporting the anthropic principle. A quantity that provides a measure for the shape of the propagation front and therefore of signal fidelity is the kurtosis, $\kappa$, of the distribution for the particle positions. A definition of kurtosis for a multivariate distributions is given by Mardia et al. \cite{Mardia74p115}, which at time $t$ is given by
\begin{equation}\label{KurtosisDef}
\kappa=\langle\left[\boldsymbol{x}-\langle\boldsymbol{x}\rangle\right]\cdot\boldsymbol{\Sigma}^{-1}\cdot\left[\boldsymbol{x}-\langle\boldsymbol{x}\rangle\right]\rangle 
\end{equation}
where $\boldsymbol{\Sigma}$ corresponds to the covariance matrix defined by the average of the dyadic product $\left[\boldsymbol{x}-\langle\boldsymbol{x}\rangle\right] \cdot\left[\boldsymbol{x}-\langle\boldsymbol{x}\rangle\right].$ For Gaussian distributions the kurtosis gives the invariant value 15, 8 and 3 in three, two and one dimensions, respectively. Thus any deviation from these values measure the departure from a Gaussian behavior either by transient effects from non-equilibrium initial distributions or by the breakdown of the fluctuation dissipation relation. In the same spirit, the kurtosis could equally characterize the shape of the distribution for which propagation wake-like effects can be identified. 

The time dependence of the kurtosis provides a mark for the temporal evolution of the distribution of the particle positions. For instance, for the three-dimensional Ornstein-Uhlenbeck process, the kurtosis of the particle position distribution deviates from its corresponding value 15, basically due to transient effects induced by non-equilibrium initial distributions, which are convoluted with the Gaussian propagator in the general solution of the corresponding Fokker-Planck equation. In the present analysis we leave aside these kind of transient effects and focus on the time dependence of the kurtosis of the corresponding Green functions for the self-propelled particles, i.e., in the distribution of the particle positions.

If no chirality effects are present, the distribution of particles are spherically distributed around an arbitrary point (the location of the initial pulse) in the plane which, without loss of generality, can be chosen as the origin. In such a case, the kurtosis acquires a simple form, namely 
\begin{equation}\label{KurtosisSymmetric}
 \kappa(t)=9\frac{\langle\boldsymbol{x}^{4}(t)\rangle}{\left[\langle\boldsymbol{x}^{2}(t)\rangle\right]^{2}},
\end{equation}
for which only the fourth an second moments are required. An analogous expression to Eq. \eqref{TotalMSD} can be found for the fourth moment, namely
\begin{multline}\label{4thMomentT}
 \langle\boldsymbol{x}^{4}(t)\rangle=60\, (D_{B}t)^{2}+12\, D_{B}t\, \langle\boldsymbol{x}^{2}(t)\rangle_{0}+\langle\boldsymbol{x}^{4}(t)\rangle_{0}\\
 +48\, D_{B}t\, \left[\frac{\sqrt{2}(2\pi)^{2}}{k}\frac{\partial}{\partial k}\hat{P}_{0}^{0}(\boldsymbol{k},t)\right]_{\boldsymbol{k}=0}.
\end{multline}
The first term in the rhs corresponds to the contribution due to translational fluctuations and $\langle\boldsymbol{x}^{2}(t)\rangle_{0}$ is given in \eqref{MSD}.

Calculation of the last two terms in expression \eqref{4thMomentT} requires the knowledge of $\sqrt{4\pi}P_{0}^{0}(\boldsymbol{x},t)$. If the $P_{2}$'s coefficients and higher order multipoles are neglected, the fourth moment is approximated by the one of the telegrapher's equation propagator \eqref{TE3D}, that leads only to an approximated expression for the time dependence of $\langle\boldsymbol{x}^{4}(t)\rangle_{0}$ \cite{SevillaPRE2014,SevillaPRE2015}. Such approximation results in a kurtosis whose time dependence gives the value 5 in the short time regime $D_{\Omega}t\ll1$, value that characterizes wave-like propagation with wake effects. As time increases, the kurtosis grows monotonically to saturate at the value 15 in the diffusive regime or long-time limit (thin-dashed line in Fig. \ref{fig:2}).

If the coupling of the $P_{2}$'s coefficients to higher multipoles are neglected, Eq. \eqref{IHTE3D} can be closed for $\hat{P}_{0}^{0}$ and can be written as (recall that $\tau_{0}=0$)
\begin{widetext}
\begin{equation}\label{MemTE3D}
\frac{d^{2}}{dt^{2}}\hat{P}_{0}^{0}(\boldsymbol{k},t)+2D_{\Omega}\frac{d}{dt}\hat{P}_{0}^{0}(\boldsymbol{k},t)
+v_{0}^{2}\boldsymbol{k}^{2}\int_{0}^{t}ds\, \phi(t-s)\hat{P}_{0}^{0}(\boldsymbol{k},s)=
\sqrt{\frac{8}{15}}\left(\frac{v_{0}}{2}\right)^{2}e^{-6D_{\Omega}t}\left[Q(\boldsymbol{k})+4\sqrt{\frac{2}{15}}\boldsymbol{k}^{2}\hat{P}_{0}^{0}(\boldsymbol{k},0)\right],
\end{equation}
where the memory function $\phi(t)$ is given explicitly by $\frac{3}{5}\delta(t)-\frac{8}{5}D_{\Omega}e^{-6D_{\Omega}t}$ and 
\begin{multline}
Q(\boldsymbol{k})=(k_{y}-ik_{x})^{2}\hat{P}_{2}^{2}(\boldsymbol{k},0)+(k_{y}+ik_{x})^{2}\hat{P}_{2}^{-2}(\boldsymbol{k},0)+\\
+2ik_{z}\left[(k_{y}-ik_{x})\hat{P}_{2}^{1}(\boldsymbol{k},0)+(k_{y}+ik_{x})\hat{P}_{2}^{-1}(\boldsymbol{k},0)\right]+\\
+\left(\frac{2}{3}\right)^{1/2}\left[(k_{x}^{2}+k_{y}^{2})-2k_{z}^{2}\right]\hat{P}_{2}^{0}(\boldsymbol{k},0)
\end{multline}
\end{widetext}
is a term that depends only on the initial conditions and that vanishes for the initial conditions chosen.

Though a mere approximation, the solution to Eq. \eqref{MemTE3D}, which in Fourier-Laplace domain is given by 
\begin{equation}\label{P00BeyondTE}
 \hat{P}_{0}^{0}(\boldsymbol{k},\epsilon)=\hat{P}_{0}^{0}(\boldsymbol{k},0)\frac{\epsilon+2D_{\Omega}+\frac{4}{15}\frac{v_{0}^{2}\boldsymbol{k}^{2}}{\epsilon+6D_{\Omega}}}{\epsilon^{2}+2D_{\Omega}\epsilon+v_{0}^{2}\boldsymbol{k}^{2}\tilde{\phi}(\epsilon)},
\end{equation}
leads to the exact time dependence of $\langle\boldsymbol{x}^{4}(t)\rangle_{0}$ and of the last term in Eq. \eqref{4thMomentT}, as is shown when compared to numerical simulations. 

The exact formula for the fourth moment is the found from the following equation 
\begin{multline} 
 \frac{d^{2}}{dt^{2}}\langle\boldsymbol{x}^{4}(t)\rangle_{0}+2D_{\Omega}\frac{d}{dt}\langle\boldsymbol{x}^{4}(t)\rangle_{0}=12v_{0}^{2}\langle\boldsymbol{x}^{2}(t)\rangle_{0}-\\
 32v_{0}^{2}D_{\Omega}\int_{0}^{t}ds\, e^{-6D_{\Omega}(t-s)}\langle\boldsymbol{x}^{2}(s)\rangle_{0},
\end{multline}
which is directly obtained from equation \eqref{MemTE3D} when multiplied by $\boldsymbol{x}^{4}$ and integrated over all space. The solution to the last equation is given in terms of the second moment of $P_{0}^{0}(\boldsymbol{x},t)$ as
\begin{multline}
 \langle\boldsymbol{x}^{4}(t)\rangle_{0}=4v_{0}^{2}\int_{0}^{t}ds\int_{0}^{s}ds^{\prime}e^{-2D_{\Omega}(s-s^{\prime})}\left[\dfrac{}{} 3\langle\boldsymbol{x}^{2}(s^{\prime})\rangle_{0}\right.\\
 -8D_{\Omega}\int_{0}^{s^{\prime}}\left.ds^{\prime\prime}e^{-6D_{\Omega}(s^{\prime}-s^{\prime\prime})}\langle\boldsymbol{x}^{2}(s^{\prime\prime})\rangle_{0} \right].
\end{multline}
After substitution of the second moment and evaluation of the integrals we get 
\begin{multline}\label{4thMomentActive}
 \langle\boldsymbol{x}^{4}(t)\rangle_{0}=\frac{v_{0}^{4}}{D_{\Omega}^{4}}\left[\frac{5}{3}(D_{\Omega}t)^{2}-\frac{26}{9}D_{\Omega}t-e^{-2D_{\Omega}t}D_{\Omega}t\right.\\
 +\left. 2\left(1-e^{-2D_{\Omega}t}\right)-\frac{1}{54}\left(1-e^{-6D_{\Omega}t}\right)\right],
\end{multline}
which gives the exact time-dependence for the fourth moment of the distribution that carries the effects of persistence. In the short time regime $\langle\boldsymbol{x}^{4}(t)\rangle_{0}$ is simplified to $v_{0}^{4}t^{4}$ and therefore a kurtosis in this regime gets the value $9$ which differs from the value $5$ for the distribution of positions corresponding to the wave-like propagation (see Fig \ref{fig:2}). It can be shown that the value 9 corresponds to a position distribution whose shape at time $t$, is a spherical shell given by $\delta^{(3)}(\vert\boldsymbol{x}\vert-ct)/4\pi\vert\boldsymbol{x}\vert^{2}$. In the asymptotic limit, $D_{\Omega}t\gg1$, expression \eqref{4thMomentActive} gives $\langle\boldsymbol{x}^{4}(t)\rangle_{0} \longrightarrow(5/3)(v_{0}/D_{\Omega})^{4}(D_{\Omega}t)^{2}$, from which the kurtosis value 15, corresponding to Gaussian distributions, is obtained. 

The factor in square brackets in the last term in the rhs of expression \eqref{4thMomentT} can now be calculated with the help of expression \eqref{P00BeyondTE}, namely, after Laplace inversion we get
\begin{multline}\label{Factor}
 \left[\frac{\sqrt{2}(2\pi)^{2}}{k}\frac{\partial}{\partial k}\hat{P}_{0}^{0}(\boldsymbol{k},t)\right]_{\boldsymbol{k}=0}=\frac{v_{0}^{2}}{6D_{\Omega}^{2}}\times\\
 \left[1-e^{-2D_{\Omega}t}
 -2D_{\Omega}t\right].
\end{multline}
\begin{figure}[t]
\includegraphics[width=0.45\textwidth,clip=true]{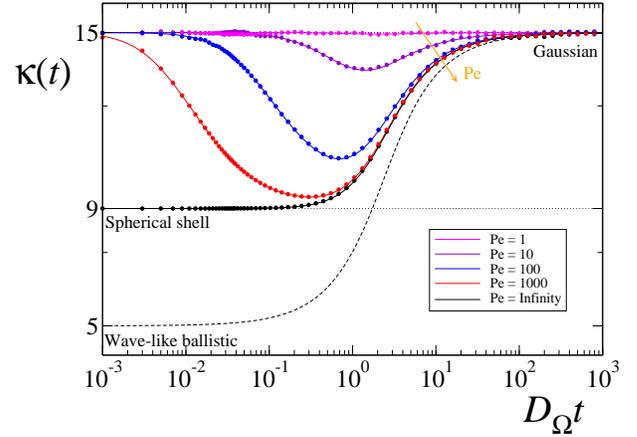}
\caption{\label{fig:3} (Color online) Time dependence of the kurtosis, $\kappa(t)$ for the rotationally invariant case, for several values of the $Pe$. Solid lines are plots of analytical expression of $\kappa(t)$ as explained in text, while squares are the values obtained from numerical simulations. The dashed line corresponds to the kurtosis when the probability density $P_{0}^{0}$ is obtained from the telegrapher's equation \eqref{TE3D} and $Pe=$ Infinity.}
\end{figure}

By collecting the results \eqref{ActiveMSD}, \eqref{4thMomentActive}, \eqref{Factor} and \eqref{TotalMSD} and putting them in expression \eqref{KurtosisSymmetric}, the time dependence of the kurtosis is obtained. In figure \ref{fig:2} such a dependence is shown for different values of $Pe$, namely 1, 10, 100, 1000 and infinity. A comparison with the numerical solutions of Eqs. \eqref{modelo} is also presented in the same figure, an excellent agreement with the analytical solution (lines) is remarkable. 

\paragraph{Effects of chirality about a fixed direction}
Consideration of chirality in the locomotion behavior of active particles is justified in many observed patterns of motion of biological organisms 
or artificial active particles \cite{OrdemannPhysicaA2003,NourhaniPRE2013}. Due to different mechanisms, chirality breaks rotational symmetry which makes diffusion anisotropic, in the simple case in which the rotational symmetry is broken about a fixed, arbitrary direction, diffusion is split into diffusion along that direction and along the perpendicular plane. We set such a direction as $\hat{\boldsymbol{z}}$ for simplicity, the diffusive approximation leads to the Eq.
\begin{multline}\label{TEChiral}
 \frac{d^{2}}{dt^{2}}\hat{P}_{0}^{0}(\boldsymbol{k},t)+2D_{\Omega}\frac{d}{dt}\hat{P}_{0}^{0}(\boldsymbol{k},t)+\frac{v_{0}^{2}}{3}\boldsymbol{k}^{2}\hat{P}_{0}^{0}(\boldsymbol{k},t)=\\
 \frac{v_{0}^{2}}{3}\boldsymbol{k}_{\perp}^{2}\tau_{0}\int_{0}^{t}ds\, \eta(t-s)\hat{P}_{0}^{0}(\boldsymbol{k},s)
 \end{multline}
where $\boldsymbol{k}_{\perp}=\left(k_{x},k_{y}\right)$ denotes the vectors in $\boldsymbol{k}$-space that span the two dimensional subspace orthogonal to the direction $k_{z}$. Last expression generalizes the telegrapher equation \eqref{TE3D}, $\eta(t)\equiv\tau_{0} e^{-2D_{\Omega}t}\sin\tau_{0}t$ being a memory function that makes evident the anisotropic effects induced by chirality. An explicit solution that considers this anisotropy can be found in the Laplace-Fourier domain, to say
\begin{multline}\label{DiffSolRotTorque}
 \widetilde{P}^{0}_{0}(\boldsymbol{k},\epsilon)=\frac{(\epsilon+2D_{\Omega})\hat{P}^{0}_{0}(\boldsymbol{k},0)}{\left(\epsilon+D_{\Omega}\right)^{2}+{\omega}_{k}^{2}-\dfrac{v_{0}^{2}}{3}\boldsymbol{k}^{2}_{\perp}\dfrac{\tau_{0}^{2}}{\left(\epsilon+2D_{\Omega}\right)^{2}+\tau_{0}^{2}}},
\end{multline}
where, as before, we have used initial conditions with vanishing probability flux, i.e. $d\hat{P}_{0}^{0}(\boldsymbol{k},0)/dt=0$, and ${\omega}_{k}^{2}$ is given in \eqref{omegak}. As is immediately clear from \eqref{DiffSolRotTorque}, the marginal probability distribution in the long-time regime along the $\hat{\boldsymbol{z}}$ direction, $P_{0}^{0}(z,t)$ obtained from \eqref{DiffSolRotTorque} when evaluating the inverse Laplace-Fourier transform with $\boldsymbol{k}_{\perp}=0$, is not affected by chirality and it satisfies the standard one-dimensional telegrapher's equation, whose integrodifferential form is given by the expression
\begin{equation}\label{TEz}
 \frac{\partial}{\partial t}P_{0}^{0}(k_{z},t)=\frac{v_{0}^{2}}{3}\int_{0}^{t}ds\, e^{-2D_{\Omega}(t-s)}\frac{\partial^{2}}{\partial k_{z}^{2}}\hat{P}_{0}^{0}(k_{z},s),
\end{equation}
and whose solution is well known to be appropriate in the long-time regime \cite{PorraPRE1997}. In contrast, the marginal probability distribution, $P_{0}^{0}(\boldsymbol{x}_{\perp},t)$, on the plane where rotational motion due to chirality take place, satisfies the continuity equation 
\begin{equation}\label{SmoluchowskiChiral}
 \frac{\partial}{\partial t}P_{0}^{0}(\boldsymbol{x}_{\perp},t)+\nabla_{\perp}\cdot\boldsymbol{J}(\boldsymbol{x}_{\perp},t)=0
\end{equation}
provided that initial conditions with vanishing probability flux are chosen and $\nabla_{\perp}\equiv(\partial/\partial x,\partial/\partial y)$. The total probability current in \eqref{SmoluchowskiChiral}, $\boldsymbol{J}(\boldsymbol{x}_{\perp},s)$, is the sum of two contributions: one that we denote with 
\begin{equation}\label{Jpersistence}
 \boldsymbol{J}_{\text{p}}(\boldsymbol{x}_{\perp},t)=-(v_{0}^{2}/3)\int_{0}^{t}ds\, e^{-2D_{\Omega}(t-s)}\nabla_{\perp}P_{0}^{0}(\boldsymbol{x}_{\perp},t),
\end{equation}
is the current generated not only by the instantaneous of the negative of the gradient of the instantaneous density inhomogeneities, but for all previous ones weighted by an exponentially decaying memory function that lead to the persistence effects. The other contribution denoted with 
\begin{multline}\label{Jchiral}
\boldsymbol{J}_{\text{ch}}(\boldsymbol{x}_{\perp},t)=(v_{0}^{2}/3)\int_{0}^{t}ds\, e^{-2D_{\Omega}(t-s)}\times\\
\int_{0}^{s}ds^{\prime}\eta(s-s^{\prime})\nabla_{\perp}P_{0}^{0}(\boldsymbol{x}_{\perp},t) 
\end{multline}
corresponds to a current in the direction of the gradient of the doubly convoluted probability density with memory functions $e^{-2D_{\Omega}t}$ and the one that incorporates the effects of chirality, $\eta(t)$ just defined above. With these considerations, combination of equation \eqref{SmoluchowskiChiral} with the constitutive relations \eqref{Jpersistence} ,\eqref{Jchiral} constitutes the long-time-regime Smoluchowski equation for chiral, active particles. As is shown in the following, this equation provides the exact time dependence of the mean squared displacement from which expressions for the effective diffusion coefficient can be derived and that have been obtained before from Langevin equations for Brownian circle swimmers \cite{VanTeeffelenPRE2008,EbbensPRE2010,WeberPRE2011}.

From expression \eqref{DiffSolRotTorque} the explicit time dependence contribution to the mean square displacement, due to active motion, can be straightforwardly obtained, namely
\begin{widetext}
\begin{multline}\label{msdRotTorque}
 \langle\boldsymbol{x}^{2}(t)\rangle_{0}=\frac{v_{0}^{2}}{D_{\Omega}}\left(\frac{D_{\Omega}^{2}+\tau_{0}^{2}/12}{D_{\Omega}^{2}+\tau_{0}^{2}/4}\right)t-\frac{1}{6}\frac{v_{0}^{2}}{D_{\Omega}^{2}}\left(1-e^{-2D_{\Omega}t}\right)+\frac{4}{3}\frac{v_{0}^{2}\tau_{0}^{2}}{\left(4D_{\Omega}^{2}+\tau_{0}^{2}\right)^{2}}\times\\
 \left[\left(1-4\frac{D_{\Omega}^{2}}{\tau_{0}^{2}}\right)\left(1-e^{-2D_{\Omega}t}\cos\tau_{0}t\right)-4\frac{D_{\Omega}}{\tau_{0}}e^{-2D_{\Omega}t}\sin\tau_{0}t\right].
\end{multline}
\end{widetext}
where the effects of chirality about the $\hat{\boldsymbol{z}}$ direction are apparent.

\begin{figure}
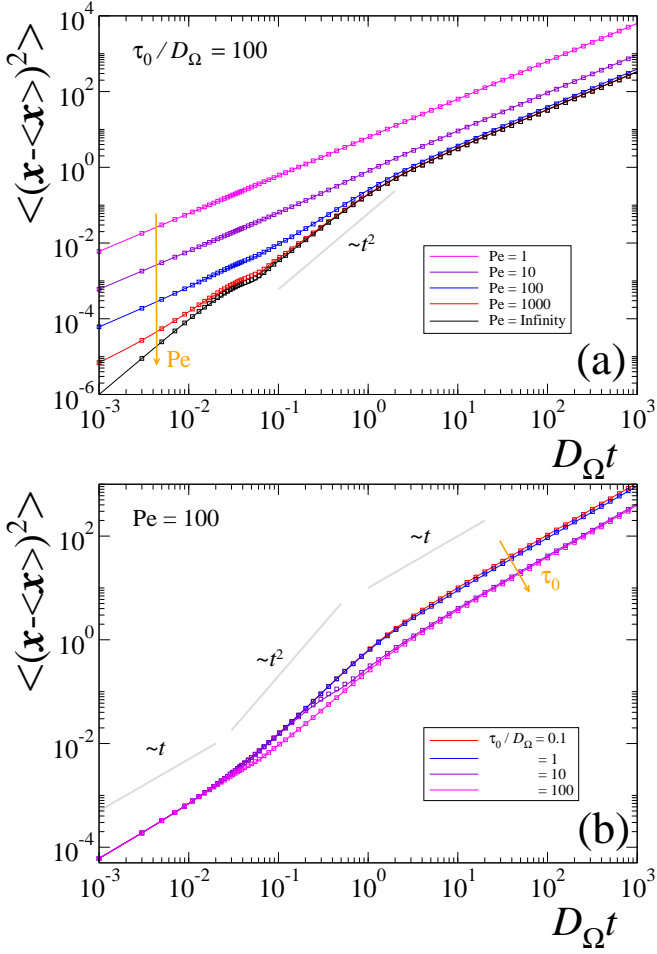

 \includegraphics[width=\columnwidth,clip=true]{Fig5a_MSD_Chirality}\\
 \includegraphics[width=\columnwidth,clip=true]{Fig5b_MSD_Chirality_DB001}
 \caption{(Color online) Mean squared displacement in units of $v_{0}^{2}/D_{\Omega}^{2}$ as function of the dimensionless time $D_{\Omega}t$, for different values of the P\'eclet number, namely 1, 10 ,100, 1000 for a large, fixed chirality $\tau_{0}/D_{\Omega}=100$ (a); and for different values of chirality, 0.1, 1 ,10, 100 and a large value of $Pe =100$ for which the effects of persistence are important (b). Squares correspond to data gathered from numerical simulations while lines are plots of expression \eqref{TotalMSD} with $\langle\boldsymbol{x}^{2}(t) \rangle_{0}$ given by \eqref{msdRotTorque}.}
 \label{FigMSDChiral}
\end{figure}

Addition of the translational component $6D_{B}t$ to last expression gives the total msd [see Eq. \eqref{TotalMSD}]. In figure \ref{FigMSDChiral} the time dependence of the total msd is shown for two different situations, firstly for a large, fixed value of chirality, namely $\tau_{0}/D_{\Omega}=100$, and different values of the P\'eclet number [panel (a)]. 
The effects of chirality are revealed in the time regime $t\sim \tau_{0}^{-1}$ for values of the P\'eclet number for which the effects of persistence are conspicuous, $Pe=1000$ and infinity. In the long time regime the effective diffusion coefficient diminishes as $Pe$ is increased, bounded from below by $D_{B}+D_{A}/3$ [see \eqref{Deff}]. 

In the long time regime normal diffusion dominates the time dependence leading to the effective diffusion coefficient \cite{SandovalPRE2013}
\begin{equation}\label{Deff}
 D_{\text{eff}}=D_{B}+\frac{1}{6}\frac{v_{0}^{2}}{D_{\Omega}}\left(\frac{D_{\Omega}^{2}+\tau_{0}^{2}/12}{D_{\Omega}^{2}+\tau_{0}^{2}/4}\right),
\end{equation}
which results in a monotonous function of both $D_{\Omega}$ and $\tau_{0}.$ For fixed P\'eclet number the effective diffusion coefficient is bounded from above by $D_{\text{eff}}^{0}$, and from below by $D_{B}+D_{A}/3$. The first order correction is quadratic in $\tau_{0}/D_{\Omega}$ when $\tau_{0}/D_{\Omega}\ll1$, namely $D_{\text{eff}}\approx D_{\text{eff}}^{0}-(D_{A}/6)\tau_{0}^{2}/D_{\Omega}^{2}$, contrarily, the first order correction when $\tau_{0}/D_{\Omega}\gg1$ is $D_{\text{eff}}\approx D_{B}+(D_{A}/3)(1+8\tau_{0}^{2}/D_{\Omega}^{2}).$

\begin{figure}
 \includegraphics[width=\columnwidth,clip=true]{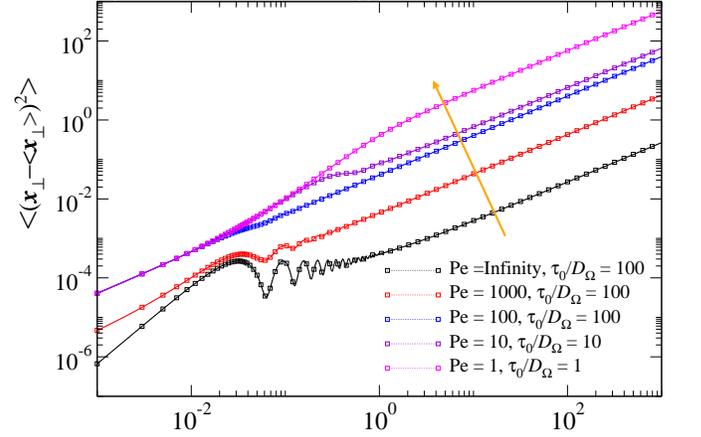}
 \caption{(Color Online) Time dependence of the mean squared displacement $\langle\left(\boldsymbol{x}_{\perp}-\langle\boldsymbol{x}_{\perp}\rangle\right)^{2}\rangle$ in units of $v^{2}/D_{\Omega}^{2}$ along the $\hat{\boldsymbol{x}}\hat{\boldsymbol{y}}$-plane perpendicular to the chirality direction, as function of the dimensionless time $D_{\Omega}t$. Analytical expression \eqref{msdRotTorquePerp} is shown in solid lines while data acquired from numerical simulations are shown by squares.}
 \label{FigMSDChiralPerp}
\end{figure}

In panel (b), the msd displacement is shown for the fixed P\'eclet number $100$, value for which the effects of persistence of active motion are important, and different values of chirality. In the short-time regime the msd is linear in $t$ with a diffusion coefficient that depends only $Pe$ and not on chirality as is apparent in the figure. At long times, in the diffusive regime, the effective diffusion coefficient diminishes as chirality is increased, bounded from below by $D_{B}+D_{A}/3$ [see \eqref{Deff}].

Due to the anisotropy induced by chirality, the motion can be split into motion along the $\hat{\boldsymbol{z}}$ direction and motion on the plane orthogonal to $\hat{\boldsymbol{z}}$.  It is straightforward to show that the mean squared displacement along the $\hat{\boldsymbol{z}}$ direction, computed from equation \eqref{TEz}, is one third of the result given in \eqref{TotalMSD}. On the other hand, we reproduce the exact time-dependence of the mean square displacement \cite{VanTeeffelenPRE2008,EbbensPRE2010} on the $\hat{\boldsymbol{x}}\hat{\boldsymbol{y}}$-plane directly from the Smoluchowski equation given by Eqs. \eqref{SmoluchowskiChiral}-\eqref{Jchiral} given explicitly by 
\begin{widetext}
\begin{equation}\label{msdRotTorquePerp}
 \langle\boldsymbol{x}_{\perp}^{2}(t)\rangle=4D_{\perp}t+\frac{4}{3}\frac{v_{0}^{2}\tau_{0}^{2}}{\left(4D_{\Omega}^{2}+\tau_{0}^{2}\right)^{2}} \left[\left(1-4\frac{D_{\Omega}^{2}}{\tau_{0}^{2}}\right)\left(1-e^{-2D_{\Omega}t}\cos\tau_{0}t\right)-4\frac{D_{\Omega}}{\tau_{0}}e^{-2D_{\Omega}t}\sin\tau_{0}t\right],
\end{equation}
\end{widetext}
and shown in Fig. \ref{FigMSDChiralPerp} for different values of $Pe$ and $\tau/D_{\Omega}$, symbols correspond to data from numerical simulations while lines to plots of the analytical expression \eqref{msdRotTorquePerp}. Notice the conspicuous oscillations for $Pe\gg1$ and $\tau_{0}/D_{\Omega}\gg1$.

If the limit $t\rightarrow\infty$ is applied to expression \eqref{msdRotTorquePerp} after dividing by $4t$, we recover previous results regarding the effective diffusion coefficient of chiral, active particles in two dimensions \cite{WeberPRE2011,MarinePRE2013}, namely  
\begin{equation}
D_{\perp}=D_{B}+\frac{2}{3}v_{0}^{2}\frac{D_{\Omega}}{4D_{\Omega}^{2}+\tau_{0}^{2}}.
\end{equation}
which is a non-monotonous function of $D_{\Omega}$ reaching its maximum value $D_{B}+v_{0}^{2}/6\tau_{0}$ at $D_{\Omega}=\tau_{0}/2$, as has been pointed out in Ref. \cite[and references therein]{WeberPRE2011} for active particle diffusing in two dimensions under the effects of a constant torque or in Ref. \cite{LarraldePRE1997} for the two-dimensional chiral random walker. 

\begin{figure}
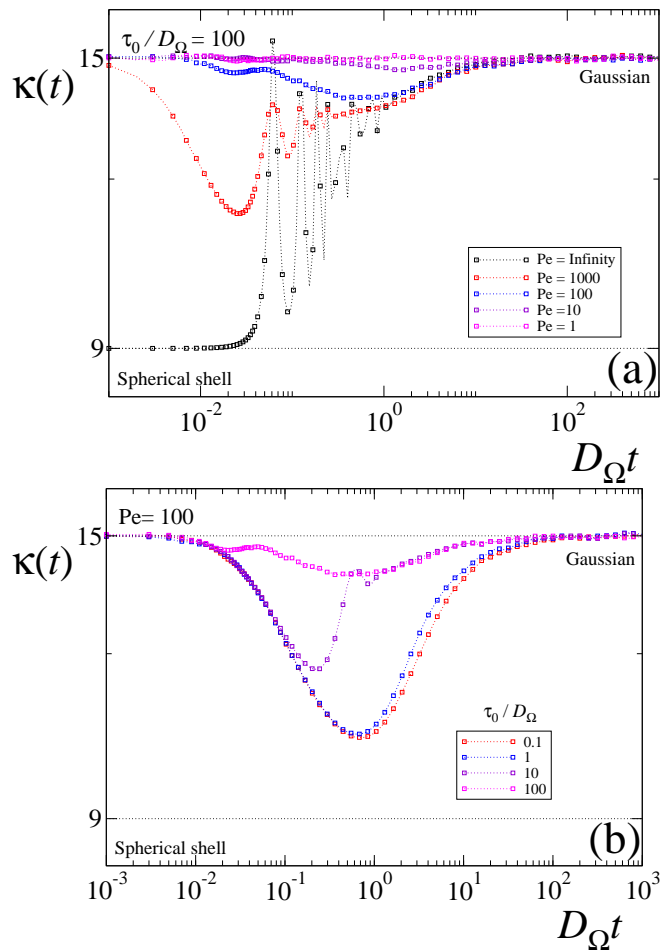

 \includegraphics[width=\columnwidth,clip=true]{Fig7a_Kurtosis2}\\
 \includegraphics[width=\columnwidth,clip=true]{Fig7b_Kurtosis2b}
 \caption{(Color online) Time dependence of the kurtosis $\kappa(t)$ as defined in \eqref{KurtosisDef} for different values of the P\'eclet number and chirality: $\tau_{0}/D_{\Omega}=100$ (a) and for different values of chirality for a fixed value of the P\'eclet number, $Pe=100$ (b). Dotted lines are just guides for the eye.}
 \label{FigKurtosis1}
\end{figure}

Another relevant aspect refers to the effects of chirality on the ``shape'' of the probability distribution of the particle positions, measured by the kurtosis \cite{hagen,ZhengPRE2013,SevillaPRE2014, SevillaPRE2015}. As ha been pointed out in the previous section, and in Refs. \cite{SevillaPRE2014, SevillaPRE2015} for the two-dimensional case, the exact, analytical time-dependence of the kurtosis is obtained by keeping the quadrupole terms, which make the calculation particularly difficult due to the anisotropy induced by chirality that makes the use of the expression \eqref{KurtosisSymmetric} useless. In the top panel of Fig. \ref{FigKurtosis1} the exact time dependence of the kurtosis, obtained from numerical simulations for $\tau_{0}/D_{\Omega}=100$ and different values $Pe$, is shown. In the short-time regime, the probability distribution is approximately Gaussian, except for the case $Pe=\infty$, for which the persistence effects are dominant leading to an expanding spherical shell ($\kappa\simeq9$) as the shape of the position distribution of the particles. Afterwards, the kurtosis diminishes due to the effects of persistence and rises again to reach a Gaussian in the asymptotic limit. Note, however that for large values of the P\'eclet number, \emph{oscillations} of the kurtosis appear in the short-time regime basically due to the helical nature of the particle trajectories. The oscillations mark periods of time where particles are tightly distributed (values close to 9) and periods of time where the particles tend to spread as a Gaussian distribution.

In the bottom panel of Fig. \ref{FigKurtosis1}, $\kappa(t)$ is shown as function of time for $Pe=100$ and different values of $\tau_{0}$, that is to say $\tau_{0}=0.1 D_{\Omega}$, $ D_{\Omega}$, $10 D_{\Omega}$ and $100 D_{\Omega}$. It is natural to expect that no traces of rotational motion are observed in the particle position distribution if the period of rotation is less or of the order of the persistence time (the lines that have a deeper minima), however if the rotation period is larger than the persistence time, oscillations are present (barely distinguishable in the case $\tau_{0}/D_{\Omega}=100$).   

\begin{figure}
 \includegraphics[width=\columnwidth]{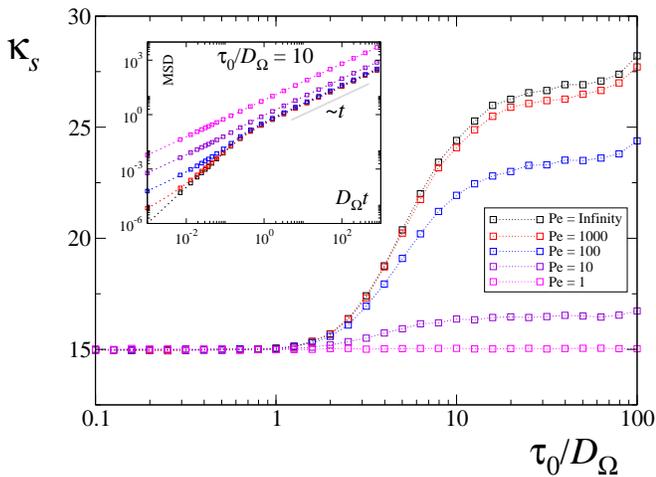}
 \caption{(Color online) Asymptotic values of the kurtosis, $\kappa_{s},$ \emph{vs} the dimensionless chirality, $\tau_{0}/D_{\Omega}$, for the values of the P\'eclet number: infinite, 1000, 100, 10, and 1. Dotted lines are guides for the eye. Inset contains the time dependence of the mean squared displacement for an ensemble of particles moving with a chirality direction uniformly distributed in the sphere and $\tau_{0}/D_{\Omega}=10$. The linear dependence with time is shown in the long time regime.}
 \label{KurtosisFig3}
\end{figure}

\paragraph{Uniformly distributed random directions of chirality: ``Anomalous, yet Brownian, diffusion''}
Lastly, we consider the case at which each chiral active particle has its ``own'' axis of rotation, constant in time, but arbitrary. We choose the simple case that corresponds to an ensemble of chiral particles whose rotation axes are uniformly distributed on the unitary sphere. In that situation, it is observed from numerical simulations (see Fig. \ref{KurtosisFig3}), that the stationary value of the kurtosis $\kappa_{s}=\lim_{t\rightarrow\infty}\kappa(t)$ of the distribution of positions departs from the Gaussian one with the intensity of chirality for P\'eclet numbers larger than 1. As shown, $\kappa_{s}$ increases for $\tau_{0}/D_{\Omega}\gtrsim1$ indicating a non-Gaussian ``flatten'' distribution of the particles positions caused by chirality. This asymptotic non-Gaussian regime is characterized by normal diffusion as has been checked from the numerical simulations (see inset in the same figure). This phenomenon has been observed in systems where tracers diffuse within complex fluids different systems and it is currently referred to as ``anomalous, yet Brownian, diffusion'' \cite{WangPNAS2009,WangNatureMat2012,BhattacharyaJPCB2013}. The phenomenon has been also observed in financial data analysis, particularly, the mean squared displacement of the logarithm of the returns of the price of an asset in a financial market, grows linearly with time, while the probability density function of the log-returns is strongly non-Gaussian due to long-range memory effects of the absolute value of the log-returns \cite{CressoniPRE2012}. The phenomenon has been also addressed theoretically in different one-dimensional models \cite{CressoniPRE2012,ChubynskyPRL2014,WangPRE2016}. In Ref. \cite{CressoniPRE2012} the effects of long-range correlations of the direction of motion, introduced by particular microscopic rules of the displacements, on a random-walk are considered. In there, the authors find a departure from the expected Gaussian distribution of the particle positions, notwithstanding the mean squared displacement being linear in time, effect they referred to as ``weakly anomalous diffusion''. Alternatively, a \emph{non-persistent} random walk model (in that there is no correlations in the displacement direction) that leads to the same phenomenon is considered in Ref. \cite{ChubynskyPRL2014}. Such a model considers a stochastic, diffusion coefficient, from which the authors recover the main features observed in the experiments \cite{WangPNAS2009,WangNatureMat2012}. More recently the phenomenon has been reported as consequence of delocalization in a model \cite{WangPRE2016} for the diffusion of energy along a anharmonic, disordered lattice at finite temperature. Our results point out that ``anomalous, yet Brownian, diffusion'' occurs in a two-dimensional model of diffusing active, chiral particles subject to memory effects in the direction of motion (persistence). A more detailed analysis on the origin of this effect is necessary and will be discussed elsewhere.

\section{\label{SectVI} Conclusions and final remarks}
The diffusion of chiral, active Brownian particles in free, three-dimensional space has been considered. Particular attention was conceded to the probability density, $P(\boldsymbol{x},t)$, of finding a particle at position $\boldsymbol{x}$ at time $t$ independently of its swimming direction, quantity that is susceptible of experimental sampling by the use of single-particle tracking techniques. A systematic method, based on the multipole expansion of the complete probability density $P(\boldsymbol{x},\hat{\boldsymbol{v}},t)$, where $\boldsymbol{v}$ denotes the particle's direction of motion, allows to find Smoluchowski-like equations for $P(\boldsymbol{x},t)$ that includes the effects of chirality for different time regimes. 

For the rotationally invariant motion, i.e. in the absence of chirality, diffusion is described by the standard telegrapher's equation which emerge from the method in the long-time regimen when the hierarchy can be cut up to the dipole terms. Notwithstanding the nature of the approximation, the telegrapher's equation provides the exact time dependence, of the mean squared displacement for arbitrary values of the P\'eclet number as was verified by numerical simulations using the corresponding Langevin equation for active Brownian particles. We found that such is the case even when the effects of chirality are taken into account, in that instance, the telegrapher's equation is modified by an extra term that carries the information about the anisotropy due to the rotational component of the motion. Previous reported expressions for the effective diffusion coefficient were recovered from our theoretical framework.

The fourth moment of $P(\boldsymbol{x},t)$ was also calculated and the kurtosis, that measures the ``shape'' of the probability density, analyzed. For this, the quadrupole terms of the expansion were included in the analysis, which resulted into a generalization of the telegrapher's equation from which analytical expressions for the fourth moment, and therefore for the kurtosis, were obtained in the rotationally invariant case. Numerical simulations were performed to verify the exactness of the time dependence of the kurtosis. In the isotropic case ($\tau_{0}=0$) $\kappa(t)$ is bounded from below by 9, value that corresponds to a spherical shell distribution, and from above by invariant value for a Gaussian distribution, 15, and exhibits a non-monotonic behavior for finite values of the P\'eclet number in the form of a global minimum which is related to the persistence effects. On the other hand, the particles trace stochastic helical trajectories along the $\hat{\boldsymbol{z}}$ direction as chirality breaks rotational invariance, making diffusion anisotropic. For large enough P\'eclet numbers the transient of the probability density shows an interesting oscillating behavior between a Gaussian shape and a spherical shell one. No analytical expressions were obtained in this case, however it is possible to obtain analytical expression of the kurtosis of the marginal distribution of the particles position in the plane orthogonal to the axis of rotation. 

The case for which a chiral active particle moves rotating along an axis of rotation uniformly distributed on the sphere is presented. A statistical analysis of the trajectories obtained from numerical simulations of the Langevin equations, indicates that the asymptotic regime presents normal diffusion described by non Gaussian distribution, revealing an instance  where ``anomalous, yet Brownian, diffusion'' is exhibited.

The results presented in this paper has proven that the method employed to obtain analytical expressions of the exact time dependence for standard experimental data, namely, the mean squared displacement and the kurtosis of the particles position distribution, is valuable and complements the common approach based only on Langevin equations, particularly for the description of the combined effects of chirality and active motion, a situation that is of interest in biological and man-made systems. Though we have restricted our analysis to the case of free diffusion it is of interest to extend the method presented in this paper to the case when particles diffuse under the action of position/velocity dependent forces.

\begin{acknowledgments}
 I kindly acknowledge support from grant UNAM-DGAPA-PAPIIT-IN113114.
\end{acknowledgments}

\appendix
\section{\label{appA}The Langevin equations for the spherical angles}
The numerical solution of the Langevin equations in three-dimensional Euclidean coordinates given by Eqs. \eqref{modelo} as such, present instabilities if direct integrators are used, basically because they fail to preserve the norm of $\hat{\boldsymbol{v}}$ during the time evolution. 

Eq. \eqref{direction} can be written in a simple form as 
\begin{equation}\label{LEqTorqueless}
 d\hat{v}_{\lambda}(t)=\left[\epsilon_{\lambda\mu\nu}\, dW_{\mu}\right]\, \hat{v}_{\nu}(t),
\end{equation}
for which the multiplicative nature of the stochastic equations is made apparent. In \eqref{LEqTorqueless} $\epsilon_{\lambda\mu\nu}$ is the completely antisymmetric or Levi-Civita tensor, $dW_{\mu}(t)=\xi_{\mathcal{R}\mu}(t)$ are Wiener process and the Einstein convention, i.e. sum over repeated index, has been used. The first factor within square parenthesis in  expression \eqref{LEqTorqueless} corresponds to the elements $\mathbb{R}_{\lambda\mu}$, of a stochastic skew-symmetric matrix $\mathbb{R}$.
The statistical properties of rotational noise $\boldsymbol{\xi}_{\mathcal{R}}(t)$ were given in section \ref{SectII}, namely $\langle\xi_{\mathcal{R}\mu}(t)\rangle=\tau_{\mu}$ and $\langle\xi_{\mathcal{R}\mu}(t)\xi_{\mathcal{R}\nu}(s)\rangle=2D_{\Omega}\delta(t-s)\delta_{\mu\nu}$.

Firstly, consider the case for which $\tau_{\mu}=0$ for each $\mu$. Since the constriction $\hat{v}_{i}\hat{v}_{i}=1$ is satisfied straightforwardly in spherical coordinates, a change of variables to such coordinate system is required, namely
\begin{subequations}\label{Spherical}
\begin{align}
 \hat{v}_{x}(t)&=\sin\theta(t)\cos\varphi(t)\label{vx}\\
 \hat{v}_{y}(t)&=\sin\theta(t)\sin\varphi(t)\label{vy}\\
 \hat{v}_{z}(t)&=\cos\theta(t)\label{vz}.
\end{align}
\end{subequations}
The corresponding Langevin equations for the azimuthal $\varphi(t)$ and polar $\theta(t)$ angles can be obtained by the use the standard It\'o interpretation of Eqs. \eqref{LEqTorqueless} as follows. Equations \eqref{vx} and \eqref{vy} can be written in the complex plane as
\begin{equation}\label{plane}
 \hat{v}_{x}(t)+i\, \hat{v}_{y}(t)=\sin\theta(t) e^{i\varphi(t)}=e^{\alpha(t)+i\varphi(t)},
\end{equation}
after application of It\'o calculus \cite{GardinerBook} to Eqs. \eqref{vz}, \eqref{plane},  and some algebraic steps we have that $\varphi(t)$ and $\theta(t)$ satisfy
\cite{Brillinger}
\begin{subequations}\label{LangevinItoAppendix}
 \begin{align}
d\theta(t)&=\frac{D_{\Omega}}{\tan\theta(t)}dt+dW_{\theta}(t)\\
d\varphi(t)&=\frac{dW_{\varphi}(t)}{\sin\theta(t)}
 \end{align}
\end{subequations}
where $dW_{\theta}(t)$, $dW_{\varphi}(t)$ are two statistically independent Wiener processes defined thorough the transformations
\begin{align}
 dW_{\theta}(t)&=\cos\varphi(t)dW_{y}(t)-\sin\varphi(t)dW_{x}(t)\\
 dW_{\varphi}(t)&=\sin\theta(t)dW_{z}(t)-\cos\theta(t)dW_{+}(t)
\end{align}
being $dW_{+}(t)=\cos\varphi(t)dW_{x}(t)+\sin\varphi(t)dW_{y}(t)$ a third independent Wiener process.	

For finite $\tau_{\mu}\neq0$ equations \eqref{LEqTorqueless} can be written as
\begin{equation}\label{LEqTorque}
 d\hat{v}_{\lambda}(t)=\left[\epsilon_{\lambda\mu\nu}\, (\tau_{\mu}+dW_{\mu})\right]\, \hat{v}_{\nu}(t),
\end{equation}
and we can apply the same procedure as before leading, after some algebra, to Eqs. \eqref{LangevinIto}. 

\section{The Fokker-Planck Equation}
The probability density function of finding a particle at $\boldsymbol{x}$ moving in the direction $\hat{\boldsymbol{v}}$ at time $t$ is defined as the ensemble average over the trajectories obtained from the Langevin equations \eqref{modelo}  of $\delta^{(3)}[{\boldsymbol{x}}-{\boldsymbol{x}}(t)]\delta^{(3)}[\hat{\boldsymbol{v}}-\hat{\boldsymbol{v}}(t)]$, that is to say $P(\boldsymbol{x},\hat{\boldsymbol{v}},t)\equiv\left\langle \delta^{(3)}[{\boldsymbol{x}}-{\boldsymbol{x}}(t)]\delta^{(3)}[\hat{\boldsymbol{v}}-\hat{\boldsymbol{v}}(t)]\right\rangle$ where $\delta^{(3)}(\boldsymbol{q})=\delta(q_{x})\delta(q_{y})\delta(q_{z})$ denotes the three-dimensional Dirac delta. 

Derivation of the corresponding Fokker-Planck equation for $P(\boldsymbol{x},\hat{\boldsymbol{v}},t)$, \eqref{FPE}, is straightforward by use of the theorem of Novikov (this is the procedure used in this paper) applied to the Langevin equations \eqref{modelo} assuming Gaussian white noises. There is however a general phenomenological derivation of related diffusion-like transport equations that has been considered in Ref. \cite{DuderstadtTransportTheory}. After differentiation of $P(\boldsymbol{x},\hat{\boldsymbol{v}},t)$ with respect to time, we get
\begin{widetext}
\begin{multline}
\frac{\partial}{\partial t}P({\boldsymbol{x}},\hat{\boldsymbol{v}},t)+v_{0}\hat{\boldsymbol{v}}\cdot \nabla P({\boldsymbol{x}},\hat{\boldsymbol{v}},t)=
\nabla_{\hat{\boldsymbol{v}}}\cdot\left(\hat{\boldsymbol{v}}\times \boldsymbol{\tau}\right)P({\boldsymbol{x}},\hat{\boldsymbol{v}},t)+
\nabla_{\hat{\boldsymbol{v}}}\cdot\left[\hat{\boldsymbol{v}}\times \left\langle \boldsymbol{\xi}_{\mathcal{R}}(t)\delta^{(3)}[{\boldsymbol{x}}-{\boldsymbol{x}}(t)]\delta^{(3)}[\hat{\boldsymbol{v}}-\hat{\boldsymbol{v}}(t)]\right\rangle\right]\\
-\nabla\cdot\left\langle \boldsymbol{\xi}_{T}(t)\delta^{(3)}[{\boldsymbol{x}}-{\boldsymbol{x}}(t)]\delta^{(3)}[\hat{\boldsymbol{v}}-\hat{\boldsymbol{v}}(t)]\right\rangle,
\end{multline}   
where explicit use of equations \eqref{modelo} has been carried out. In the same spirit of the previous appendix,  we make use of a better notation to write, using the Einstein convention,
\begin{multline}\label{FPE0}
\frac{\partial}{\partial t}P(x,\hat{v},t)+v_{0}\hat{v}_{\mu}\frac{\partial}{\partial x_{\mu}} P(x,\hat{v},t)=
\frac{\partial}{\partial \hat{v}_{\mu}}\epsilon_{\mu\nu\lambda}\hat{v}_{\nu}\tau_{\lambda}P(x,\hat{v},t)+\\
\frac{\partial}{\partial \hat{v}_{\mu}}\epsilon_{\mu\nu\lambda}\hat{v}_{\nu}\left\langle 
\xi_{\mathcal{R}\lambda}(t)\prod_{\sigma}\delta[x_{\sigma}-x_{\sigma}(t)]\delta[\hat{v}_{\sigma}-\hat{v}_{\sigma}(t)]\right\rangle
-\frac{\partial}{\partial x_{\mu}}\left\langle\xi_{T\mu}(t)\prod_{\sigma}\delta[x_{\sigma}-{x_{\sigma}}(t)]\delta[\hat{v}_{\sigma}-\hat{v}_{\sigma}(t)]\right\rangle.
 \end{multline}   
Novikov's theorem \cite{MasoliverPRE1995,frank} allows to write
\begin{equation*}
 \left\langle\xi_{\mathcal{R}\lambda}(t)\prod_{\sigma}\delta[x_{\sigma}-x_{\sigma}(t)]\delta[\hat{v}_{\sigma}-\hat{v}_{\sigma}(t)]\right\rangle =
 D_{\Omega}\left\langle\frac{\delta}{\delta\xi_{\mathcal{R}\lambda}}\prod_{\sigma}\delta[x_{\sigma}-x_{\sigma}(t)]\delta[\hat{v}_{\sigma}-\hat{v}_{\sigma}(t)]\right\rangle
\end{equation*}
and
\begin{equation*}
\left\langle \xi_{\mathcal{T}\lambda}(t)\prod_{\sigma}\delta[x_{\sigma}-x_{\sigma}(t)]\delta[\hat{v}_{\sigma}-\hat{v}_{\sigma}(t)]\right\rangle =
D_{B}\left\langle\frac{\delta}{\delta\xi_{\mathcal{T}\lambda}}\prod_{\sigma}\delta[x_{\sigma}-x_{\sigma}(t)]\delta[\hat{v}_{\sigma}-\hat{v}_{\sigma}(t)]\right\rangle
\end{equation*}
and a direct calculation leads to
\begin{align*}
 \left\langle\frac{\delta}{\delta\xi_{\mathcal{R}\lambda}}\prod_{\sigma}\delta[x_{\sigma}-x_{\sigma}(t)]\delta[\hat{v}_{\sigma}-\hat{v}_{\sigma}(t)]\right\rangle&=-\epsilon_{\lambda \nu \mu}\, \hat{v}_{\nu}\frac{\partial}{\partial\hat{v}_{\mu}}P(x,\hat{v},t)\\
 \left\langle\frac{\delta}{\delta\xi_{\mathcal{T}\lambda}}\prod_{\sigma}\delta({\boldsymbol{x}}-{\boldsymbol{x}}(t))\delta(\hat{\boldsymbol{v}}-\hat{\boldsymbol{v}}(t))\right\rangle&=- \frac{\partial}{\partial x_{i}}P(x,\hat{v},t),
\end{align*}
respectively, where Eqs. \eqref{LEqTorque} were used explicitly. By substitution of these results into \eqref{FPE0} we get the Fokker-Planck equation
\begin{multline*}
\frac{\partial}{\partial t}P(x,\hat{v},t)+v_{0}\hat{v}_{\mu}\frac{\partial}{\partial x_{\mu}} P(x,\hat{v},t)=
\frac{\partial}{\partial \hat{v}_{\mu}}\epsilon_{\mu\nu\lambda}\hat{v}_{\nu}\tau_{\lambda}P(x,\hat{v},t)+\\
-D_{\Omega}\frac{\partial}{\partial \hat{v}_{\mu}}\epsilon_{\mu\nu\lambda}\hat{v}_{\nu}\epsilon_{\lambda\sigma\rho}\hat{v}_{\sigma}\frac{\partial}{\partial \hat{v}_\rho}P(x,\hat{v},t)
+D_{B}\frac{\partial}{\partial x_{\mu}}\frac{\partial}{\partial x_{\mu}}P(x,\hat{v},t).
\end{multline*}   
which by the use of the relation $\epsilon_{\mu\nu\lambda}\epsilon_{\lambda\sigma\rho}=\delta_{\mu\sigma}\delta_{\nu\rho}-\delta_{\mu\rho}\delta_{\nu\sigma}$ and that $\hat{v}_{i}\hat{v}_{i}=1$, last equation can be rewritten as
\begin{multline*}
\frac{\partial}{\partial t}P({\boldsymbol{x}},\hat{\boldsymbol{v}},t)+v_{0}\hat{\boldsymbol{v}}\cdot \nabla P({\boldsymbol{x}},\hat{\boldsymbol{v}},t)=
 \nabla_{\hat{\boldsymbol{v}}}\cdot\left(\hat{\boldsymbol{v}}\times\boldsymbol{\tau}\right)P({\boldsymbol{x}},\hat{\boldsymbol{v}},t)+D_{B}\nabla^{2}P({\boldsymbol{x}},\hat{\boldsymbol{v}},t)+\\
D_{\Omega}\left[\nabla_{\hat{\boldsymbol{v}}}^{2}-\hat{\boldsymbol{v}}\cdot\nabla_{\hat{\boldsymbol{v}}}-\left(	\hat{\boldsymbol{v}}\cdot\nabla_{\hat{\boldsymbol{v}}}\right)^{2}\right]P({\boldsymbol{x}},\hat{\boldsymbol{v}},t). 
\end{multline*}   
In spherical coordinates, $\theta$, $\varphi$, that specify the direction of $\hat{\boldsymbol{v}}$ in the unit sphere, it is satisfied that $\hat{\boldsymbol{v}}\cdot\nabla_{\hat{\boldsymbol{v}}}=0$ since $\nabla_{\hat{\boldsymbol{v}}}= \boldsymbol{\hat{\theta}}\,{\partial_{\theta}}
+ \boldsymbol{\hat{\varphi}}\, \frac{1}{\sin\theta}{\partial_{\varphi}}$ where $\boldsymbol{\hat{\theta}}$ and $\boldsymbol{\hat{\varphi}}$ are unit vectors of the spherical coordinates. Thus we get the Fokker-Planck
\begin{multline}\label{FPEappendix}
\frac{\partial}{\partial t}P({\boldsymbol{x}},\hat{\boldsymbol{v}},t)+v_{0}\hat{\boldsymbol{v}}\cdot \nabla P({\boldsymbol{x}},\hat{\boldsymbol{v}},t)=D_{B}\nabla^{2}P({\boldsymbol{x}},\hat{\boldsymbol{v}},t)
+\frac{1}{\sin\theta}\frac{\partial}{\partial\varphi}\left[\left(\hat{\boldsymbol{v}}\times\boldsymbol{\tau}\right)\cdot\hat{\boldsymbol{\varphi}}P({\boldsymbol{x}},\hat{\boldsymbol{v}},t)\right]\\
+\frac{1}{\sin\theta}\frac{\partial}{\partial\theta}\left[\sin\theta\left(\hat{\boldsymbol{v}}\times\boldsymbol{\tau}\right)\cdot\hat{\boldsymbol{\varphi}}P({\boldsymbol{x}},\hat{\boldsymbol{v}},t)\right]
+\mathcal{L}(\hat{\boldsymbol{v}})P({\boldsymbol{x}},\hat{\boldsymbol{v}},t). 
\end{multline}   
where $\mathcal{L}(\hat{\boldsymbol{v}})$ is the Laplace-Beltrami or rotational diffusion operator, explicitly given by
\begin{equation}
 \mathcal{L}(\hat{\boldsymbol{v}})=D_{\Omega}\left[\frac{1}{\sin\theta}\frac{\partial}{\partial\theta}\left(\sin\theta\frac{\partial}{\partial\theta}\right)+\frac{1}{\sin^{2}\theta}\frac{\partial^{2}}{\partial\varphi^{2}}\right].
\end{equation}

\section{The matrix elements ${I_{\mu}}^{m,m^{\prime}}_{n,n^{\prime}}$}
The matrix elements ${I_{\mu}}^{m,m^{\prime}}_{n,n^{\prime}}$ defined in expressions \eqref{MatrixElementsI} can be computed directly in a standard fashion by the use of the explicit expression of the spherical harmonics $Y_{m}^{n}(\hat{\boldsymbol{v}})=(-1)^{m}\sqrt{\frac{(2n+1)}{4\pi}\frac{(n-m)!}{(n+m)!}}  \, P_n^m (\cos{\theta}) \, e^{im\varphi}$, and the following recurrence relations for the associated Legendre polynomials 
\begin{align*}
 (2n+1)\sin\theta\, P_n^m (\cos{\theta})&=P_{n+1}^{m+1}(\cos{\theta})-P_{n-1}^{m+1} (\cos{\theta}),\\
 (2n+1)\cos\theta\, P_n^m (\cos{\theta})&=(n+m)P_{n-1}^{m}(\cos{\theta})+(n-m+1)P_{n+1}^{m} (\cos{\theta}),
\end{align*}
after some simple algebra we get
\begin{subequations}
\begin{multline}
   {I_{x}}^{m,m^{\prime}}_{n,n^{\prime}}=\frac{1}{2}\delta_{n^{\prime},n+1}\left\{\delta_{m,m^{\prime}+1}\left[\frac{(n^{\prime}-m^{\prime}-1)(n^{\prime}-m^{\prime})}{(2n^{\prime}-1)(2n^{\prime}+1)}\right]^{1/2}-\delta_{m,m^{\prime}-1}\left[\frac{(n^{\prime}+m^{\prime}-1)(n^{\prime}+m^{\prime})}{(2n^{\prime}-1)(2n^{\prime}+1)}\right]^{1/2}\right\}\\
   +\frac{1}{2}\delta_{n^{\prime},n-1}\left\{\delta_{m,m^{\prime}-1}\left[\frac{(n^{\prime}-m^{\prime}+2)(n^{\prime}-m^{\prime}-1)}{(2n^{\prime}+1)(2n^{\prime}+3)}\right]^{1/2}-\delta_{m,m^{\prime}+1}\left[\frac{(n^{\prime}+m^{\prime}+2)(n^{\prime}+m^{\prime}+1)}{(2n^{\prime}+1)(2n^{\prime}+3)}\right]^{1/2}\right\},
\end{multline}
\begin{multline}
   {I_{y}}^{m,m^{\prime}}_{n,n^{\prime}}=\frac{1}{2i}\delta_{n^{\prime},n+1}\left\{-\delta_{m,m^{\prime}+1}\left[\frac{(n^{\prime}-m^{\prime}-1)(n^{\prime}-m^{\prime})}{(2n^{\prime}-1)(2n^{\prime}+1)}\right]^{1/2}-\delta_{m,m^{\prime}-1}\left[\frac{(n^{\prime}+m^{\prime}-1)(n^{\prime}+m^{\prime})}{(2n^{\prime}-1)(2n^{\prime}+1)}\right]^{1/2}\right\}\\
   +\frac{1}{2i}\delta_{n^{\prime},n-1}\left\{\delta_{m,m^{\prime}+1}\left[\frac{(n^{\prime}+m^{\prime}+2)(n^{\prime}+m^{\prime}+1)}{(2n^{\prime}+1)(2n^{\prime}+3)}\right]^{1/2}-\delta_{m,m^{\prime}+1}\left[\frac{(n^{\prime}-m^{\prime}+2)(n^{\prime}-m^{\prime}+1)}{(2n^{\prime}+1)(2n^{\prime}+3)}\right]^{1/2}\right\},
\end{multline}
\begin{equation}
   {I_{z}}^{m,m^{\prime}}_{n,n^{\prime}}=\delta_{n^{\prime},n+1}\delta_{m,m^{\prime}}\left[\frac{(n^{\prime}-m^{\prime})(n^{\prime}+m^{\prime})}{(2n^{\prime}-1)(2n^{\prime}+1)}\right]^{1/2}+\delta_{n^{\prime},n-1}\delta_{m,m^{\prime}}\left[\frac{(n^{\prime}+m^{\prime}+1)(n^{\prime}-m^{\prime}+1)}{(2n^{\prime}+1)(2n^{\prime}+3)}\right]^{1/2}.
\end{equation}
\end{subequations}
\end{widetext}

\section{The multipole expansion}
The expansion \eqref{Expansion} is akin to the expansions in powers of the $\hat{\boldsymbol{v}}$ introduced in Ref. \cite{DuderstadtTransportTheory} and used in Ref. \cite{CatesEPL2013} in the context of active particles. In Fourier space, the expansion \eqref{Expansion} can be written in terms of powers of $\hat{\boldsymbol{v}}$ by gathering terms of the same order in $l$ as 
\begin{multline*}\label{Exp}
\hat{P}(\boldsymbol{k},\hat{\boldsymbol{v}},t)=\tilde{\varrho}(\boldsymbol{k},t)+e^{-2D_{\Omega}t}\, \widetilde{\boldsymbol{V}}(\boldsymbol{k},t)\cdot\hat{\boldsymbol{v}}+\\
e^{-6D_{\Omega}t}\, \hat{\boldsymbol{v}}\cdot\widetilde{\boldsymbol{\mathcal{Q}}}(\boldsymbol{k},t)\cdot\hat{\boldsymbol{v}}+\ldots.
\end{multline*}
where 
\begin{equation*}
 \tilde{\varrho}(\boldsymbol{k},t)=e^{-D_{B}\boldsymbol{k}^{2}t}P_{0}^{0}(\boldsymbol{k},t)/\sqrt{4\pi}
\end{equation*}
is interpreted as the Fourier transform of the density of particles, and is related with the uniform distribution of the direction of motion on the sphere (monopole) which is the only term that remains in the asymptotic limit ($t\rightarrow\infty$) of free diffusion. The next term, $\widetilde{\boldsymbol{V}}(\boldsymbol{k},t)\cdot\hat{\boldsymbol{v}}$, is identified with $e^{-D_{B}\boldsymbol{k}^{2}t}\sum_{m}e^{-i\tau_{0}mt}\, \hat{P}_{1}^{m}(\boldsymbol{k},t)\, Y_{1}^{m}(\hat{\boldsymbol{v}})$ and it refers to the dipole distribution of the direction of motion of the particles, in the context of the fluctuating hydrodynamics, it refers to the Fourier transform of the dimensionless velocity field $\widetilde{\boldsymbol{V}}(\boldsymbol{k},t)$ whose components are given explicitly by
\begin{align*}
 \widetilde{V}_{x}(\boldsymbol{k},t)&=\sqrt{\frac{3}{8\pi}}e^{-D_{B}\boldsymbol{k}^{2}t}\left[e^{i\tau_{0}t}\hat{P}_{1}^{-1}(\boldsymbol{k},t)-e^{-i\tau_{0}t}\hat{P}_{1}^{1}(\boldsymbol{k},t)\right],\\
 \widetilde{V}_{y}(\boldsymbol{k},t)&=-i\sqrt{\frac{3}{8\pi}}e^{-D_{B}\boldsymbol{k}^{2}t}\left[e^{i\tau_{0}t}\hat{P}_{1}^{-1}(\boldsymbol{k},t)+e^{-i\tau_{0}t}\hat{P}_{1}^{1}(\boldsymbol{k},t)\right],\\
 \widetilde{V}_{z}(\boldsymbol{k},t)&=\sqrt{\frac{3}{4\pi}}e^{-D_{B}\boldsymbol{k}^{2}t}\hat{P}_{1}^{0}(\boldsymbol{k},t).
\end{align*}
Analogously, the next multipole term $\hat{\boldsymbol{v}}\cdot\widetilde{\boldsymbol{\mathcal{Q}}}(\boldsymbol{k},t)\cdot\hat{\boldsymbol{v}}$ that corresponds to quadrupole distribution of the particle direction of motion, is identified with the sum of all the terms that contain the $l=2$ spherical harmonics, i.e.  $\sum_{m}\hat{P}_{2}^{m}(\boldsymbol{k},t)Y_{2}^{m},$ from which the symmetric, traceless tensor $\widetilde{\boldsymbol{\mathcal{Q}}}(\boldsymbol{k},t)$ can be recognized, namely
\begin{widetext}
\begin{equation*}
\widetilde{\boldsymbol{\mathcal{Q}}}(\boldsymbol{k},t)=\sqrt{\frac{15}{32\pi}}e^{-D_{B}\boldsymbol{k}^{2}t}
\left(
\begin{array}{ccc}
e^{i2\tau_{0}t}\hat{P}_{2}^{-2}+e^{-i2\tau_{0}t}\hat{P}_{2}^{2}-\sqrt{\frac{2}{3}}\hat{P}_{2}^{0} & i\left[e^{-i2\tau_{0}t}\hat{P}_{2}^{2}-e^{i2\tau_{0}t}\hat{P}_{2}^{-2}\right] & e^{i\tau_{0}t}\hat{P}_{2}^{-1}-e^{-i\tau_{0}t}\hat{P}_{2}^{1}\\
i\left[e^{-i2\tau_{0}t}\hat{P}_{2}^{2}-e^{i2\tau_{0}t}\hat{P}_{2}^{-2}\right] & -e^{i2\tau_{0}t}\hat{P}_{2}^{-2}-e^{-i2\tau_{0}t}\hat{P}_{2}^{2}-\sqrt{\frac{2}{3}}\hat{P}_{2}^{0} & -i\left[e^{i\tau_{0}t}\hat{P}_{2}^{-1}+e^{-i\tau_{0}t}\hat{P}_{2}^{1}\right]\\
e^{i\tau_{0}t}\hat{P}_{2}^{-1}-e^{-i\tau_{0}t}\hat{P}_{2}^{1} & -i\left[e^{i\tau_{0}t}\hat{P}_{2}^{-1}+e^{-i\tau_{0}t}\hat{P}_{2}^{1}\right] & 2\sqrt{\frac{2}{3}}\hat{P}_{2}^{0}
\end{array}
\right)
\end{equation*}
\end{widetext}
where the arguments of the $\hat{P}$'s have been omitted for the sake of writing. 

In the case of free diffusion, case analyzed in this paper, the dipole and higher multipoles vanish asymptotically with time, leaving the rotationally symmetric monopole, however this would not be the case if the particle diffuses under the influence of velocity-dependent forces. One situation of interest correspond when the particles are under the effects of polar or nematic aligning forces \cite{HancockPRE2015}.


\begin{thebibliography}{96}
\expandafter\ifx\csname natexlab\endcsname\relax\def\natexlab#1{#1}\fi
\expandafter\ifx\csname bibnamefont\endcsname\relax
  \def\bibnamefont#1{#1}\fi
\expandafter\ifx\csname bibfnamefont\endcsname\relax
  \def\bibfnamefont#1{#1}\fi
\expandafter\ifx\csname citenamefont\endcsname\relax
  \def\citenamefont#1{#1}\fi
\expandafter\ifx\csname url\endcsname\relax
  \def\url#1{\texttt{#1}}\fi
\expandafter\ifx\csname urlprefix\endcsname\relax\def\urlprefix{URL }\fi
\providecommand{\bibinfo}[2]{#2}
\providecommand{\eprint}[2][]{\url{#2}}

\bibitem[{\citenamefont{Golestanian et~al.}(2007)\citenamefont{Golestanian,
  Liverpool, and Ajdari}}]{GolestanianNJP2007}
\bibinfo{author}{\bibfnamefont{R.}~\bibnamefont{Golestanian}},
  \bibinfo{author}{\bibfnamefont{T.~B.} \bibnamefont{Liverpool}},
  \bibnamefont{and} \bibinfo{author}{\bibfnamefont{A.}~\bibnamefont{Ajdari}},
  \bibinfo{journal}{New Journal of Physics} \textbf{\bibinfo{volume}{9}},
  \bibinfo{pages}{126} (\bibinfo{year}{2007}),
  \urlprefix\url{http://stacks.iop.org/1367-2630/9/i=5/a=126}.

\bibitem[{\citenamefont{Abbott et~al.}(2009)\citenamefont{Abbott, Peyer,
  Lagomarsino, Zhang, Dong, Kaliakatsos, and Nelson}}]{abbot}
\bibinfo{author}{\bibfnamefont{J.~J.} \bibnamefont{Abbott}},
  \bibinfo{author}{\bibfnamefont{K.~E.} \bibnamefont{Peyer}},
  \bibinfo{author}{\bibfnamefont{M.~C.} \bibnamefont{Lagomarsino}},
  \bibinfo{author}{\bibfnamefont{L.}~\bibnamefont{Zhang}},
  \bibinfo{author}{\bibfnamefont{L.}~\bibnamefont{Dong}},
  \bibinfo{author}{\bibfnamefont{I.~K.} \bibnamefont{Kaliakatsos}},
  \bibnamefont{and} \bibinfo{author}{\bibfnamefont{B.~J.}
  \bibnamefont{Nelson}}, \bibinfo{journal}{Int. J. Robot. Res.}
  \textbf{\bibinfo{volume}{28}}, \bibinfo{pages}{1434} (\bibinfo{year}{2009}).

\bibitem[{\citenamefont{Mirkovic et~al.}(2010)\citenamefont{Mirkovic, Zacharia,
  Scholes, and Ozin}}]{MirkovicACSNano2010}
\bibinfo{author}{\bibfnamefont{T.}~\bibnamefont{Mirkovic}},
  \bibinfo{author}{\bibfnamefont{N.~S.} \bibnamefont{Zacharia}},
  \bibinfo{author}{\bibfnamefont{G.~D.} \bibnamefont{Scholes}},
  \bibnamefont{and} \bibinfo{author}{\bibfnamefont{G.~A.} \bibnamefont{Ozin}},
  \bibinfo{journal}{ACS Nano} \textbf{\bibinfo{volume}{4}},
  \bibinfo{pages}{1782} (\bibinfo{year}{2010}), \bibinfo{note}{pMID: 20420469},
  \eprint{http://dx.doi.org/10.1021/nn100669h},
  \urlprefix\url{http://dx.doi.org/10.1021/nn100669h}.

\bibitem[{\citenamefont{{Sanchez Tim} et~al.}(2012)\citenamefont{{Sanchez Tim},
  {Chen Daniel T. N.}, {DeCamp Stephen J.}, {Heymann Michael}, and {Dogic
  Zvonimir}}}]{SanchezNature2012}
\bibinfo{author}{\bibnamefont{{Sanchez Tim}}},
  \bibinfo{author}{\bibnamefont{{Chen Daniel T. N.}}},
  \bibinfo{author}{\bibnamefont{{DeCamp Stephen J.}}},
  \bibinfo{author}{\bibnamefont{{Heymann Michael}}}, \bibnamefont{and}
  \bibinfo{author}{\bibnamefont{{Dogic Zvonimir}}}, \bibinfo{journal}{Nature}
  \textbf{\bibinfo{volume}{491}}, \bibinfo{pages}{431} (\bibinfo{year}{2012}),
  ISSN \bibinfo{issn}{0028-0836}, \bibinfo{note}{10.1038/nature11591},
  \urlprefix\url{http://www.nature.com/nature/journal/v491/n7424/abs/nature11591.html\#supplementary-information}.

\bibitem[{\citenamefont{Kosa et~al.}(2012)\citenamefont{Kosa, Jakab, Szekely,
  and Hata}}]{kosa}
\bibinfo{author}{\bibfnamefont{G.}~\bibnamefont{Kosa}},
  \bibinfo{author}{\bibfnamefont{P.}~\bibnamefont{Jakab}},
  \bibinfo{author}{\bibfnamefont{G.}~\bibnamefont{Szekely}}, \bibnamefont{and}
  \bibinfo{author}{\bibfnamefont{N.}~\bibnamefont{Hata}},
  \bibinfo{journal}{Biomed. Microdevices} \textbf{\bibinfo{volume}{14}},
  \bibinfo{pages}{165} (\bibinfo{year}{2012}).

\bibitem[{\citenamefont{Soto and Golestanian}(2014)}]{SotoPRL2014}
\bibinfo{author}{\bibfnamefont{R.}~\bibnamefont{Soto}} \bibnamefont{and}
  \bibinfo{author}{\bibfnamefont{R.}~\bibnamefont{Golestanian}},
  \bibinfo{journal}{Phys. Rev. Lett.} \textbf{\bibinfo{volume}{112}},
  \bibinfo{pages}{068301} (\bibinfo{year}{2014}),
  \urlprefix\url{http://link.aps.org/doi/10.1103/PhysRevLett.112.068301}.

\bibitem[{\citenamefont{Gao et~al.}(2014)\citenamefont{Gao, Dong,
  Thamphiwatana, Li, Gao, Zhang, and Wang}}]{GaoNano2014}
\bibinfo{author}{\bibfnamefont{W.}~\bibnamefont{Gao}},
  \bibinfo{author}{\bibfnamefont{R.}~\bibnamefont{Dong}},
  \bibinfo{author}{\bibfnamefont{S.}~\bibnamefont{Thamphiwatana}},
  \bibinfo{author}{\bibfnamefont{J.}~\bibnamefont{Li}},
  \bibinfo{author}{\bibfnamefont{W.}~\bibnamefont{Gao}},
  \bibinfo{author}{\bibfnamefont{L.}~\bibnamefont{Zhang}}, \bibnamefont{and}
  \bibinfo{author}{\bibfnamefont{J.}~\bibnamefont{Wang}}, \bibinfo{journal}{ACS
  nano}  (\bibinfo{year}{2014}).

\bibitem[{\citenamefont{Cates}(2012)}]{CatesRepProgPhys2012}
\bibinfo{author}{\bibfnamefont{M.~E.} \bibnamefont{Cates}},
  \bibinfo{journal}{Reports on Progress in Physics}
  \textbf{\bibinfo{volume}{75}}, \bibinfo{pages}{042601}
  (\bibinfo{year}{2012}),
  \urlprefix\url{http://stacks.iop.org/0034-4885/75/i=4/a=042601}.

\bibitem[{\citenamefont{Dossetti and Sevilla}(2015)}]{DossettiPRL2015}
\bibinfo{author}{\bibfnamefont{V.}~\bibnamefont{Dossetti}} \bibnamefont{and}
  \bibinfo{author}{\bibfnamefont{F.~J.} \bibnamefont{Sevilla}},
  \bibinfo{journal}{Phys. Rev. Lett.} \textbf{\bibinfo{volume}{115}},
  \bibinfo{pages}{058301} (\bibinfo{year}{2015}),
  \urlprefix\url{http://link.aps.org/doi/10.1103/PhysRevLett.115.058301}.

\bibitem[{\citenamefont{Bazazi et~al.}(2008)\citenamefont{Bazazi, Buhl, Hale,
  Anstey, Sword, Simpson, and Couzin}}]{BazaziCurrBio2008}
\bibinfo{author}{\bibfnamefont{S.}~\bibnamefont{Bazazi}},
  \bibinfo{author}{\bibfnamefont{J.}~\bibnamefont{Buhl}},
  \bibinfo{author}{\bibfnamefont{J.~J.} \bibnamefont{Hale}},
  \bibinfo{author}{\bibfnamefont{M.~L.} \bibnamefont{Anstey}},
  \bibinfo{author}{\bibfnamefont{G.~A.} \bibnamefont{Sword}},
  \bibinfo{author}{\bibfnamefont{S.~J.} \bibnamefont{Simpson}},
  \bibnamefont{and} \bibinfo{author}{\bibfnamefont{I.~D.}
  \bibnamefont{Couzin}}, \bibinfo{journal}{Current Biology}
  \textbf{\bibinfo{volume}{18}}, \bibinfo{pages}{735} (\bibinfo{year}{2008}).

\bibitem[{\citenamefont{Bazazi et~al.}(2011)\citenamefont{Bazazi, Romanczuk,
  Thomas, Schimansky-Geier, Hale, Miller, Sword, Simpson, and
  Couzin}}]{BazaziProcRSocLondon2011}
\bibinfo{author}{\bibfnamefont{S.}~\bibnamefont{Bazazi}},
  \bibinfo{author}{\bibfnamefont{P.}~\bibnamefont{Romanczuk}},
  \bibinfo{author}{\bibfnamefont{S.}~\bibnamefont{Thomas}},
  \bibinfo{author}{\bibfnamefont{L.}~\bibnamefont{Schimansky-Geier}},
  \bibinfo{author}{\bibfnamefont{J.~J.} \bibnamefont{Hale}},
  \bibinfo{author}{\bibfnamefont{G.~A.} \bibnamefont{Miller}},
  \bibinfo{author}{\bibfnamefont{G.~A.} \bibnamefont{Sword}},
  \bibinfo{author}{\bibfnamefont{S.~J.} \bibnamefont{Simpson}},
  \bibnamefont{and} \bibinfo{author}{\bibfnamefont{I.~D.}
  \bibnamefont{Couzin}}, \bibinfo{journal}{Proceedings of the Royal Society B:
  Biological Sciences} \textbf{\bibinfo{volume}{278}}, \bibinfo{pages}{356}
  (\bibinfo{year}{2011}).

\bibitem[{\citenamefont{B{\"o}deker et~al.}(2010)\citenamefont{B{\"o}deker,
  Beta, Frank, and Bodenschatz}}]{BodekerELett2010}
\bibinfo{author}{\bibfnamefont{H.~U.} \bibnamefont{B{\"o}deker}},
  \bibinfo{author}{\bibfnamefont{C.}~\bibnamefont{Beta}},
  \bibinfo{author}{\bibfnamefont{T.~D.} \bibnamefont{Frank}}, \bibnamefont{and}
  \bibinfo{author}{\bibfnamefont{E.}~\bibnamefont{Bodenschatz}},
  \bibinfo{journal}{EPL (Europhysics Letters)} \textbf{\bibinfo{volume}{90}},
  \bibinfo{pages}{28005} (\bibinfo{year}{2010}).

\bibitem[{\citenamefont{Edwards et~al.}(2007)\citenamefont{Edwards, Phillips,
  Watkins, Freeman, Murphy, Afanasyev, Buldyrev, da~Luz, Raposo, Stanley
  et~al.}}]{EdwardsNature2007}
\bibinfo{author}{\bibfnamefont{A.~M.} \bibnamefont{Edwards}},
  \bibinfo{author}{\bibfnamefont{R.~A.} \bibnamefont{Phillips}},
  \bibinfo{author}{\bibfnamefont{N.~W.} \bibnamefont{Watkins}},
  \bibinfo{author}{\bibfnamefont{M.~P.} \bibnamefont{Freeman}},
  \bibinfo{author}{\bibfnamefont{E.~J.} \bibnamefont{Murphy}},
  \bibinfo{author}{\bibfnamefont{V.}~\bibnamefont{Afanasyev}},
  \bibinfo{author}{\bibfnamefont{S.~V.} \bibnamefont{Buldyrev}},
  \bibinfo{author}{\bibfnamefont{M.~G.~E.} \bibnamefont{da~Luz}},
  \bibinfo{author}{\bibfnamefont{E.~P.} \bibnamefont{Raposo}},
  \bibinfo{author}{\bibfnamefont{H.~E.} \bibnamefont{Stanley}},
  \bibnamefont{et~al.}, \bibinfo{journal}{Nature}
  \textbf{\bibinfo{volume}{449}}, \bibinfo{pages}{1044} (\bibinfo{year}{2007}).

\bibitem[{\citenamefont{Gautrais et~al.}(2009)\citenamefont{Gautrais, Jost,
  Soria, Campo, Motsch, Fournier, Blanco, and
  Theraulaz}}]{GautraisJMathBio2009}
\bibinfo{author}{\bibfnamefont{J.}~\bibnamefont{Gautrais}},
  \bibinfo{author}{\bibfnamefont{C.}~\bibnamefont{Jost}},
  \bibinfo{author}{\bibfnamefont{M.}~\bibnamefont{Soria}},
  \bibinfo{author}{\bibfnamefont{A.}~\bibnamefont{Campo}},
  \bibinfo{author}{\bibfnamefont{S.}~\bibnamefont{Motsch}},
  \bibinfo{author}{\bibfnamefont{R.}~\bibnamefont{Fournier}},
  \bibinfo{author}{\bibfnamefont{S.}~\bibnamefont{Blanco}}, \bibnamefont{and}
  \bibinfo{author}{\bibfnamefont{G.}~\bibnamefont{Theraulaz}},
  \bibinfo{journal}{Journal of mathematical biology}
  \textbf{\bibinfo{volume}{58}}, \bibinfo{pages}{429} (\bibinfo{year}{2009}).

\bibitem[{\citenamefont{Li et~al.}(2011)\citenamefont{Li, Cox, and
  Flyvbjerg}}]{LiPhysBiol2011}
\bibinfo{author}{\bibfnamefont{L.}~\bibnamefont{Li}},
  \bibinfo{author}{\bibfnamefont{E.~C.} \bibnamefont{Cox}}, \bibnamefont{and}
  \bibinfo{author}{\bibfnamefont{H.}~\bibnamefont{Flyvbjerg}},
  \bibinfo{journal}{Physical biology} \textbf{\bibinfo{volume}{8}},
  \bibinfo{pages}{046006} (\bibinfo{year}{2011}).

\bibitem[{\citenamefont{Viswanathan et~al.}(2011)\citenamefont{Viswanathan,
  da~Luz, Raposo, and Stanley}}]{ViswanathanBook}
\bibinfo{author}{\bibfnamefont{G.}~\bibnamefont{Viswanathan}},
  \bibinfo{author}{\bibfnamefont{M.}~\bibnamefont{da~Luz}},
  \bibinfo{author}{\bibfnamefont{E.}~\bibnamefont{Raposo}}, \bibnamefont{and}
  \bibinfo{author}{\bibfnamefont{H.}~\bibnamefont{Stanley}},
  \emph{\bibinfo{title}{{The physics of foraging: an introduction to random
  searches and biological encounters}}} (\bibinfo{publisher}{Cambridge
  University Press}, \bibinfo{year}{2011}).

\bibitem[{\citenamefont{Howse et~al.}(2007)\citenamefont{Howse, Jones, Ryan,
  Gough, Vafabakhsh, and Golestanian}}]{HowsePRL2007}
\bibinfo{author}{\bibfnamefont{J.~R.} \bibnamefont{Howse}},
  \bibinfo{author}{\bibfnamefont{R.~A.~L.} \bibnamefont{Jones}},
  \bibinfo{author}{\bibfnamefont{A.~J.} \bibnamefont{Ryan}},
  \bibinfo{author}{\bibfnamefont{T.}~\bibnamefont{Gough}},
  \bibinfo{author}{\bibfnamefont{R.}~\bibnamefont{Vafabakhsh}},
  \bibnamefont{and}
  \bibinfo{author}{\bibfnamefont{R.}~\bibnamefont{Golestanian}},
  \bibinfo{journal}{Phys. Rev. Lett.} \textbf{\bibinfo{volume}{99}},
  \bibinfo{pages}{048102} (\bibinfo{year}{2007}).

\bibitem[{\citenamefont{Jiang et~al.}(2010)\citenamefont{Jiang, Yoshinaga, and
  Sano}}]{JiangPRL2010}
\bibinfo{author}{\bibfnamefont{H.-R.} \bibnamefont{Jiang}},
  \bibinfo{author}{\bibfnamefont{N.}~\bibnamefont{Yoshinaga}},
  \bibnamefont{and} \bibinfo{author}{\bibfnamefont{M.}~\bibnamefont{Sano}},
  \bibinfo{journal}{Phys. Rev. Lett.} \textbf{\bibinfo{volume}{105}},
  \bibinfo{pages}{268302} (\bibinfo{year}{2010}).

\bibitem[{\citenamefont{Schnitzer}(1993)}]{SchnitzerPRE1993}
\bibinfo{author}{\bibfnamefont{M.~J.} \bibnamefont{Schnitzer}},
  \bibinfo{journal}{Phys. Rev. E} \textbf{\bibinfo{volume}{48}},
  \bibinfo{pages}{2553} (\bibinfo{year}{1993}),
  \urlprefix\url{http://link.aps.org/doi/10.1103/PhysRevE.48.2553}.

\bibitem[{\citenamefont{{M. Schienbein} and Gruler}(1993)}]{Schienbein1993}
\bibinfo{author}{\bibnamefont{{M. Schienbein}}} \bibnamefont{and}
  \bibinfo{author}{\bibfnamefont{H.}~\bibnamefont{Gruler}},
  \bibinfo{journal}{Bulletin of Mathematical Biology}
  \textbf{\bibinfo{volume}{55}}, \bibinfo{pages}{585} (\bibinfo{year}{1993}).

\bibitem[{\citenamefont{Bartumeus et~al.}(2008)\citenamefont{Bartumeus,
  Catalan, Viswanathan, Raposo, and da~Luz}}]{BartumeusJTB2008}
\bibinfo{author}{\bibfnamefont{F.}~\bibnamefont{Bartumeus}},
  \bibinfo{author}{\bibfnamefont{J.}~\bibnamefont{Catalan}},
  \bibinfo{author}{\bibfnamefont{G.}~\bibnamefont{Viswanathan}},
  \bibinfo{author}{\bibfnamefont{E.}~\bibnamefont{Raposo}}, \bibnamefont{and}
  \bibinfo{author}{\bibfnamefont{M.}~\bibnamefont{da~Luz}},
  \bibinfo{journal}{Journal of Theoretical Biology}
  \textbf{\bibinfo{volume}{252}}, \bibinfo{pages}{43 } (\bibinfo{year}{2008}),
  ISSN \bibinfo{issn}{0022-5193},
  \urlprefix\url{http://www.sciencedirect.com/science/article/pii/S0022519308000180}.

\bibitem[{\citenamefont{Codling et~al.}(2008)\citenamefont{Codling, Plank, and
  Benhamou}}]{CodlingJRoySocInterface2008}
\bibinfo{author}{\bibfnamefont{E.~A.} \bibnamefont{Codling}},
  \bibinfo{author}{\bibfnamefont{M.~J.} \bibnamefont{Plank}}, \bibnamefont{and}
  \bibinfo{author}{\bibfnamefont{S.}~\bibnamefont{Benhamou}},
  \bibinfo{journal}{Journal of the Royal Society Interface}
  \textbf{\bibinfo{volume}{5}}, \bibinfo{pages}{813} (\bibinfo{year}{2008}).

\bibitem[{\citenamefont{Sevilla and Nava}(2014)}]{SevillaPRE2014}
\bibinfo{author}{\bibfnamefont{F.}~\bibnamefont{Sevilla}} \bibnamefont{and}
  \bibinfo{author}{\bibfnamefont{L.~G.} \bibnamefont{Nava}},
  \bibinfo{journal}{Physical Review E} \textbf{\bibinfo{volume}{90}},
  \bibinfo{pages}{022130} (\bibinfo{year}{2014}).

\bibitem[{\citenamefont{Taktikos et~al.}(2014)\citenamefont{Taktikos, Stark,
  and Zaburdaev}}]{TaktikosPlos2014}
\bibinfo{author}{\bibfnamefont{J.}~\bibnamefont{Taktikos}},
  \bibinfo{author}{\bibfnamefont{H.}~\bibnamefont{Stark}}, \bibnamefont{and}
  \bibinfo{author}{\bibfnamefont{V.}~\bibnamefont{Zaburdaev}},
  \bibinfo{journal}{PLoS ONE} \textbf{\bibinfo{volume}{8}}, \bibinfo{pages}{1}
  (\bibinfo{year}{2014}),
  \urlprefix\url{http://dx.doi.org/10.1371%2Fjournal.pone.0081936}.

\bibitem[{\citenamefont{Sevilla and Sandoval}(2015)}]{SevillaPRE2015}
\bibinfo{author}{\bibfnamefont{F.~J.} \bibnamefont{Sevilla}} \bibnamefont{and}
  \bibinfo{author}{\bibfnamefont{M.}~\bibnamefont{Sandoval}},
  \bibinfo{journal}{Phys. Rev. E} \textbf{\bibinfo{volume}{91}},
  \bibinfo{pages}{052150} (\bibinfo{year}{2015}),
  \urlprefix\url{http://link.aps.org/doi/10.1103/PhysRevE.91.052150}.

\bibitem[{\citenamefont{Berg and Turner}(1990)}]{BergBioPhysJ1990}
\bibinfo{author}{\bibfnamefont{H.}~\bibnamefont{Berg}} \bibnamefont{and}
  \bibinfo{author}{\bibfnamefont{L.}~\bibnamefont{Turner}},
  \bibinfo{journal}{Biophysical Journal} \textbf{\bibinfo{volume}{58}},
  \bibinfo{pages}{919 } (\bibinfo{year}{1990}), ISSN \bibinfo{issn}{0006-3495},
  \urlprefix\url{http://www.sciencedirect.com/science/article/pii/S000634959082436X}.

\bibitem[{\citenamefont{Crenshaw}(1996)}]{CrenshawAmZoo1996}
\bibinfo{author}{\bibfnamefont{H.~C.} \bibnamefont{Crenshaw}},
  \bibinfo{journal}{American Zoologist} \textbf{\bibinfo{volume}{36}},
  \bibinfo{pages}{608} (\bibinfo{year}{1996}), ISSN \bibinfo{issn}{0003-1569},
  \eprint{http://az.oxfordjournals.org/content/36/6/608.full.pdf},
  \urlprefix\url{http://az.oxfordjournals.org/content/36/6/608}.

\bibitem[{\citenamefont{Crenshaw}(1993{\natexlab{a}})}]{Crenshaw1BullMathBio1993}
\bibinfo{author}{\bibfnamefont{H.~C.} \bibnamefont{Crenshaw}},
  \bibinfo{journal}{Bulletin of Mathematical Biology}
  \textbf{\bibinfo{volume}{55}}, \bibinfo{pages}{197 }
  (\bibinfo{year}{1993}{\natexlab{a}}), ISSN \bibinfo{issn}{0092-8240},
  \urlprefix\url{http://www.sciencedirect.com/science/article/pii/S0092824005800692}.

\bibitem[{\citenamefont{Crenshaw and
  Edelstein-Keshet}(1993)}]{Crenshaw2BullMathBio1993}
\bibinfo{author}{\bibfnamefont{H.~C.} \bibnamefont{Crenshaw}} \bibnamefont{and}
  \bibinfo{author}{\bibfnamefont{L.}~\bibnamefont{Edelstein-Keshet}},
  \bibinfo{journal}{Bulletin of Mathematical Biology}
  \textbf{\bibinfo{volume}{55}}, \bibinfo{pages}{213} (\bibinfo{year}{1993}),
  ISSN \bibinfo{issn}{1522-9602},
  \urlprefix\url{http://dx.doi.org/10.1007/BF02460303}.

\bibitem[{\citenamefont{Crenshaw}(1993{\natexlab{b}})}]{Crenshaw3BullMathBio1993}
\bibinfo{author}{\bibfnamefont{H.~C.} \bibnamefont{Crenshaw}},
  \bibinfo{journal}{Bulletin of Mathematical Biology}
  \textbf{\bibinfo{volume}{55}}, \bibinfo{pages}{231}
  (\bibinfo{year}{1993}{\natexlab{b}}), ISSN \bibinfo{issn}{1522-9602},
  \urlprefix\url{http://dx.doi.org/10.1007/BF02460304}.

\bibitem[{\citenamefont{Woolley}(2003)}]{WoolleyReproduction2003}
\bibinfo{author}{\bibfnamefont{D.}~\bibnamefont{Woolley}},
  \bibinfo{journal}{Reproduction} \textbf{\bibinfo{volume}{126}},
  \bibinfo{pages}{259} (\bibinfo{year}{2003}),
  \eprint{http://www.reproduction-online.org/content/126/2/259.full.pdf+html},
  \urlprefix\url{http://www.reproduction-online.org/content/126/2/259.abstract}.

\bibitem[{\citenamefont{DiLuzio et~al.}(2005)\citenamefont{DiLuzio, Turner,
  Mayer, Garstecki, Weibel, Berg, and Whitesides}}]{DiLuzioNature2005}
\bibinfo{author}{\bibfnamefont{W.~R.} \bibnamefont{DiLuzio}},
  \bibinfo{author}{\bibfnamefont{L.}~\bibnamefont{Turner}},
  \bibinfo{author}{\bibfnamefont{M.}~\bibnamefont{Mayer}},
  \bibinfo{author}{\bibfnamefont{P.}~\bibnamefont{Garstecki}},
  \bibinfo{author}{\bibfnamefont{D.~B.} \bibnamefont{Weibel}},
  \bibinfo{author}{\bibfnamefont{H.~C.} \bibnamefont{Berg}}, \bibnamefont{and}
  \bibinfo{author}{\bibfnamefont{G.~M.} \bibnamefont{Whitesides}},
  \bibinfo{journal}{Nature} \textbf{\bibinfo{volume}{435}},
  \bibinfo{pages}{1271} (\bibinfo{year}{2005}).

\bibitem[{\citenamefont{Riedel et~al.}(2005)\citenamefont{Riedel, Kruse, and
  Howard}}]{RiedelScience2005}
\bibinfo{author}{\bibfnamefont{I.~H.} \bibnamefont{Riedel}},
  \bibinfo{author}{\bibfnamefont{K.}~\bibnamefont{Kruse}}, \bibnamefont{and}
  \bibinfo{author}{\bibfnamefont{J.}~\bibnamefont{Howard}},
  \bibinfo{journal}{Science} \textbf{\bibinfo{volume}{309}},
  \bibinfo{pages}{300} (\bibinfo{year}{2005}).

\bibitem[{\citenamefont{Lauga et~al.}(2006)\citenamefont{Lauga, DiLuzio,
  Whitesides, and Stone}}]{LaugaBioPhysJ2006}
\bibinfo{author}{\bibfnamefont{E.}~\bibnamefont{Lauga}},
  \bibinfo{author}{\bibfnamefont{W.~R.} \bibnamefont{DiLuzio}},
  \bibinfo{author}{\bibfnamefont{G.~M.} \bibnamefont{Whitesides}},
  \bibnamefont{and} \bibinfo{author}{\bibfnamefont{H.~A.} \bibnamefont{Stone}},
  \bibinfo{journal}{Biophysical Journal} \textbf{\bibinfo{volume}{90}},
  \bibinfo{pages}{400} (\bibinfo{year}{2006}), ISSN \bibinfo{issn}{0006-3495},
  \urlprefix\url{http://www.sciencedirect.com/science/article/pii/S0006349506722214}.

\bibitem[{\citenamefont{Hill et~al.}(2007)\citenamefont{Hill, Kalkanci,
  McMurry, and Koser}}]{HillPRL2007}
\bibinfo{author}{\bibfnamefont{J.}~\bibnamefont{Hill}},
  \bibinfo{author}{\bibfnamefont{O.}~\bibnamefont{Kalkanci}},
  \bibinfo{author}{\bibfnamefont{J.~L.} \bibnamefont{McMurry}},
  \bibnamefont{and} \bibinfo{author}{\bibfnamefont{H.}~\bibnamefont{Koser}},
  \bibinfo{journal}{Phys. Rev. Lett.} \textbf{\bibinfo{volume}{98}},
  \bibinfo{pages}{068101} (\bibinfo{year}{2007}),
  \urlprefix\url{http://link.aps.org/doi/10.1103/PhysRevLett.98.068101}.

\bibitem[{\citenamefont{Shenoy et~al.}(2007)\citenamefont{Shenoy, Tambe,
  Prasad, and Theriot}}]{ShenoyPNAS2007}
\bibinfo{author}{\bibfnamefont{V.~B.} \bibnamefont{Shenoy}},
  \bibinfo{author}{\bibfnamefont{D.~T.} \bibnamefont{Tambe}},
  \bibinfo{author}{\bibfnamefont{A.}~\bibnamefont{Prasad}}, \bibnamefont{and}
  \bibinfo{author}{\bibfnamefont{J.~A.} \bibnamefont{Theriot}},
  \bibinfo{journal}{Proceedings of the National Academy of Sciences}
  \textbf{\bibinfo{volume}{104}}, \bibinfo{pages}{8229} (\bibinfo{year}{2007}),
  \eprint{http://www.pnas.org/content/104/20/8229.full.pdf},
  \urlprefix\url{http://www.pnas.org/content/104/20/8229.abstract}.

\bibitem[{\citenamefont{Friedrich and J\"ulicher}(2008)}]{FriedrichNJP2008}
\bibinfo{author}{\bibfnamefont{B.~M.} \bibnamefont{Friedrich}}
  \bibnamefont{and}
  \bibinfo{author}{\bibfnamefont{F.}~\bibnamefont{J\"ulicher}},
  \bibinfo{journal}{New Journal of Physics} \textbf{\bibinfo{volume}{10}},
  \bibinfo{pages}{123025} (\bibinfo{year}{2008}),
  \urlprefix\url{http://stacks.iop.org/1367-2630/10/i=12/a=123025}.

\bibitem[{\citenamefont{Friedrich and J\"ulicher}(2009)}]{FriedrichPRL2009}
\bibinfo{author}{\bibfnamefont{B.~M.} \bibnamefont{Friedrich}}
  \bibnamefont{and}
  \bibinfo{author}{\bibfnamefont{F.}~\bibnamefont{J\"ulicher}},
  \bibinfo{journal}{Phys. Rev. Lett.} \textbf{\bibinfo{volume}{103}},
  \bibinfo{pages}{068102} (\bibinfo{year}{2009}),
  \urlprefix\url{http://link.aps.org/doi/10.1103/PhysRevLett.103.068102}.

\bibitem[{\citenamefont{Su et~al.}(2012)\citenamefont{Su, Xue, and
  Ozcan}}]{SuPNAS2012}
\bibinfo{author}{\bibfnamefont{T.-W.} \bibnamefont{Su}},
  \bibinfo{author}{\bibfnamefont{L.}~\bibnamefont{Xue}}, \bibnamefont{and}
  \bibinfo{author}{\bibfnamefont{A.}~\bibnamefont{Ozcan}},
  \bibinfo{journal}{Proc. Natl. Acad. Sci. USA} \textbf{\bibinfo{volume}{109}},
  \bibinfo{pages}{16018} (\bibinfo{year}{2012}),
  \eprint{http://www.pnas.org/content/109/40/16018.full.pdf},
  \urlprefix\url{http://www.pnas.org/content/109/40/16018.abstract}.

\bibitem[{\citenamefont{Nakata et~al.}(1997)\citenamefont{Nakata, Iguchi, Ose,
  Kuboyama, Ishii, and Yoshikawa}}]{NakataLangmuir1997}
\bibinfo{author}{\bibfnamefont{S.}~\bibnamefont{Nakata}},
  \bibinfo{author}{\bibfnamefont{Y.}~\bibnamefont{Iguchi}},
  \bibinfo{author}{\bibfnamefont{S.}~\bibnamefont{Ose}},
  \bibinfo{author}{\bibfnamefont{M.}~\bibnamefont{Kuboyama}},
  \bibinfo{author}{\bibfnamefont{T.}~\bibnamefont{Ishii}}, \bibnamefont{and}
  \bibinfo{author}{\bibfnamefont{K.}~\bibnamefont{Yoshikawa}},
  \bibinfo{journal}{Langmuir} \textbf{\bibinfo{volume}{13}},
  \bibinfo{pages}{4454} (\bibinfo{year}{1997}),
  \eprint{http://dx.doi.org/10.1021/la970196p},
  \urlprefix\url{http://dx.doi.org/10.1021/la970196p}.

\bibitem[{\citenamefont{Dreyfus et~al.}(2005)\citenamefont{Dreyfus, Baudry,
  Roper, Fermigier, Stone, and Bibette}}]{DreyfusNatureLetters2005}
\bibinfo{author}{\bibfnamefont{R.}~\bibnamefont{Dreyfus}},
  \bibinfo{author}{\bibfnamefont{J.}~\bibnamefont{Baudry}},
  \bibinfo{author}{\bibfnamefont{M.}~\bibnamefont{Roper}},
  \bibinfo{author}{\bibfnamefont{M.}~\bibnamefont{Fermigier}},
  \bibinfo{author}{\bibfnamefont{H.}~\bibnamefont{Stone}}, \bibnamefont{and}
  \bibinfo{author}{\bibfnamefont{J.}~\bibnamefont{Bibette}},
  \bibinfo{journal}{Nature Letters} \textbf{\bibinfo{volume}{437}},
  \bibinfo{pages}{862 } (\bibinfo{year}{2005}).

\bibitem[{\citenamefont{{P. Dhar} et~al.}(2006)\citenamefont{{P. Dhar}, {Th. M.
  Fischer}, {Y. Wang}, {T. E. Mallouk}, {W. F. Paxton}, and
  Sen}}]{DharNanoLet2006}
\bibinfo{author}{\bibnamefont{{P. Dhar}}}, \bibinfo{author}{\bibnamefont{{Th.
  M. Fischer}}}, \bibinfo{author}{\bibnamefont{{Y. Wang}}},
  \bibinfo{author}{\bibnamefont{{T. E. Mallouk}}},
  \bibinfo{author}{\bibnamefont{{W. F. Paxton}}}, \bibnamefont{and}
  \bibinfo{author}{\bibfnamefont{A.}~\bibnamefont{Sen}}, \bibinfo{journal}{Nano
  Letters} \textbf{\bibinfo{volume}{6}}, \bibinfo{pages}{66}
  (\bibinfo{year}{2006}), \bibinfo{note}{pMID: 16402789},
  \eprint{http://dx.doi.org/10.1021/nl052027s},
  \urlprefix\url{http://dx.doi.org/10.1021/nl052027s}.

\bibitem[{\citenamefont{Schmidt et~al.}(2008)\citenamefont{Schmidt, van~der
  Gucht, Biesheuvel, Weinkamer, Helfer, and Fery}}]{SchmidtEBioPhysJ2008}
\bibinfo{author}{\bibfnamefont{S.}~\bibnamefont{Schmidt}},
  \bibinfo{author}{\bibfnamefont{J.}~\bibnamefont{van~der Gucht}},
  \bibinfo{author}{\bibfnamefont{P.~M.} \bibnamefont{Biesheuvel}},
  \bibinfo{author}{\bibfnamefont{R.}~\bibnamefont{Weinkamer}},
  \bibinfo{author}{\bibfnamefont{E.}~\bibnamefont{Helfer}}, \bibnamefont{and}
  \bibinfo{author}{\bibfnamefont{A.}~\bibnamefont{Fery}},
  \bibinfo{journal}{European Biophysics Journal} \textbf{\bibinfo{volume}{37}},
  \bibinfo{pages}{1361} (\bibinfo{year}{2008}), ISSN \bibinfo{issn}{1432-1017},
  \urlprefix\url{http://dx.doi.org/10.1007/s00249-008-0340-x}.

\bibitem[{\citenamefont{Walther and Muller}(2008)}]{WaltherSoftMat2008}
\bibinfo{author}{\bibfnamefont{A.}~\bibnamefont{Walther}} \bibnamefont{and}
  \bibinfo{author}{\bibfnamefont{A.~H.~E.} \bibnamefont{Muller}},
  \bibinfo{journal}{Soft Matter} \textbf{\bibinfo{volume}{4}},
  \bibinfo{pages}{663} (\bibinfo{year}{2008}),
  \urlprefix\url{http://dx.doi.org/10.1039/B718131K}.

\bibitem[{\citenamefont{Marine et~al.}(2013)\citenamefont{Marine, Wheat, Ault,
  and Posner}}]{MarinePRE2013}
\bibinfo{author}{\bibfnamefont{N.~A.} \bibnamefont{Marine}},
  \bibinfo{author}{\bibfnamefont{P.~M.} \bibnamefont{Wheat}},
  \bibinfo{author}{\bibfnamefont{J.}~\bibnamefont{Ault}}, \bibnamefont{and}
  \bibinfo{author}{\bibfnamefont{J.~D.} \bibnamefont{Posner}},
  \bibinfo{journal}{Phys. Rev. E} \textbf{\bibinfo{volume}{87}},
  \bibinfo{pages}{052305} (\bibinfo{year}{2013}),
  \urlprefix\url{http://link.aps.org/doi/10.1103/PhysRevE.87.052305}.

\bibitem[{\citenamefont{Shum et~al.}(2010)\citenamefont{Shum, Gaffney, and
  Smith}}]{ShumPRSocLonA2010}
\bibinfo{author}{\bibfnamefont{H.}~\bibnamefont{Shum}},
  \bibinfo{author}{\bibfnamefont{E.~A.} \bibnamefont{Gaffney}},
  \bibnamefont{and} \bibinfo{author}{\bibfnamefont{D.~J.} \bibnamefont{Smith}},
  \bibinfo{journal}{Proceedings of the Royal Society of London A: Mathematical,
  Physical and Engineering Sciences} \textbf{\bibinfo{volume}{466}},
  \bibinfo{pages}{1725} (\bibinfo{year}{2010}), ISSN \bibinfo{issn}{1364-5021},
  \eprint{http://rspa.royalsocietypublishing.org/content/466/2118/1725.full.pdf},
  \urlprefix\url{http://rspa.royalsocietypublishing.org/content/466/2118/1725}.

\bibitem[{\citenamefont{Leoni and Liverpool}(2010)}]{LeoniEPL2010}
\bibinfo{author}{\bibfnamefont{M.}~\bibnamefont{Leoni}} \bibnamefont{and}
  \bibinfo{author}{\bibfnamefont{T.~B.} \bibnamefont{Liverpool}},
  \bibinfo{journal}{EPL (Europhysics Letters)} \textbf{\bibinfo{volume}{92}},
  \bibinfo{pages}{64004} (\bibinfo{year}{2010}),
  \urlprefix\url{http://stacks.iop.org/0295-5075/92/i=6/a=64004}.

\bibitem[{\citenamefont{Ledesma-Aguilar
  et~al.}(2012)\citenamefont{Ledesma-Aguilar, L{\"o}wen, and
  Yeomans}}]{Ledesma-AguilarEPJE2012}
\bibinfo{author}{\bibfnamefont{R.}~\bibnamefont{Ledesma-Aguilar}},
  \bibinfo{author}{\bibfnamefont{H.}~\bibnamefont{L{\"o}wen}},
  \bibnamefont{and} \bibinfo{author}{\bibfnamefont{J.~M.}
  \bibnamefont{Yeomans}}, \bibinfo{journal}{The European Physical Journal E}
  \textbf{\bibinfo{volume}{35}}, \bibinfo{pages}{1} (\bibinfo{year}{2012}),
  ISSN \bibinfo{issn}{1292-895X},
  \urlprefix\url{http://dx.doi.org/10.1140/epje/i2012-12070-5}.

\bibitem[{\citenamefont{Dunstan et~al.}(2012)\citenamefont{Dunstan, Mi\~o,
  Clement, and Soto}}]{DunstanPhysFluids2012}
\bibinfo{author}{\bibfnamefont{J.}~\bibnamefont{Dunstan}},
  \bibinfo{author}{\bibfnamefont{G.}~\bibnamefont{Mi\~o}},
  \bibinfo{author}{\bibfnamefont{E.}~\bibnamefont{Clement}}, \bibnamefont{and}
  \bibinfo{author}{\bibfnamefont{R.}~\bibnamefont{Soto}},
  \bibinfo{journal}{Physics of Fluids} \textbf{\bibinfo{volume}{24}},
  \bibinfo{eid}{011901} (\bibinfo{year}{2012}),
  \urlprefix\url{http://scitation.aip.org/content/aip/journal/pof2/24/1/10.1063/1.3676245}.

\bibitem[{\citenamefont{van Teeffelen and {L.
  owen}}(2008)}]{VanTeeffelenPRE2008}
\bibinfo{author}{\bibfnamefont{S.}~\bibnamefont{van Teeffelen}}
  \bibnamefont{and} \bibinfo{author}{\bibfnamefont{H.}~\bibnamefont{{L.
  owen}}}, \bibinfo{journal}{Phys. Rev. E} \textbf{\bibinfo{volume}{78}},
  \bibinfo{pages}{020101} (\bibinfo{year}{2008}),
  \urlprefix\url{http://link.aps.org/doi/10.1103/PhysRevE.78.020101}.

\bibitem[{\citenamefont{K\"ummel et~al.}(2013)\citenamefont{K\"ummel, ten
  Hagen, Wittkowski, Buttinoni, Eichhorn, Volpe, L\"owen, and
  Bechinger}}]{KummelPRL2013}
\bibinfo{author}{\bibfnamefont{F.}~\bibnamefont{K\"ummel}},
  \bibinfo{author}{\bibfnamefont{B.}~\bibnamefont{ten Hagen}},
  \bibinfo{author}{\bibfnamefont{R.}~\bibnamefont{Wittkowski}},
  \bibinfo{author}{\bibfnamefont{I.}~\bibnamefont{Buttinoni}},
  \bibinfo{author}{\bibfnamefont{R.}~\bibnamefont{Eichhorn}},
  \bibinfo{author}{\bibfnamefont{G.}~\bibnamefont{Volpe}},
  \bibinfo{author}{\bibfnamefont{H.}~\bibnamefont{L\"owen}}, \bibnamefont{and}
  \bibinfo{author}{\bibfnamefont{C.}~\bibnamefont{Bechinger}},
  \bibinfo{journal}{Phys. Rev. Lett.} \textbf{\bibinfo{volume}{110}},
  \bibinfo{pages}{198302} (\bibinfo{year}{2013}),
  \urlprefix\url{http://link.aps.org/doi/10.1103/PhysRevLett.110.198302}.

\bibitem[{\citenamefont{K\"ummel et~al.}(2014)\citenamefont{K\"ummel, ten
  Hagen, Wittkowski, Takagi, Buttinoni, Eichhorn, Volpe, L\"owen, and
  Bechinger}}]{KummelReplyPRL2014}
\bibinfo{author}{\bibfnamefont{F.}~\bibnamefont{K\"ummel}},
  \bibinfo{author}{\bibfnamefont{B.}~\bibnamefont{ten Hagen}},
  \bibinfo{author}{\bibfnamefont{R.}~\bibnamefont{Wittkowski}},
  \bibinfo{author}{\bibfnamefont{D.}~\bibnamefont{Takagi}},
  \bibinfo{author}{\bibfnamefont{I.}~\bibnamefont{Buttinoni}},
  \bibinfo{author}{\bibfnamefont{R.}~\bibnamefont{Eichhorn}},
  \bibinfo{author}{\bibfnamefont{G.}~\bibnamefont{Volpe}},
  \bibinfo{author}{\bibfnamefont{H.}~\bibnamefont{L\"owen}}, \bibnamefont{and}
  \bibinfo{author}{\bibfnamefont{C.}~\bibnamefont{Bechinger}},
  \bibinfo{journal}{Phys. Rev. Lett.} \textbf{\bibinfo{volume}{113}},
  \bibinfo{pages}{029802} (\bibinfo{year}{2014}),
  \urlprefix\url{http://link.aps.org/doi/10.1103/PhysRevLett.113.029802}.

\bibitem[{\citenamefont{Nourhani et~al.}(2013)\citenamefont{Nourhani, Lammert,
  Borhan, and Crespi}}]{NourhaniPRE2013}
\bibinfo{author}{\bibfnamefont{A.}~\bibnamefont{Nourhani}},
  \bibinfo{author}{\bibfnamefont{P.~E.} \bibnamefont{Lammert}},
  \bibinfo{author}{\bibfnamefont{A.}~\bibnamefont{Borhan}}, \bibnamefont{and}
  \bibinfo{author}{\bibfnamefont{V.~H.} \bibnamefont{Crespi}},
  \bibinfo{journal}{Phys. Rev. E} \textbf{\bibinfo{volume}{87}},
  \bibinfo{pages}{050301} (\bibinfo{year}{2013}),
  \urlprefix\url{http://link.aps.org/doi/10.1103/PhysRevE.87.050301}.

\bibitem[{\citenamefont{Kline et~al.}(2005)\citenamefont{Kline, Paxton,
  Mallouk, and Sen}}]{KlineAnChem2005}
\bibinfo{author}{\bibfnamefont{T.~R.} \bibnamefont{Kline}},
  \bibinfo{author}{\bibfnamefont{W.~F.} \bibnamefont{Paxton}},
  \bibinfo{author}{\bibfnamefont{T.~E.} \bibnamefont{Mallouk}},
  \bibnamefont{and} \bibinfo{author}{\bibfnamefont{A.}~\bibnamefont{Sen}},
  \bibinfo{journal}{Angewandte Chemie} \textbf{\bibinfo{volume}{117}},
  \bibinfo{pages}{754} (\bibinfo{year}{2005}), ISSN \bibinfo{issn}{1521-3757},
  \urlprefix\url{http://dx.doi.org/10.1002/ange.200461890}.

\bibitem[{\citenamefont{Weber et~al.}(2011)\citenamefont{Weber, Radtke,
  Schimansky-Geier, and H{\"a}nggi}}]{WeberPRE2011}
\bibinfo{author}{\bibfnamefont{C.}~\bibnamefont{Weber}},
  \bibinfo{author}{\bibfnamefont{P.~K.} \bibnamefont{Radtke}},
  \bibinfo{author}{\bibfnamefont{L.}~\bibnamefont{Schimansky-Geier}},
  \bibnamefont{and}
  \bibinfo{author}{\bibfnamefont{P.}~\bibnamefont{H{\"a}nggi}},
  \bibinfo{journal}{Physical Review E} \textbf{\bibinfo{volume}{84}},
  \bibinfo{pages}{011132} (\bibinfo{year}{2011}).

\bibitem[{\citenamefont{Weber et~al.}(2012)\citenamefont{Weber, Sokolov, and
  Schimansky-Geier}}]{WeberPRE2012}
\bibinfo{author}{\bibfnamefont{C.}~\bibnamefont{Weber}},
  \bibinfo{author}{\bibfnamefont{I.~M.} \bibnamefont{Sokolov}},
  \bibnamefont{and}
  \bibinfo{author}{\bibfnamefont{L.}~\bibnamefont{Schimansky-Geier}},
  \bibinfo{journal}{Physical Review E} \textbf{\bibinfo{volume}{85}},
  \bibinfo{pages}{052101} (\bibinfo{year}{2012}).

\bibitem[{\citenamefont{Radtke and Schimansky-Geier}(2012)}]{RadtkePRE2012}
\bibinfo{author}{\bibfnamefont{P.~K.} \bibnamefont{Radtke}} \bibnamefont{and}
  \bibinfo{author}{\bibfnamefont{L.}~\bibnamefont{Schimansky-Geier}},
  \bibinfo{journal}{Physical Review E} \textbf{\bibinfo{volume}{85}},
  \bibinfo{pages}{051110} (\bibinfo{year}{2012}).

\bibitem[{\citenamefont{Ao et~al.}(2015)\citenamefont{Ao, Ghosh, Li, Schmid,
  H{\"a}nggi, and Marchesoni}}]{AoEPL2015}
\bibinfo{author}{\bibfnamefont{X.}~\bibnamefont{Ao}},
  \bibinfo{author}{\bibfnamefont{P.~K.} \bibnamefont{Ghosh}},
  \bibinfo{author}{\bibfnamefont{Y.}~\bibnamefont{Li}},
  \bibinfo{author}{\bibfnamefont{G.}~\bibnamefont{Schmid}},
  \bibinfo{author}{\bibfnamefont{P.}~\bibnamefont{H{\"a}nggi}},
  \bibnamefont{and}
  \bibinfo{author}{\bibfnamefont{F.}~\bibnamefont{Marchesoni}},
  \bibinfo{journal}{EPL (Europhysics Letters)} \textbf{\bibinfo{volume}{109}},
  \bibinfo{pages}{10003} (\bibinfo{year}{2015}),
  \urlprefix\url{http://stacks.iop.org/0295-5075/109/i=1/a=10003}.

\bibitem[{\citenamefont{Larralde}(1997)}]{LarraldePRE1997}
\bibinfo{author}{\bibfnamefont{H.}~\bibnamefont{Larralde}},
  \bibinfo{journal}{Phys. Rev. E} \textbf{\bibinfo{volume}{56}},
  \bibinfo{pages}{5004} (\bibinfo{year}{1997}),
  \urlprefix\url{http://link.aps.org/doi/10.1103/PhysRevE.56.5004}.

\bibitem[{\citenamefont{Sandoval}(2013)}]{SandovalPRE2013}
\bibinfo{author}{\bibfnamefont{M.}~\bibnamefont{Sandoval}},
  \bibinfo{journal}{Physical Review E} \textbf{\bibinfo{volume}{87}},
  \bibinfo{pages}{032708} (\bibinfo{year}{2013}).

\bibitem[{\citenamefont{Larralde and Leyvraz}(2015)}]{LarraldeJPhysA2015}
\bibinfo{author}{\bibfnamefont{H.}~\bibnamefont{Larralde}} \bibnamefont{and}
  \bibinfo{author}{\bibfnamefont{F.}~\bibnamefont{Leyvraz}},
  \bibinfo{journal}{Journal of Physics A: Mathematical and Theoretical}
  \textbf{\bibinfo{volume}{48}}, \bibinfo{pages}{265001}
  (\bibinfo{year}{2015}),
  \urlprefix\url{http://stacks.iop.org/1751-8121/48/i=26/a=265001}.

\bibitem[{\citenamefont{Wittkowski and {L. owen}}(2012)}]{WittkowskiPRE2012}
\bibinfo{author}{\bibfnamefont{R.}~\bibnamefont{Wittkowski}} \bibnamefont{and}
  \bibinfo{author}{\bibfnamefont{H.}~\bibnamefont{{L. owen}}},
  \bibinfo{journal}{Phys. Rev. E} \textbf{\bibinfo{volume}{85}},
  \bibinfo{pages}{021406} (\bibinfo{year}{2012}),
  \urlprefix\url{http://link.aps.org/doi/10.1103/PhysRevE.85.021406}.

\bibitem[{\citenamefont{Zheng et~al.}(2013)\citenamefont{Zheng, ten Hagen,
  Kaiser, Wu, Cui, Silber-Li, and Loewen}}]{ZhengPRE2013}
\bibinfo{author}{\bibfnamefont{X.}~\bibnamefont{Zheng}},
  \bibinfo{author}{\bibfnamefont{B.}~\bibnamefont{ten Hagen}},
  \bibinfo{author}{\bibfnamefont{A.}~\bibnamefont{Kaiser}},
  \bibinfo{author}{\bibfnamefont{M.}~\bibnamefont{Wu}},
  \bibinfo{author}{\bibfnamefont{H.}~\bibnamefont{Cui}},
  \bibinfo{author}{\bibfnamefont{Z.}~\bibnamefont{Silber-Li}},
  \bibnamefont{and} \bibinfo{author}{\bibfnamefont{H.}~\bibnamefont{Loewen}},
  \bibinfo{journal}{Physical Review E} \textbf{\bibinfo{volume}{88}},
  \bibinfo{pages}{032304} (\bibinfo{year}{2013}).

\bibitem[{\citenamefont{Takatori et~al.}(2014)\citenamefont{Takatori, Yan, and
  Brady}}]{TakatoriPRL2014}
\bibinfo{author}{\bibfnamefont{S.~C.} \bibnamefont{Takatori}},
  \bibinfo{author}{\bibfnamefont{W.}~\bibnamefont{Yan}}, \bibnamefont{and}
  \bibinfo{author}{\bibfnamefont{J.~F.} \bibnamefont{Brady}},
  \bibinfo{journal}{Phys. Rev. Lett.} \textbf{\bibinfo{volume}{113}},
  \bibinfo{pages}{028103} (\bibinfo{year}{2014}),
  \urlprefix\url{http://link.aps.org/doi/10.1103/PhysRevLett.113.028103}.

\bibitem[{\citenamefont{Romanczuk and
  Schimansky-Geier}(2011)}]{RomanczukPRL2011}
\bibinfo{author}{\bibfnamefont{P.}~\bibnamefont{Romanczuk}} \bibnamefont{and}
  \bibinfo{author}{\bibfnamefont{L.}~\bibnamefont{Schimansky-Geier}},
  \bibinfo{journal}{Phys. Rev. Lett.} \textbf{\bibinfo{volume}{106}},
  \bibinfo{pages}{230601} (\bibinfo{year}{2011}),
  \urlprefix\url{http://link.aps.org/doi/10.1103/PhysRevLett.106.230601}.

\bibitem[{\citenamefont{Gardiner}(1985)}]{GardinerBook}
\bibinfo{author}{\bibfnamefont{C.}~\bibnamefont{Gardiner}}
  (\bibinfo{publisher}{Springer-Verlag}, \bibinfo{year}{1985}),
  \bibinfo{edition}{second edition} ed.

\bibitem[{\citenamefont{Masoliver and Wang}(1995{\natexlab{a}})}]{gang}
\bibinfo{author}{\bibfnamefont{J.}~\bibnamefont{Masoliver}} \bibnamefont{and}
  \bibinfo{author}{\bibfnamefont{K.-G.} \bibnamefont{Wang}},
  \bibinfo{journal}{Phys. Rev. E} \textbf{\bibinfo{volume}{51}},
  \bibinfo{pages}{2987} (\bibinfo{year}{1995}{\natexlab{a}}),
  \urlprefix\url{http://link.aps.org/doi/10.1103/PhysRevE.51.2987}.

\bibitem[{\citenamefont{Horsthemke and Lefever}(1984)}]{HorsthemkeBook}
\bibinfo{author}{\bibfnamefont{W.~W.} \bibnamefont{Horsthemke}}
  \bibnamefont{and} \bibinfo{author}{\bibfnamefont{R.}~\bibnamefont{Lefever}},
  \emph{\bibinfo{title}{Noise-induced transitions : theory and applications in
  physics, chemistry, and biology}}, Springer series in synergetics
  (\bibinfo{publisher}{Springer-Verlag}, \bibinfo{address}{Berlin, New York},
  \bibinfo{year}{1984}), ISBN \bibinfo{isbn}{0-387-11359-2},
  \urlprefix\url{http://opac.inria.fr/record=b1091778}.

\bibitem[{\citenamefont{van Kampen}(1986)}]{vanKampenJStatPhys1986}
\bibinfo{author}{\bibfnamefont{N.~G.} \bibnamefont{van Kampen}},
  \bibinfo{journal}{Journal of Statistical Physics}
  \textbf{\bibinfo{volume}{44}}, \bibinfo{pages}{1} (\bibinfo{year}{1986}),
  ISSN \bibinfo{issn}{1572-9613},
  \urlprefix\url{http://dx.doi.org/10.1007/BF01010902}.

\bibitem[{\citenamefont{Duderstadt and
  Martin}(1979)}]{DuderstadtTransportTheory}
\bibinfo{author}{\bibfnamefont{J.~J.} \bibnamefont{Duderstadt}}
  \bibnamefont{and} \bibinfo{author}{\bibfnamefont{W.~R.}
  \bibnamefont{Martin}}, \emph{\bibinfo{title}{{Transport theory.}}},
  vol.~\bibinfo{volume}{1} (\bibinfo{year}{1979}).

\bibitem[{\citenamefont{Cates and Tailleur}(2013)}]{CatesEPL2013}
\bibinfo{author}{\bibfnamefont{M.~E.} \bibnamefont{Cates}} \bibnamefont{and}
  \bibinfo{author}{\bibfnamefont{J.}~\bibnamefont{Tailleur}},
  \bibinfo{journal}{EPL (Europhysics Letters)} \textbf{\bibinfo{volume}{101}},
  \bibinfo{pages}{20010} (\bibinfo{year}{2013}),
  \urlprefix\url{http://stacks.iop.org/0295-5075/101/i=2/a=20010}.

\bibitem[{\citenamefont{Goldstein}(1951)}]{GoldsteinQJMAM1951}
\bibinfo{author}{\bibfnamefont{S.}~\bibnamefont{Goldstein}},
  \bibinfo{journal}{The Quarterly Journal of Mechanics and Applied Mathematics}
  \textbf{\bibinfo{volume}{4}}, \bibinfo{pages}{129} (\bibinfo{year}{1951}),
  \eprint{http://qjmam.oxfordjournals.org/content/4/2/129.full.pdf+html},
  \urlprefix\url{http://qjmam.oxfordjournals.org/content/4/2/129.abstract}.

\bibitem[{\citenamefont{Bourret}(1960)}]{BourretCJP1960}
\bibinfo{author}{\bibfnamefont{R.}~\bibnamefont{Bourret}},
  \bibinfo{journal}{Canadian Journal of Physics} \textbf{\bibinfo{volume}{38}},
  \bibinfo{pages}{665} (\bibinfo{year}{1960}),
  \eprint{http://dx.doi.org/10.1139/p60-072},
  \urlprefix\url{http://dx.doi.org/10.1139/p60-072}.

\bibitem[{\citenamefont{Bourret}(1961)}]{BourretCJP1961}
\bibinfo{author}{\bibfnamefont{R.~C.} \bibnamefont{Bourret}},
  \bibinfo{journal}{Canadian Journal of Physics} \textbf{\bibinfo{volume}{39}},
  \bibinfo{pages}{133} (\bibinfo{year}{1961}),
  \eprint{http://dx.doi.org/10.1139/p61-010},
  \urlprefix\url{http://dx.doi.org/10.1139/p61-010}.

\bibitem[{\citenamefont{Kenkre and Sevilla}(2007)}]{KenkreSevilla2007}
\bibinfo{author}{\bibfnamefont{V.}~\bibnamefont{Kenkre}} \bibnamefont{and}
  \bibinfo{author}{\bibfnamefont{F.~J.} \bibnamefont{Sevilla}}, in
  \emph{\bibinfo{booktitle}{{Contributions to Mathematical Physics: a Tribute
  to Gerard G. Emch TS. Ali, KB. Sinha, eds.}}} (\bibinfo{publisher}{{Hindustan
  Book Agency, New Delhi}}, \bibinfo{year}{2007}), pp.
  \bibinfo{pages}{147--160}.

\bibitem[{\citenamefont{Barrow}(1983)}]{BarrowPhilTransRoySocLond1983}
\bibinfo{author}{\bibfnamefont{J.~D.} \bibnamefont{Barrow}},
  \bibinfo{journal}{Philosophical Transactions of the Royal Society of London.
  Series A, Mathematical and Physical Sciences} \textbf{\bibinfo{volume}{310}},
  \bibinfo{pages}{337} (\bibinfo{year}{1983}).

\bibitem[{\citenamefont{Tailleur and Cates}(2009)}]{TailleurEPL2009}
\bibinfo{author}{\bibfnamefont{J.}~\bibnamefont{Tailleur}} \bibnamefont{and}
  \bibinfo{author}{\bibfnamefont{M.~E.} \bibnamefont{Cates}},
  \bibinfo{journal}{EPL (Europhysics Letters)} \textbf{\bibinfo{volume}{86}},
  \bibinfo{pages}{60002} (\bibinfo{year}{2009}),
  \urlprefix\url{http://stacks.iop.org/0295-5075/86/i=6/a=60002}.

\bibitem[{\citenamefont{Enculescu and Stark}(2011)}]{encu}
\bibinfo{author}{\bibfnamefont{M.}~\bibnamefont{Enculescu}} \bibnamefont{and}
  \bibinfo{author}{\bibfnamefont{H.}~\bibnamefont{Stark}},
  \bibinfo{journal}{Phys. Rev. Lett.} \textbf{\bibinfo{volume}{107}},
  \bibinfo{pages}{058301} (\bibinfo{year}{2011}),
  \urlprefix\url{http://link.aps.org/doi/10.1103/PhysRevLett.107.058301}.

\bibitem[{\citenamefont{Palacci et~al.}(2010)\citenamefont{Palacci,
  Cottin-Bizonne, Ybert, and Bocquet}}]{PalacciPRL2010}
\bibinfo{author}{\bibfnamefont{J.}~\bibnamefont{Palacci}},
  \bibinfo{author}{\bibfnamefont{C.}~\bibnamefont{Cottin-Bizonne}},
  \bibinfo{author}{\bibfnamefont{C.}~\bibnamefont{Ybert}}, \bibnamefont{and}
  \bibinfo{author}{\bibfnamefont{L.}~\bibnamefont{Bocquet}},
  \bibinfo{journal}{Phys. Rev. Lett.} \textbf{\bibinfo{volume}{105}},
  \bibinfo{pages}{088304} (\bibinfo{year}{2010}).

\bibitem[{\citenamefont{Maggi et~al.}(2013)\citenamefont{Maggi, Lepore, Solari,
  Rizzo, and Di~Leonardo}}]{MaggiSoftMatt2013}
\bibinfo{author}{\bibfnamefont{C.}~\bibnamefont{Maggi}},
  \bibinfo{author}{\bibfnamefont{A.}~\bibnamefont{Lepore}},
  \bibinfo{author}{\bibfnamefont{J.}~\bibnamefont{Solari}},
  \bibinfo{author}{\bibfnamefont{A.}~\bibnamefont{Rizzo}}, \bibnamefont{and}
  \bibinfo{author}{\bibfnamefont{R.}~\bibnamefont{Di~Leonardo}},
  \bibinfo{journal}{Soft Matter} \textbf{\bibinfo{volume}{9}},
  \bibinfo{pages}{10885} (\bibinfo{year}{2013}),
  \urlprefix\url{http://dx.doi.org/10.1039/C3SM51223A}.

\bibitem[{\citenamefont{Bechinger et~al.}(2016)\citenamefont{Bechinger,
  Di~Leonardo, L{\"o}wen, Reichhardt, Volpe, and Volpe}}]{BechingerArXiv2016}
\bibinfo{author}{\bibfnamefont{C.}~\bibnamefont{Bechinger}},
  \bibinfo{author}{\bibfnamefont{R.}~\bibnamefont{Di~Leonardo}},
  \bibinfo{author}{\bibfnamefont{H.}~\bibnamefont{L{\"o}wen}},
  \bibinfo{author}{\bibfnamefont{C.}~\bibnamefont{Reichhardt}},
  \bibinfo{author}{\bibfnamefont{G.}~\bibnamefont{Volpe}}, \bibnamefont{and}
  \bibinfo{author}{\bibfnamefont{G.}~\bibnamefont{Volpe}},
  \bibinfo{journal}{arXiv preprint arXiv:1602.00081}  (\bibinfo{year}{2016}).

\bibitem[{\citenamefont{Mardia}(1974)}]{Mardia74p115}
\bibinfo{author}{\bibfnamefont{K.~V.} \bibnamefont{Mardia}},
  \bibinfo{journal}{Sankhy{\=a}: The Indian Journal of Statistics, Series B}
  pp. \bibinfo{pages}{115--128} (\bibinfo{year}{1974}).

\bibitem[{\citenamefont{Ordemann et~al.}(2003)\citenamefont{Ordemann, Balazsi,
  and Moss}}]{OrdemannPhysicaA2003}
\bibinfo{author}{\bibfnamefont{A.}~\bibnamefont{Ordemann}},
  \bibinfo{author}{\bibfnamefont{G.}~\bibnamefont{Balazsi}}, \bibnamefont{and}
  \bibinfo{author}{\bibfnamefont{F.}~\bibnamefont{Moss}},
  \bibinfo{journal}{Physica A: Statistical Mechanics and its Applications}
  \textbf{\bibinfo{volume}{325}}, \bibinfo{pages}{260 } (\bibinfo{year}{2003}),
  ISSN \bibinfo{issn}{0378-4371}, \bibinfo{note}{stochastic Systems: From
  Randomness to Complexity},
  \urlprefix\url{http://www.sciencedirect.com/science/article/pii/S0378437103002048}.

\bibitem[{\citenamefont{Porra et~al.}(1997)\citenamefont{Porra, Masoliver, and
  Weiss}}]{PorraPRE1997}
\bibinfo{author}{\bibfnamefont{J.~M.} \bibnamefont{Porra}},
  \bibinfo{author}{\bibfnamefont{J.}~\bibnamefont{Masoliver}},
  \bibnamefont{and} \bibinfo{author}{\bibfnamefont{G.~H.} \bibnamefont{Weiss}},
  \bibinfo{journal}{Physical Review E} \textbf{\bibinfo{volume}{55}},
  \bibinfo{pages}{7771} (\bibinfo{year}{1997}).

\bibitem[{\citenamefont{Ebbens et~al.}(2010)\citenamefont{Ebbens, Jones, Ryan,
  Golestanian, and Howse}}]{EbbensPRE2010}
\bibinfo{author}{\bibfnamefont{S.}~\bibnamefont{Ebbens}},
  \bibinfo{author}{\bibfnamefont{R.~A.~L.} \bibnamefont{Jones}},
  \bibinfo{author}{\bibfnamefont{A.~J.} \bibnamefont{Ryan}},
  \bibinfo{author}{\bibfnamefont{R.}~\bibnamefont{Golestanian}},
  \bibnamefont{and} \bibinfo{author}{\bibfnamefont{J.~R.} \bibnamefont{Howse}},
  \bibinfo{journal}{Phys. Rev. E} \textbf{\bibinfo{volume}{82}},
  \bibinfo{pages}{015304} (\bibinfo{year}{2010}),
  \urlprefix\url{http://link.aps.org/doi/10.1103/PhysRevE.82.015304}.

\bibitem[{\citenamefont{ten Hagen et~al.}(2011)\citenamefont{ten Hagen, van
  Teeffelen, and Lowen}}]{hagen}
\bibinfo{author}{\bibfnamefont{B.}~\bibnamefont{ten Hagen}},
  \bibinfo{author}{\bibfnamefont{S.}~\bibnamefont{van Teeffelen}},
  \bibnamefont{and} \bibinfo{author}{\bibfnamefont{H.}~\bibnamefont{Lowen}},
  \bibinfo{journal}{J. Phys. Condens. Matter} \textbf{\bibinfo{volume}{23}},
  \bibinfo{pages}{194119} (\bibinfo{year}{2011}).

\bibitem[{\citenamefont{Wang et~al.}(2009)\citenamefont{Wang, Anthony, Bae, and
  Granick}}]{WangPNAS2009}
\bibinfo{author}{\bibfnamefont{B.}~\bibnamefont{Wang}},
  \bibinfo{author}{\bibfnamefont{S.~M.} \bibnamefont{Anthony}},
  \bibinfo{author}{\bibfnamefont{S.~C.} \bibnamefont{Bae}}, \bibnamefont{and}
  \bibinfo{author}{\bibfnamefont{S.}~\bibnamefont{Granick}},
  \bibinfo{journal}{Proceedings of the National Academy of Sciences}
  \textbf{\bibinfo{volume}{106}}, \bibinfo{pages}{15160}
  (\bibinfo{year}{2009}),
  \eprint{http://www.pnas.org/content/106/36/15160.full.pdf},
  \urlprefix\url{http://www.pnas.org/content/106/36/15160.abstract}.

\bibitem[{\citenamefont{{Wang Bo} et~al.}(2012)\citenamefont{{Wang Bo}, {Kuo
  James}, {Bae Sung Chul}, and {Granick Steve}}}]{WangNatureMat2012}
\bibinfo{author}{\bibnamefont{{Wang Bo}}}, \bibinfo{author}{\bibnamefont{{Kuo
  James}}}, \bibinfo{author}{\bibnamefont{{Bae Sung Chul}}}, \bibnamefont{and}
  \bibinfo{author}{\bibnamefont{{Granick Steve}}}, \bibinfo{journal}{Nat Mater}
  \textbf{\bibinfo{volume}{11}}, \bibinfo{pages}{481} (\bibinfo{year}{2012}),
  ISSN \bibinfo{issn}{1476-1122}, \bibinfo{note}{10.1038/nmat3308}.

\bibitem[{\citenamefont{Bhattacharya et~al.}(2013)\citenamefont{Bhattacharya,
  Sharma, Saurabh, De, Sain, Nandi, and Chowdhury}}]{BhattacharyaJPCB2013}
\bibinfo{author}{\bibfnamefont{S.}~\bibnamefont{Bhattacharya}},
  \bibinfo{author}{\bibfnamefont{D.~K.} \bibnamefont{Sharma}},
  \bibinfo{author}{\bibfnamefont{S.}~\bibnamefont{Saurabh}},
  \bibinfo{author}{\bibfnamefont{S.}~\bibnamefont{De}},
  \bibinfo{author}{\bibfnamefont{A.}~\bibnamefont{Sain}},
  \bibinfo{author}{\bibfnamefont{A.}~\bibnamefont{Nandi}}, \bibnamefont{and}
  \bibinfo{author}{\bibfnamefont{A.}~\bibnamefont{Chowdhury}},
  \bibinfo{journal}{The Journal of Physical Chemistry B}
  \textbf{\bibinfo{volume}{117}}, \bibinfo{pages}{7771} (\bibinfo{year}{2013}),
  \bibinfo{note}{pMID: 23777572}, \eprint{http://dx.doi.org/10.1021/jp401704e},
  \urlprefix\url{http://dx.doi.org/10.1021/jp401704e}.

\bibitem[{\citenamefont{Cressoni et~al.}(2012)\citenamefont{Cressoni,
  Viswanathan, Ferreira, and da~Silva}}]{CressoniPRE2012}
\bibinfo{author}{\bibfnamefont{J.~C.} \bibnamefont{Cressoni}},
  \bibinfo{author}{\bibfnamefont{G.~M.} \bibnamefont{Viswanathan}},
  \bibinfo{author}{\bibfnamefont{A.~S.} \bibnamefont{Ferreira}},
  \bibnamefont{and} \bibinfo{author}{\bibfnamefont{M.~A.~A.}
  \bibnamefont{da~Silva}}, \bibinfo{journal}{Phys. Rev. E}
  \textbf{\bibinfo{volume}{86}}, \bibinfo{pages}{022103}
  (\bibinfo{year}{2012}),
  \urlprefix\url{http://link.aps.org/doi/10.1103/PhysRevE.86.022103}.

\bibitem[{\citenamefont{Chubynsky and Slater}(2014)}]{ChubynskyPRL2014}
\bibinfo{author}{\bibfnamefont{M.~V.} \bibnamefont{Chubynsky}}
  \bibnamefont{and} \bibinfo{author}{\bibfnamefont{G.~W.}
  \bibnamefont{Slater}}, \bibinfo{journal}{Phys. Rev. Lett.}
  \textbf{\bibinfo{volume}{113}}, \bibinfo{pages}{098302}
  (\bibinfo{year}{2014}),
  \urlprefix\url{http://link.aps.org/doi/10.1103/PhysRevLett.113.098302}.

\bibitem[{\citenamefont{Wang et~al.}(2016)\citenamefont{Wang, Zhang, and
  Zhao}}]{WangPRE2016}
\bibinfo{author}{\bibfnamefont{J.}~\bibnamefont{Wang}},
  \bibinfo{author}{\bibfnamefont{Y.}~\bibnamefont{Zhang}}, \bibnamefont{and}
  \bibinfo{author}{\bibfnamefont{H.}~\bibnamefont{Zhao}},
  \bibinfo{journal}{Phys. Rev. E} \textbf{\bibinfo{volume}{93}},
  \bibinfo{pages}{032144} (\bibinfo{year}{2016}),
  \urlprefix\url{http://link.aps.org/doi/10.1103/PhysRevE.93.032144}.

\bibitem[{\citenamefont{Brillinger}(2012)}]{Brillinger}
\bibinfo{author}{\bibfnamefont{D.}~\bibnamefont{Brillinger}}, in
  \emph{\bibinfo{booktitle}{{Selected Works of David Brillinger}}}, edited by
  \bibinfo{editor}{\bibfnamefont{P.}~\bibnamefont{Guttorp}} \bibnamefont{and}
  \bibinfo{editor}{\bibfnamefont{D.}~\bibnamefont{Brillinger}}
  (\bibinfo{publisher}{Springer New York}, \bibinfo{year}{2012}), {Selected
  Works in Probability and Statistics}, pp. \bibinfo{pages}{73--87}, ISBN
  \bibinfo{isbn}{978-1-4614-1343-1},
  \urlprefix\url{http://dx.doi.org/10.1007/978-1-4614-1344-8_7}.

\bibitem[{\citenamefont{Masoliver and
  Wang}(1995{\natexlab{b}})}]{MasoliverPRE1995}
\bibinfo{author}{\bibfnamefont{J.}~\bibnamefont{Masoliver}} \bibnamefont{and}
  \bibinfo{author}{\bibfnamefont{K.-G.} \bibnamefont{Wang}},
  \bibinfo{journal}{Phys. Rev. E} \textbf{\bibinfo{volume}{51}},
  \bibinfo{pages}{2987} (\bibinfo{year}{1995}{\natexlab{b}}),
  \urlprefix\url{http://link.aps.org/doi/10.1103/PhysRevE.51.2987}.

\bibitem[{\citenamefont{Frank}(2005)}]{frank}
\bibinfo{author}{\bibfnamefont{T.~D.} \bibnamefont{Frank}},
  \bibinfo{journal}{Phys. Rev. E} \textbf{\bibinfo{volume}{72}},
  \bibinfo{pages}{011112} (\bibinfo{year}{2005}),
  \urlprefix\url{http://link.aps.org/doi/10.1103/PhysRevE.72.011112}.

\bibitem[{\citenamefont{Hancock and Baskaran}(2015)}]{HancockPRE2015}
\bibinfo{author}{\bibfnamefont{B.}~\bibnamefont{Hancock}} \bibnamefont{and}
  \bibinfo{author}{\bibfnamefont{A.}~\bibnamefont{Baskaran}},
  \bibinfo{journal}{Phys. Rev. E} \textbf{\bibinfo{volume}{92}},
  \bibinfo{pages}{052143} (\bibinfo{year}{2015}),
  \urlprefix\url{http://link.aps.org/doi/10.1103/PhysRevE.92.052143}.

\end{thebibliography}

\providecommand{\noopsort}[1]{}\providecommand{\singleletter}[1]{#1}%

\end{document}